\documentclass[%
  journal=jctcce,%
  manuscript=article,%
  layout=twocolumn,%
]{achemso}
\setkeys{acs}{maxauthors=10, etalmode=truncate}

\usepackage[T1]{fontenc} % Use modern font encodings
\usepackage{graphicx}
\usepackage{amssymb}
\usepackage{booktabs}
\usepackage{array}
\usepackage{mathtools}
\usepackage[range-phrase=\text{--}, range-units=single]{siunitx}
\DeclareSIUnit\angstrom{\text{Å}}
\usepackage{comment}

\usepackage{caption}
\usepackage{ifthen}
\makeatletter
\newcommand{\isSingleColumn}{\equal{\acs@layout}{traditional}}
\newcommand{\isSuppinfo}{\equal{\acs@manuscript}{suppinfo}}

\ifthenelse{\isSingleColumn}%
    {% if single column
        \usepackage{setspace}
        \AtBeginDocument{\singlespacing}

        \renewcommand{\maketitle}{\section*{\@title}}
        \usepackage[fontsize=11pt]{fontsize}
    }{% else (double column)
	    \usepackage[fontsize=9pt]{fontsize}
    }%

\ifthenelse{\isSuppinfo}%
  {%
  	\setlength{\bibsep}{4pt plus 0.3ex}
  	\DeclareCaptionTextFormat{baseline}{\baselineskip 4.5mm #1}
  }{%
    \setlength{\bibsep}{0pt plus 0.3ex}%
    
    \DeclareCaptionTextFormat{baseline}{\baselineskip 9.5pt #1}
  }%

\makeatother

\newcommand{\dir}{../Bib}
\newcommand{\dirfig}{./Figures}
\newcommand{\vr}{{\mbox{\boldmath $r$}}}
\usepackage[usenames]{color}

\usepackage[normalem]{ulem}

\newcommand{\SMTabContacts}{S1}
\newcommand{\SMTabTransMatrix}{S2}
\newcommand{\SMTabPaths}{S3}

\newcommand{\SMcoopTraj}{S1}
\newcommand{\SMdihedrals}{S2}
\newcommand{\SMPCA}{S3}
\newcommand{\SMclustering}{S4}
\newcommand{\SMCKtest}{S5}
\newcommand{\SMPreviousModels}{S6}
\newcommand{\SMfiltering}{S7}
\author{Daniel Nagel}
\author{Sofia Sartore}
\author{Gerhard Stock}
\email{stock@physik.uni-freiburg.de}
\affiliation{Biomolecular Dynamics, Institute of Physics,
	University of Freiburg, 79104 Freiburg, Germany.}
\date{\today}

\title[Selecting Features for Markov Modeling: A Case Study on HP35]%
  {Selecting Features for Markov Modeling: A Case Study on HP35}

\begin{document}

\begin{abstract}
  Markov state models represent a popular means to
  interpret molecular dynamics trajectories in terms of memoryless
  transitions between metastable conformational states. To provide a
  mechanistic understanding of the considered biomolecular process,
  these states should reflect structurally distinct conformations and
  ensure a timescale separation between fast intrastate and slow
  interstate dynamics. Adopting the folding of villin headpiece (HP35)
  as a well-established model problem, here we discuss the selection
  of suitable input coordinates or `features', such as backbone
  dihedral angles and interresidue distances. We show that dihedral
  angles account accurately for the structure of the native energy
  basin of HP35, while the unfolded region of the free energy
  landscape and the folding process are best described by tertiary
  contacts of the protein. To construct a contact-based model,
  we consider various ways to define and select contact distances, and
  introduce a low-pass filtering of the feature trajectory as well
  as a correlation-based characterization of states. Relying on
  input data that faithfully account for the mechanistic origin of the
  studied process, the states of the resulting Markov model are
  clearly discriminated by the features, describe consistently the
  hierarchical structure of the free energy landscape, and---as a
  consequence---correctly reproduce the slow timescales of the process.
\end{abstract}

\section{Introduction}

Classical molecular dynamics (MD) simulations facilitate the
microscopic study of structure, dynamics and function of biomolecular
systems.\cite{Berendsen07} To obtain a concise
interpretation of the ever-growing amount of simulation data, we
typically want to construct a coarse-grained model of the considered
process, such as a Langevin equation\cite{Lange06b, Hegger09, Ayaz21}
or a Markov state model (MSM). \cite{Buchete08, Bowman09,
  Prinz11,Bowman13a,Wang17a}
In particular, MSMs have become popular for many practitioners of MD
simulations, as they provide a generally accepted state-of-the-art
analysis of the dynamics, promise to predict long-time dynamics from
short trajectories, and are straightforward to build using open-source
packages such as PyEmma \cite{Scherer15} and
MSMBuilder.\cite{MSMBuilder} The usual workflow to construct an MSM
consists of (i) selection of suitable input coordinates, also called
`features', (ii) dimensionality reduction from the high-dimensional
feature space to some low-dimensional space of collective variables,
(iii) geometrical clustering of these low-dimensional data into
microstates, (iv) dynamical clustering of these microstates into
metastable conformational states, and (v) estimation of the transition
matrix associated with these states.
In principle, all these steps can be optimized by employing a
variational principle that states that the MSM producing the slowest
timescales represents the best model.\cite{Nueske14,Wu20} While this
is conceptually similar to the well-known variational principle of
quantum mechanics where the exact ground-state wave function produces
the lowest energy, in practice the analogy is only in part
applicable. In particular, it rests on the assumption that the
considered process is sufficiently sampled (to ensure that the MD data
are statistically meaningful) and is appropriately described by the
chosen input coordinates.

Since `you get what you put in', it is hard to overstate the
importance of identifying suitable and relevant input coordinates for
the analysis.\cite{Sittel18,Scherer19,Ravindra20,Konovalov21,Husic16}
Due to inevitable mixing of overall and internal motion, Cartesian
coordinates are in general not suited for dimensionality
reduction.\cite{note2,Mu05,Sittel14,Scherer19} Internal coordinates
such as dihedral angles and interatomic distances, on the other hand,
are by definition not plagued by this problem and also represent a
natural choice, because the molecular force field is given in terms of
internal coordinates. While $(\phi, \psi)$ backbone dihedral angles
have been shown to accurately describe the conformation of secondary
structures, \cite{Altis08,Maisuradze09a,Riccardi09,Fenwick14}
interresidue distances appear to be well suited to also characterize
the overall structure of a protein.\cite{Laetzer08,Hori09,Kalgin13} A
drawback of using interresidue distances is that their number scales
quadratically with the number of residues. To avoid this
overrepresentation, it has been suggested to restrict the analysis on
distances reflecting interresidue contacts such as hydrogen bonds,
salt bridges, and hydrophobic contacts.\cite{Swope04II,Best05,
  Ernst15,Ernst17,Oide22} In this way, we focus on the very
interactions that cause the studied conformational transition, and
consider the long-distance motions as a consequence of these contact
changes.

Irrespective of the type of features, it turns out to be important to
first exclude irrelevant motions from the analysis.  This may include
coordinates that do not change during the functional motion (e.g.,
stable contacts), coordinates that change randomly (e.g., wildly
dangling terminal residues that exhibit large amplitude motion), and
coordinates describing slow but non-functional motions (e.g.,
transitions between right- and left-handed helices, where the latter
are hardly populated). Apart from merely corrupting the
signal-to-noise ratio, such motion may be deceptive for the subsequent
dimensionality reduction. For example, principal component analysis
maximizes the variance of the first principal
components,\cite{Amadei93} and is therefore deceived by irrelevant
large-amplitude motion. Time-lagged independent component
analysis\cite{Perez-Hernandez13} (and other timescale optimizing
approaches\cite{Nueske14,Wu20}) aims at maximizing the timescales of
the first components and is therefore deceived by functionally
irrelevant slow motion (such as left to right-handed
transitions).\cite{Sittel18} To discriminate collective motions
underlying functional dynamics from uncorrelated motion, we recently
proposed a correlation analysis termed MoSAIC,\cite{Diez22} which
block-diagonalizes the correlation matrix of the considered
coordinates.

\begin{figure}[t!]
	\centering
	\includegraphics{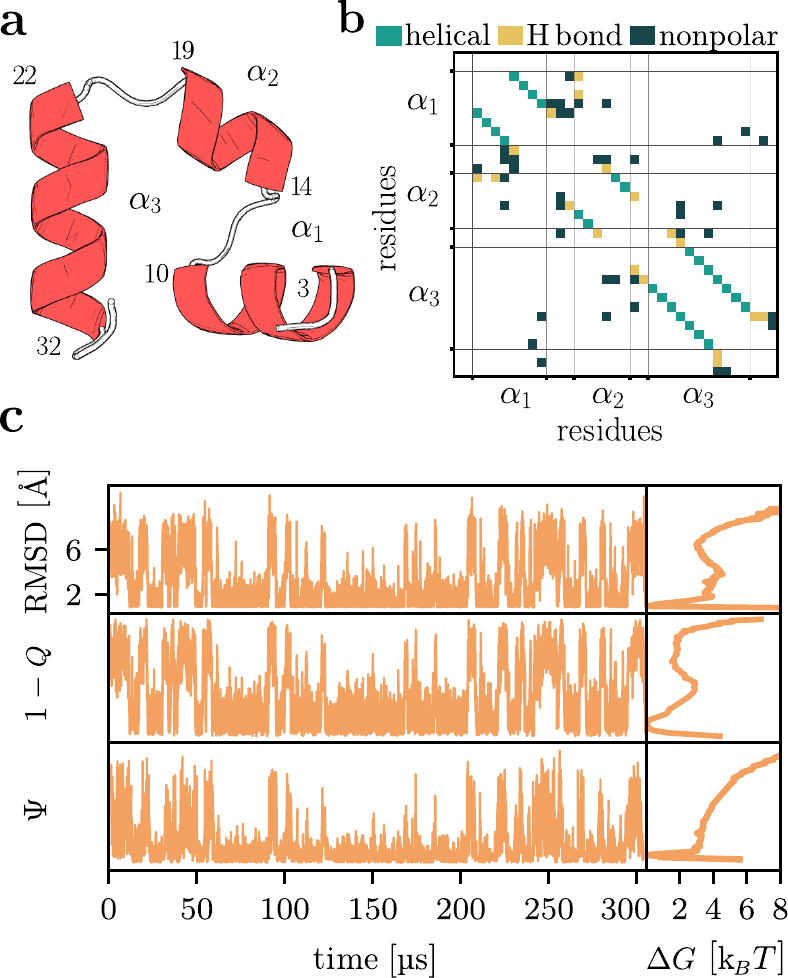}
	\caption{
		The folding of villin headpiece
		(HP35). (a) Molecular structure, consisting of three
		$\alpha$-helices (residues 3--10, 14--19 and 22--32) that are
		connected by two loops. (b) Contact map showing the 42 native
		contacts of HP35. (c) Time
		evolution of the RMSD (with respect to the C$_\alpha$-atoms of
		residues 3 to 33), compared to $1-Q$ (with $Q$ being the
		fraction of native contacts) and $\Psi$ representing
		the sum over the backbone dihedral angles $\psi_i$ of the three
		$\alpha$-helices. The right side shows the corresponding free
		energy curves (in units of $k_{\rm B}T$) along these
		coordinates.}
	\label{fig:HP35}
\end{figure}

In this work we present a detailed study on the virtues and
shortcomings of using contact distances or backbone dihedral angles,
for short `contacts' and `dihedrals'. As both sets of coordinates
appear to describe the protein structure reasonably well, and since
metastable molecular structures are thought to represent a physical
property of the system, one would assume that this choice of
coordinates should not significantly affect the outcome of Markov
modeling. On the other hand, because dihedrals and contacts reflect
different aspects of the structural dynamics and due to limited
sampling, they may result in different collective variables and
metastable states, leading to MSMs with different timescales and
pathways.

Since such a study depends significantly on the considered system and
the specific MD data, here we focus on a well-established model
problem, that is, the folding of villin headpiece
(HP35),\cite{Chiu05,Brewer07,Kubelka06, Kubelka08,Reiner10,
  Duan98,Snow02, Ensign07,Rajan10, Beauchamp11a,Piana12,Jain14,
  Nagel19,Nagel20,Sormani20,Damjanovic21,Klem22,Chong21} see
Fig.~\ref{fig:HP35}a. In particular, we employ a
\SI{300}{\micro\second}-long MD trajectory of HP35 produced by Piana
et al.,\cite{Piana12} which is publicly available from D.~E.~Shaw
Research. Showing the time evolution of the root-mean-square deviation
(RMSD) of the MD trajectory from the crystal structure,
Fig.~\ref{fig:HP35}c reveals that the system undergoes reversible
folding and unfolding on a microsecond timescale, which compares well
to experimental data.\cite{Chiu05,Brewer07, Kubelka06,
  Kubelka08,Reiner10} A comparison to the fraction of native contacts
$Q$ as well as to the sum $\Psi$ over the backbone dihedral angles
$\psi_i$ of the three $\alpha$-helices reveals that both contacts and
dihedrals appear to monitor the overall structural evolution of HP35.

The paper starts with a detailed discussion of the choice of suitable
features for the folding of HP35. In particular, we consider various
ways to define and select contact distances, and introduce a
correlation-based characterization of states in terms of contact
clusters. Following the description of our workflow to construct the
MSM, we discuss the structural and dynamical properties of the
metastable conformational states obtained for contacts and dihedrals,
and illustrate the insights on the folding process gained directly
from the features and from the MSM.
The simulation data and all intermediate results, including scripts
and detailed descriptions can be downloaded from our Github page
https://github.com/moldyn/HP35.

%
%%%%%%%%%%%%%%%%%%%%%%%%%%%%%%%%%%%%%%%%%%%%%%%%%%%%%%%%%%%%%%%%%%%%%%
%
\section{Feature selection}

All analyses done in this work are based on the
$\SI{300}{\micro\second}$-long MD simulation ($1.5 \times 10^6$ data
points) of the fast folding Lys24Nle/Lys29Nle mutant of HP35 at
$T=\SI{360}{\kelvin}$ by Piana et al.,\cite{Piana12} using the Amber
ff99SB*-ILDN force-field \cite{Hornak06, Best09, Lindorff-Larsen10}
and the TIP3P water model.\cite{Jorgensen83}

\subsection{Definition of contacts}\label{sec:contacts}

The definition of interresidue protein contacts includes
\vspace{-2mm}
\begin{itemize}
\item the conditions when a contact is established, e.g., via a
  distance cutoff,\vspace{-2mm}
\item the choice of the molecular structures from which contacts are to
  be identified, e.g., a single crystal-structure or an ensemble of MD
  structures, and \vspace{-2mm}
\item the definition of the distance $d_{ij}$ between two residues,
  e.g., the distance between the C$_\alpha$-atoms or between the
  closest heavy atoms of each residue.
\end{itemize}
Extending previous work,\cite{Ernst15} here we assume a contact to be
formed if (1) the distance $d_{ij}$ between the closest non-hydrogen
atoms of residues $i$ and $j$ is shorter than \SI{4.5}{\angstrom}, (2)
the residues are more than three residues apart, and (3) the
contact is populated more than \SI{30}{\%} of the simulation time
($P_{ij} \ge 0.3$), which ensures that we focus on native contacts.
Applied to the MD trajectory of HP35 by Piana et al.,\cite{Piana12}
this results in total in 42 native contacts, which include 13
helix-stabilizing ($n, n+4$) contacts, 20 hydrophobic contacts, and 9
hydrogen bonds, see the contact map shown
in Fig.~\ref{fig:HP35}b.\cite{note4}

Let us discuss the justifications and implications of the above
choices. To begin with, the distance cutoff
$d_\text{c}=\SI{4.5}{\angstrom}$ rests on studies of the distance
distribution $P(d_{ij})$ of various proteins, whose prominent peak at
short $d_{ij}$ clearly indicates a contact.\cite{Heringa91,Yao19} In
our experience, this definition covers the vast majority of polar and
nonpolar contacts,\cite{Ernst15,Ernst17} and also covers the
commonly applied criteria for hydrogen bonds.\cite{Swope04II,Best05}
Restricting ourselves to residues that are more than three residues
apart, we exclude---apart from irrelevant contacts with next and second
next residues---($n, n+3$) contacts that occur, e.g., in
$3_{10}$-helices. While these contacts certainly exist in the unfolded
ensemble of HP35, they are very short-lived and therefore of little
interest.

Discussing protein folding, it is advantageous to focus on native
contacts, because they have been shown to largely determine the
folding pathways.\cite{Sali94,Wolynes95,Best13} Moreover, it is well
established that the fraction of native contacts $Q$ is highly
correlated with the RMSD of the folding trajectory
(Fig.~\ref{fig:HP35}c) and therefore represents a well-defined
one-dimensional reaction
coordinate.\cite{Best13,Best10} To this end, we request
that the contacts occur at least $P_\text{c}=\SI{30}{\%}$ of the simulation
time, which excludes nonnative contacts that are typically infrequent
and short-lived.
For somewhat smaller population cutoffs we obtain a higher number of
still native contacts (e.g., 56 instead of 42 for $P_\text{c}=\SI{10}{\%}$),
while an increasing number of non-native contacts arise only for
significantly lower cutoffs (e.g., in total 122 contacts for
$P_\text{c}=\SI{1}{\%}$).
While the threshold $P_\text{c}=0.3$ proves suitable for the folding
of HP35, in general it needs to be adapted to suit the problem under
consideration.
We note that the resulting contacts obtained from the MD trajectory
are a different choice of native contacts than the one determined from
the crystal structure\cite{Kubelka06} (PDB 2f4k). The latter exhibits
contacts of the N-terminus due to the packing in the crystal, which do
not occur in solution (and therefore also not in the MD
trajectory). Moreover, these spurious contacts impede the formation of
true native contacts such as Asp3-Thr13 and Asp5-Arg14, which turn out
important to discriminate various co-existing folded states. See
Table~\SMTabContacts{} for a list of all native contacts found in the
crystal structure and the MD simulation.

Apart from the choice of contacts, the appropriate calculation of the
contact distance $d_{ij}$ along the trajectory is crucial for the
subsequent modeling. While this is straightforward if the contacted
atoms roughly remain the same throughout the simulation, it becomes more
involved when we consider multicenter contacts between hydrophobic
residues involving many and rapidly changing atom pairs. Employing
$\text{C}_\alpha$-distances, this problem may be circumvented, however, at the
cost of neglecting most microscopic details of the making and breaking
of contacts, which typically leads to structurally not well-defined
conformational states.\cite{Ernst15} Alternatively, we may consider
the distance $d_{ij}$ between the closest non-hydrogen atoms of
residues $i$ and $j$, defined by\cite{Ernst15}
\begin{equation} \label{eq:MinDist}
d_{ij}(t) = \min_{n, m}|\vr_{i,n}(t)-\vr_{j,m}(t)|,
\end{equation}
where the indices $n$ and $m$ run over all heavy atoms of the selected
residue pair $(i,j)$. Listing population probabilities $P_{ij}$ and
life times of all native contacts, Table~\SMTabContacts{} reveals that
this definition of minimum distances leads to somewhat shorter life times but
significantly higher populations than those found when only a single
atom pair for each contact is used.

When we consider hydrophobic residues with extended and flexible side
chains (e.g., Phe, Lys, Met and Nle), however, we find that a large
number (say, tens) of connecting atom pairs may occur. In particular,
this is the case for protein folding, where during the formation of
the hydrophobic core many unusual atom combinations with distances
below the cutoff $d_\text{c}$ may exist for some short time. Hence, by
using the definition in Eq.~\eqref{eq:MinDist} along the trajectory, we
instantaneously choose the atom pair with the minimal distance,
regardless how exotic and short-lived this contact might be. Resulting
in frequent hopping between multiple atom pairs, this renders the
contact definition somewhat ill-defined, because significantly
different side-chain conformations may result in a similar contact
distance.
Hence, when calculating shortest distances between residues, we
propose to exclude atom pairs $(n,m)$ that do not meet the population
cutoff $P_{nm} \ge 0.3$. This leads to a new definition of minimum
distances
\begin{equation} \label{eq:MinDist2}
d_{ij}(t) = \min_{\substack{n, m \\ P_{nm} \ge 0.3}} |\vr_{i,n}(t)-\vr_{j,m}(t)|,
\end{equation}
which will be used throughout this work.
As expected, the new criterion significantly reduces the number of
considered atom pairs, that is, typically only 2-6 (out of tens)
survive. Moreover, the population probability $P_{ij}$ of most
contacts is reduced (Table~\SMTabContacts), such that 8 (out of 50) of
the weaker contacts fall below the overall population cutoff
$P_{ij} \ge 0.3$, and are therefore not considered as contacts anymore. (Since
redundant contacts between neighboring residues exist,
no information is lost.)  On the other hand, we obtain
considerably longer life times of the contacts, which leads to
conformational states of higher metastability.

Since Eq.~\eqref{eq:MinDist2} requires the calculation of all atom
  pair distances of all possible contacts, the approach is
  computationally expensive and takes a few CPU hours on a standard
  desktop computer already for a small protein such as HP35. It is
  important to note, however, that the computation of contact
  distances generally requires only to consider residues that are
  nearby to a given residue. Implementing a local search routine, the
  computational effort will therefore scale approximately linear with
  the number of residues $N$, rendering the computational effort to a
  few CPU days for proteins with $N\!\approx\!10^3$ residues.

%\newpage
%
%%%%%%%%%%%%%%%%%%%%%%%%%%%%%%%%%%%%%%%%%%%%%%%%%%%%%%%%%%%%%%%%%%%%%%
%
\subsection{Correlation analysis of contacts}
\vspace{-2mm}

To characterize the above defined contacts and detect their
interdependencies, it is instructive to calculate the linear
correlation matrix
\begin{equation} \label{eq:sigma}
\rho_{ij,kl}(t) = \frac{\langle  \delta d_{ij} \delta d_{kl}
    \rangle}{\sigma_{ij}\sigma_{kl}},
\end{equation}
where $\delta d_{ij} = d_{ij} - \langle d_{ij}\rangle$ and
$\sigma_{ij}$ is the standard deviation of $d_{ij}$. (For scalar
variables, nonlinear correlation measures typically do not provide
essential new information.\cite{Lange06,Diez22}) Since the ordering of
the contacts is {\em per se} arbitrary, we block-diagonalize $\rho$ in
order to associate the resulting blocks or clusters with functional
motions. Following Diez et al., \cite{Diez22} this is achieved via a
community detection technique called Leiden clustering,\cite{Traag19}
using the constant Potts model as objective function and a Leiden
resolution parameter $\gamma = 0.78$. All results reported below were
produced using the Python package MoSAIC.\cite{Diez22}

\begin{figure}[t!]
    \centering
    \includegraphics{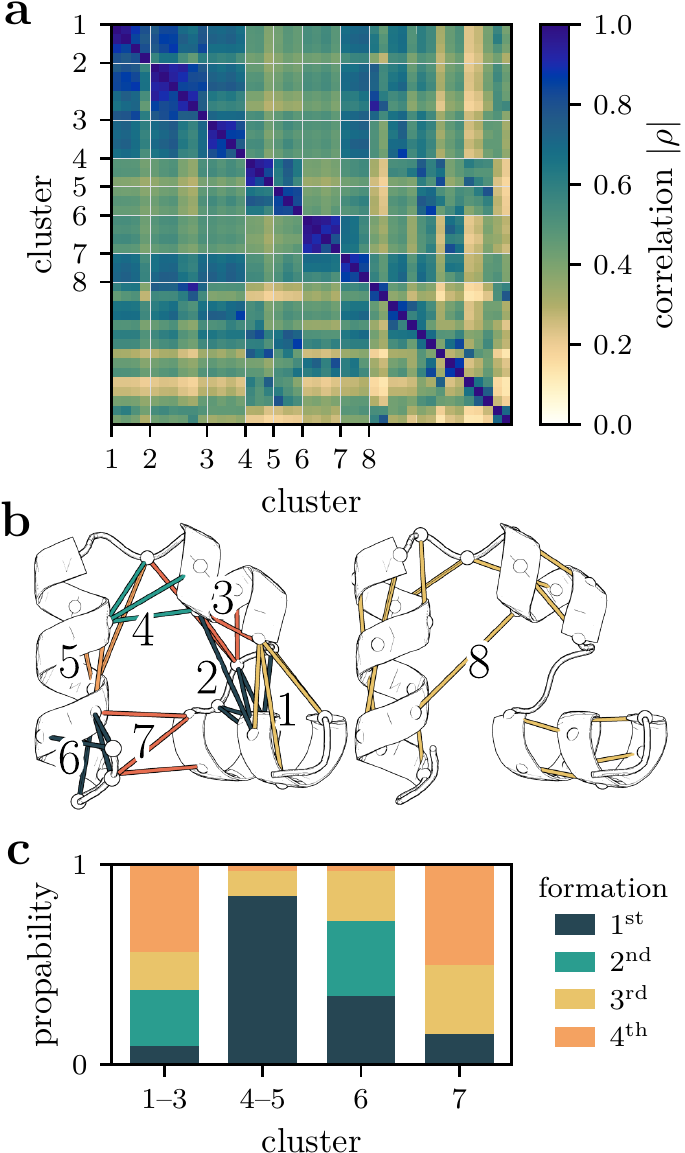}
    \caption{
      (a) Block-diagonalized correlation
      matrix of contact distances, revealing seven main clusters and
      various mini-clusters comprised in cluster\,8. Contacts
      of main clusters are
      1:\! $d_{3,14}$, $d_{3,13}$, $d_{6,14}$, $d_{5,14}$;
      2:\! $d_{7, 12}$, $d_{7, 13}$, $d_{6, 12}$, $d_{7, 11}$, $d_{6, 11}$, $d_{6, 17}$;
      3:\! $d_{12, 17}$, $d_{12, 16}$, $d_{12, 20}$, $d_{13, 17}$;
      4:\! $d_{18, 25}$, $d_{17, 25}$, $d_{20, 25}$;
      5:\! $d_{24, 28}$, $d_{20, 28}$, $d_{25, 29}$;
      6:\! $d_{29, 35}$, $d_{29, 34}$, $d_{30, 35}$, $d_{29, 33}$;
      7:\! $d_{10, 34}$, $d_{9, 32}$,  $d_{10, 29}$;
      (b) Structural illustration of the interresidue contacts of
      clusters. (c) Order of clusters formed during a folding event.}
    \label{fig:mosaic}
\end{figure}
%\vspace{-2mm}

Figure~\ref{fig:mosaic}a shows the modulus of the resulting
block-diagonal correlation matrix $\{|\rho_{ij,kl}|\}$, which
reveals seven main clusters. Within such a cluster, the contacts are
highly correlated (i.e., on average $|\rho_{ij,kl}| \ge \gamma$),
while the correlation between different clusters is comparatively low
(i.e., $|\rho_{ij,kl}| < \gamma$). As illustrated in Fig.~\SMcoopTraj{}
for the first few folding events, high intracluster
correlation means that the contacts of a cluster typically change
together, i.e., in a cooperative manner. Moreover, there are contacts (or
mini-clusters with $< 3$ contacts) that are not a member of a main
cluster but still exhibit moderate correlation with some of the main
clusters; they are collected in cluster\,8.

To illustrate the contacts contained in the main clusters,
Fig.~\ref{fig:mosaic}b displays the corresponding contact distances
inserted into the structure of HP35. We notice that the contact
clusters follow nicely the protein backbone from the N- to the
C-terminus, such that each cluster features contacts that are located
in the same region. In this way, the clusters can be employed to
characterize the structure of the various conformational states of the
protein, see Fig.~\ref{fig:StateRep} below. On the other hand,
cluster\,8 is found to mostly represent helix-stabilizing contacts
along the protein backbone. Interestingly, we find that lifetimes of
these helical contacts are typically shorter ($\sim \SI{100}{\nano\second}$) than
the contacts of the main clusters ($\sim \SI{1}{\micro\second}$), see
Tab.~\SMTabContacts. As discussed by Buchenberg et al.,\cite{Buchenberg15}
this timescale separation indicates hierarchical dynamics, where the
fast opening and closing of helix-stabilizing contacts is a
prerequisite of overall conformational change to occur.

Since we consider well-populated native contacts and because we study
protein folding (which involves the making and breaking of virtually
all contacts), there are no coordinates that correlate only weakly
with few other coordinates. (This would be the case, e.g., for overall
stable contacts and contacts that form and break frequently, which are
excluded here.)  This is in contrast to the study of functional motion
in a folded protein, where \SIrange{60}{90}{\%} of all contacts were found to be
only weakly correlated and could be therefore discarded in the further
analysis.\cite{Post22a,Ali22}

%\newpage
%
%%%%%%%%%%%%%%%%%%%%%%%%%%%%%%%%%%%%%%%%%%%%%%%%%%%%%%%%%%%%%%%%%%%%%%
%
\subsection{Selection of dihedral angles}
\vspace{-2mm}

Various authors have employed ($\phi_i,\psi_i$) backbone dihedral
angles of residues $i$ to describe the folding of
HP35.\cite{Jain14,Nagel19,Nagel20,Sormani20,Damjanovic21,Klem22} The
number of dihedral angles scales linearly with the number of residues
and are valuable conformational descriptors that directly indicate
whether the protein forms helices, sheets or loops.  While dihedral
angles are readily obtained from the MD trajectory and do not require
a particular definition (as contact distances do), their periodic
nature needs to be taken into account in a statistical
analysis.\cite{Altis07,Sittel17,Zoubouloglou22} For example, we may
convert all angles $\vartheta$ to sine/cosine-transformed coordinates
($x_1\!= \!\cos \vartheta, \; x_2\!=\! \sin \vartheta$) in order to
obtain a linear coordinate space with the usual Euclidean distance as
metric,\cite{Mu05,Altis07,Riccardi09} and perform the statistical
analysis in this space.  To avoid the inherent doubling of variables
($\vartheta \rightarrow x_1, x_2$) and the nonlinear nature of the
transformation, we may alternatively exploit the well-known
Ramachandran plot which demonstrates that protein backbone dihedral
angles do not cover the full angular space $(-\pi, \pi]$ but are
limited to specific regions due to steric hindrance. Hence, natural
cuts between sampled regions can be defined, and by shifting the
original data to align the periodic border to this `maximal gap' in
sampling, statistical analyses can be directly performed on the
dihedral angles in a standard manner.\cite{Sittel17} Combined with a
principal component analyses, this approach was termed dPCA+.

\begin{figure}[t!]
    \centering
    \includegraphics{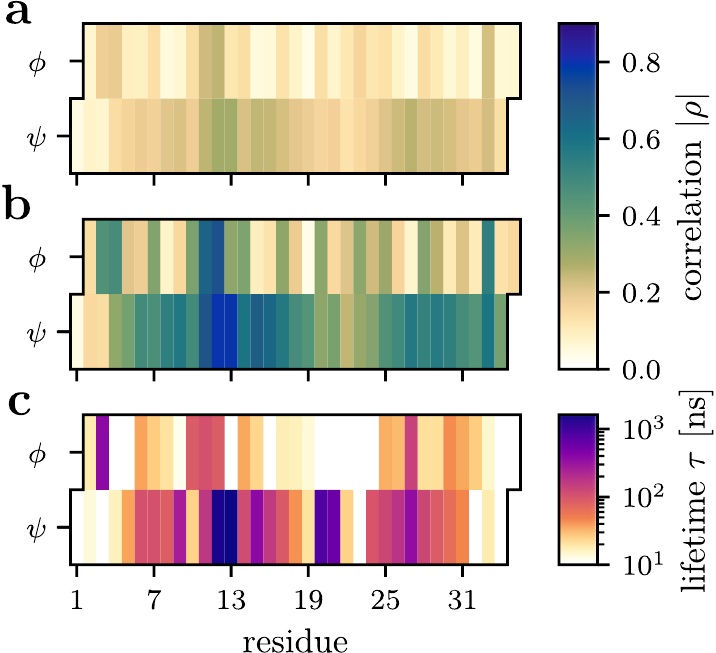}
    \caption{
      Correlation analysis of ($\phi,\psi$)
      backbone dihedral angles. Shown are (a) the mean correlation of
      all dihedrals, (b) their correlation with the RMSD shown in
      Fig.~\ref{fig:HP35}, and (c) the decay time of their
      autocorrelation function. The terminal angles $\phi_1$ and
      $\psi_{35}$ are not properly defined and therefore not shown.}
    \label{fig:dihedrals}
\end{figure}

When we calculate the correlation matrix of the maximal gap-shifted
($\phi,\psi$) dihedral angles, the checkerboard pattern of the matrix
indicates that the $\psi$ angles of HP35 are typically much more
correlated than the $\phi$ angles (Fig.~\SMdihedrals). Showing the
mean correlation (i.e., the average correlation of an angle with all
other angles) of all dihedral angles, Fig.~\ref{fig:dihedrals}a
confirms this finding. It also shows that the rapidly fluctuating
dihedral angles of the first and last two residue hardly correlate
with any other angle and therefore should be excluded from the further
analysis.
(If such uncorrelated motions show transitions between two states, we
will get a trivial doubling of states.)
Unlike the correlation analysis of contacts, a
block-diagonalization of the correlation matrix via Leiden clustering
is less instructive (Fig.~\SMdihedrals), because backbone dihedral angles
naturally proceed with the protein sequence.

Considering the correlation of the dihedral angles with the RMSD of
the folding trajectory, Fig.~\ref{fig:dihedrals}b shows that the
$\psi$ angles correlate strongly with the folding dynamics of
HP35. The finding is in line with the observation that the time evolution of
the RMSD and of the sum of the $\psi$ angles are highly correlated
(Fig.~\ref{fig:HP35}c). This is expected, because $\psi$ angles
decrease significantly in the Ramachandran plot when the
conformation changes from extended to helical structures, and
therefore account directly for the helicity of the system. The
importance of the $\psi$ angles is also apparent from the long decay
times of their autocorrelation functions shown in
Fig.~\ref{fig:dihedrals}c. This suggests that, except for the terminal
residues 1--2 and 34--35, we want to include all $\psi$-angles in the
further analysis.

While most $\phi$ angles change only little when the residue changes
from extended to helical structures, there are prominent exceptions to
residues in the termini and the two loops. Apart from the glycines
Gly11 and Gly33, in particular the $\phi$ angle of Asp3 is found to
coexist in left- and right-handed conformations, which leads to a
splitting of the states in the native energy basin,\cite{Ernst15,
  Sittel16, Nagel19} see below. These residues reveal also slowly
decaying autocorrelation functions, while the majority of the $\phi$
motions is rather short-lived. Since the helical $\phi$-angles
contribute only minor to the first PCs, we decided to include all
of them (except for the terminal residues) in the further analysis.
In total, this results in 62 dihedral angles as features.

%\newpage
%
%%%%%%%%%%%%%%%%%%%%%%%%%%%%%%%%%%%%%%%%%%%%%%%%%%%%%%%%%%%%%%%%%%%%%%
%%%%%%%%%%%%%%%%%%%%%%%%%%%%%%%%%%%%%%%%%%%%%%%%%%%%%%%%%%%%%%%%%%%%%%
%
\section{Construction of metastable states}\label{sec:metastable_states}
\vspace{-2mm}

To identify metastable conformational states from the above defined
feature trajectories, the following protocol is used. First we employ
a Gaussian low-pass filter that eliminates high-frequent fluctuation
of the feature trajectory. We then use principal
component analysis\cite{Amadei93} (PCA) in order to convert the
high-dimensional feature variables to low-dimensional ($\lesssim 5$)
collective variables.\cite{Rohrdanz13, Wang20,Glielmo21} The low
dimensionality facilitates robust density-based
clustering,\cite{Sittel16} which is used to generate (typically
hundreds of) microstates. Using the most probable path
algorithm\cite{Jain12} (MPP) to lump the microstates into a few
macrostates, we obtain the desired set of metastable conformational
states.
%Finally, we use the projection method of Hummer and
%Szabo,\cite{Hummer15} to construct the transition matrix of the
%metastable states, and provide a detailed characterization of their
%structural and dynamical properties.
%
All analyses shown in this paper were performed using our open-source
Python package msmhelper, which can be downloaded from
https://github.com/moldyn. To facilitate the reproduction and analysis
of our results, we furthermore provide trajectories of all
intermediate steps.
Using an Intel Core i9-10900 processor, the complete above described
MSM workflow (filtering, PCA, clustering) applied to $1.5 \times 10^6$
data points required a wall clock time of about
\SI{16}{\minute}. (Employing two NVIDIA GeForce GTX 680, robust
density-based clustering\cite{Sittel16} took \SI{7}{\minute} of this
time; without GPU acceleration it takes about \SI{3.5}{\hour}.)
Moreover, we spent about $\SI{2}{\hour}$ on contact definition,
\SI{3}{\minute} on angle definition, $\SI{30}{\minute}$ for the MSM
analyses shown in Fig.~\ref{fig:timescales}, and in total
\SI{9.5}{\hour} for the XGBoost analyses shown in
Fig.~\ref{fig:xgbAnalysis}.

%
%%%%%%%%%%%%%%%%%%%%%%%%%%%%%%%%%%%%%%%%%%%%%%%%%%%%%%%%%%%%%%%%%%%%%%
%
\subsection{Data filtering} \label{sec:filter}
\vspace{-1mm}

While the typical lifetime of the selected contact and dihedral
variables of HP35 is between 0.1 and \SI{1}{\micro\second}
(Tab.~\SMTabContacts{} and Fig.~\ref{fig:dihedrals}c), the variables are
found to fluctuate on a picosecond timescale, reflecting fast moving
atoms in the vicinity. Since we eventually want to define metastable
conformational states from the variables, we face the problem that the
resulting state trajectory may also fluctuate rapidly between various
states when the contacts are close to the distance cutoff. This is in
contrast to the fact that state changes are associated with rare
transitions over free energy barriers (cf.\ Fig.~\ref{fig:HP35}c),
i.e., they are expected to occur infrequently and without spurious
back-transitions. The problem is caused by the projection of the
high-dimensional protein dynamics onto low-dimensional variables
(i.e., the contact distances or dihedral angles). This may lead to
misclassification of the data points in the transition
regions,\cite{Nagel19} which are notoriously undersampled in unbiased
MD simulations.

As a simple but effective remedy, it has been suggested that a
transition from one state to another must reach the core region of the
other state; otherwise, it is not counted as a
transition. \cite{Buchete08,Schuette11,Lemke16} Alternatively, we may
request that the trajectory spends a minimum time in the new state for
the transition to be counted.\cite{Jain14,Nagel19} While this
`dynamic coring' is capable of correcting spurious transitions, it
cannot correct a wrong state assignment caused by them.
This means, for example, that if two states
overlap due to suboptimally chosen collective variables and can
therefore no longer be discriminated, the correct state assignment
cannot be reconstructed from coring. The same holds for alternative
approaches of dynamic correction that are subsequent to
clustering.\cite{Buchete08,Schuette11,Lemke16}

As a new approach, we propose to perform the dynamic correction as the
first step in the workflow. Here we use a Gaussian
low-pass filter to smoothen the high-frequency fluctuations of the
feature variables $d(t)$, i.e.,
\begin{align}
    d(t) &\to \sum_j g(t,t_j) d(t_j) \label{eq:Gauss1}, \\
    g(t,t_j) &= \frac{1}{\sqrt{2 \pi \sigma^2}} \exp\left[\frac{-
                (t_j-t)^2}{2\sigma^2}\right]\;. \label{eq:Gauss2}
\end{align}
Choosing $\sigma=\SI{2}{\nano\second}$, the Gaussian filtering
suppresses all sub-ns fluctuations and makes a subsequent coring of
the resulting conformational states obsolete.
We will discuss the effects of the filtering ansatz in comparison to
dynamic coring at the end of Sec.\ \ref{sec:DynPropStates}.

%
%%%%%%%%%%%%%%%%%%%%%%%%%%%%%%%%%%%%%%%%%%%%%%%%%%%%%%%%%%%%%%%%%%%%%%
%
\subsection{Dimensionality reduction}
\vspace{-1mm}

PCA represents a linear transformation that diagonalizes the
correlation matrix and thus removes the instantaneous linear
correlations among the variables. Ordering the eigenvalues of the
resulting eigenvectors decreasingly, the first principal components
(PCs, i.e., the projection of the input coordinates on the
eigenvectors) account for the directions of the largest correlation of the
data set.\cite{Amadei93}
While PCA becomes exact if sufficiently many PCs are
included, we aim to obtain a low-dimensional representation by
truncating the number of PCs according to the following criteria:
Non-quadratic appearance of the free energy curves along the PCs,
decay times of their autocorrelation function, and
explained percentage of the total correlation.\cite{Altis08}

Using contact distances, we find that the first 5 PCs exhibit a
multimodal structure of their free energy curves, reveal the slowest
timescales ($\sim \SIrange{0.1}{2}{\micro\second}$), and explain $\sim
\SI{80}{\%}$ of the total correlation (Fig.~\SMPCA). The first PC mostly
reflects the fraction of native contacts $Q$ (Fig.~\ref{fig:HP35}c), which
represents a well-established reaction coordinate.\cite{Best13,Best10} The
higher PCs account for linear combinations of contact changes in various MoSAIC
clusters (Fig.~\ref{fig:mosaic}).
Considering dihedral angles, only the first 4 PCs exhibit nontrivial
free energy curves. They account for slowest timescales
($\sim \SIrange{0.1}{2}{\micro\second}$), and explain $\sim  \SI{50}{\%}$ of the
total correlation (Fig.~\SMPCA). The first PC represents the sum $\Psi$ of
all $\psi$ angles (Fig.~\ref{fig:HP35}c), which describes the overall
helicity of the protein. The higher PCs mainly account for dihedral
angle changes of helix\,1 and helix\,3.

%
%%%%%%%%%%%%%%%%%%%%%%%%%%%%%%%%%%%%%%%%%%%%%%%%%%%%%%%%%%%%%%%%%%%%%%
%
\subsection{Clustering}
%\subsection{Density-based clustering}

In a first step, we use robust density-based
clustering,\cite{Sittel16, Nagel19} which computes a local free energy
estimate for every frame of the trajectory by counting all other
structures inside a hypersphere of fixed radius $R$. When we then
reorder all structures from low to high free energy, the minima of the
free energy landscape can be identified. By iteratively increasing an energy
threshold, all structures with a free energy below that
threshold that are closer than a certain lumping radius $r_l$ will be
assigned to the same cluster, until all clusters meet at their energy
barriers. In this way, all data points are assigned to a cluster as
one branch of the iteratively created tree.
Considering contact distances, we used a hypersphere
$R =r_l\approx 0.124$ %0.12404$
equaling the lumping radius. For dihedral angles, we choose $R
=r_l\approx \SI{0.072}{rad}$. % 0.07194
Figure \SMclustering{a} shows the resulting total number of
microstates obtained as a function of the minimal population
$P_\text{min}$ a state must contain. Here we chose
$P_\text{min}=\SI{0.01}{\%}\,\widehat{=}\,\SI{153}{frames}$, resulting in 522 and 330
microstates for contact distances and dihedral angles, respectively.

In a second step, we adopt the MPP algorithm \cite{Jain12} to construct
a small number of macrostates. Starting with the above defined
microstates, MPP first calculates the transition matrix of these
states, using a lag time $\tau_\text{MPP}=\SI{10}{\nano\second}$. (The
choice of $\tau_\text{MPP}$ is explained in the discussion of
Fig.~\ref{fig:timescales}a below.) If the self-transition probability of a given
state is lower than a certain metastability criterion $Q_\text{min} \in (0, 1]$,
the state will be lumped with the state to which the transition
probability is the highest. This procedure is reiterated, until there
are no more transitions for a given $Q_\text{min}$. Repeating the
procedure for increasing $Q_\text{min}$, we construct a dendrogram
that shows how the various metastable states merge into energy basins, thus
illustrating the topology and the hierarchical structure of the free
energy landscape. To facilitate the handling of many states, we use an automatic
branch detection scheme.\cite{note3}

\begin{figure}[t!]
    \centering
   \includegraphics{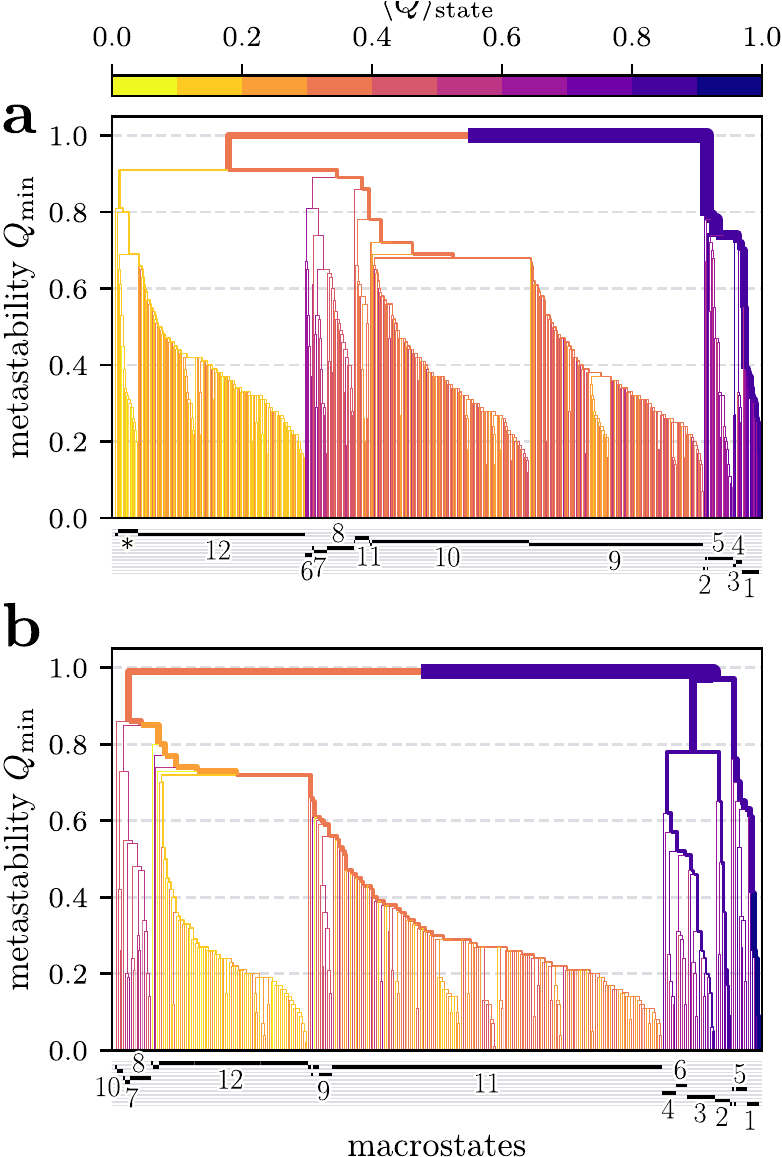}
    \caption{
      MPP dendrogram illustrating the clustering of microstates into
      metastable states upon increasing the requested minimum
      metastability criterion $Q_\text{min}$ of a state. Results are
      shown for (a) contacts and (b) dihedral angles. The states
      are colored according to their mean number of native contacts
      $\langle Q \rangle_{\rm state}$ from yellow (unfolded) via
      orange to native (purple). Black horizontal bars at the bottom
      indicate which microstates are contained in a metastable
      state.}
    \label{fig:dendrogram}
\end{figure}

\begin{figure*}%[t!]
    \centering
    \includegraphics{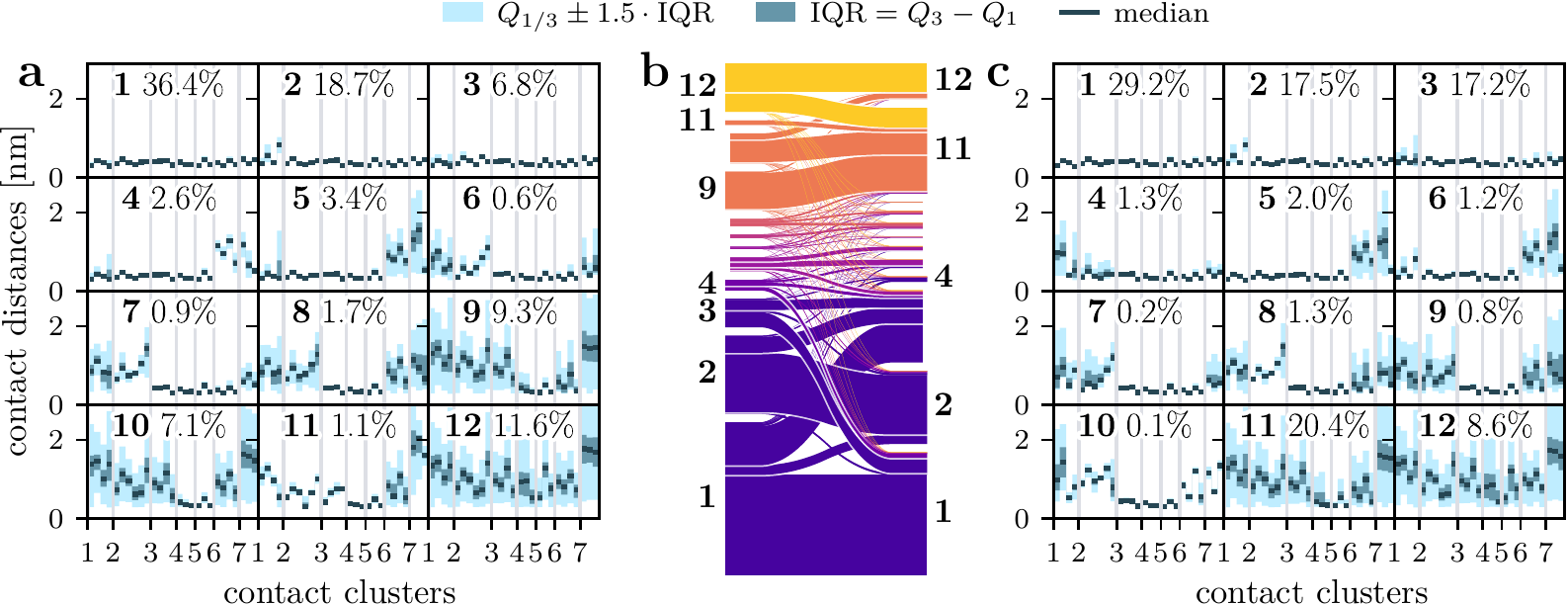}
    \caption{
      Structural characterization of the
      twelve metastable states of HP35, obtained for (a) contacts and
      (c) dihedral angles. The states are ordered by decreasing
      fraction of native contacts $Q$, the contacts are ordered
      according to the seven main MoSAIC clusters
      (Fig.~\ref{fig:mosaic}). For each state, the distribution of
      contact distances are represented by the median $Q_2$, the
      interquartile range $\text{IQR}=Q_3- Q_1$ and the lower (upper)
      bound as the smallest (largest) data point in
      $Q_{1/3} \pm 1.5\cdot\text{IQR}$. (b) Sankey diagram depicting the
      relation between the two state definitions. The color code (see
      Fig.~\ref{fig:dendrogram}) reflects the fraction of native
      contacts $Q$.}
    \label{fig:StateRep}
\end{figure*}

Figure~\ref{fig:dendrogram} shows the resulting dendrograms obtained
for (a) contacts and (b) dihedral angles. For
$Q_\text{min} \gtrsim 0.9$, we obtain only two macrostates, as all
microstates are assigned either to the native or the unfolded energy
basin of HP35. Coloring the states according to their mean number of
native contacts, the native states are drawn in purple and the
unfolded in yellow to orange. Although the unfolded basin clearly
contains more microstates, the native basin is higher populated
($\approx \SI{68}{\%}$).
% Contacts P(S1-S5)=67.77%
% Dihedrals P(S1-S6)=68.46%
By decreasing the requested metastability $Q_\text{min}$, we
in effect decrease the requested minimum barrier height between
separated states, such that the two main basins split up in an
increasing number of metastable states. Requesting that a metastable
state should have at least $Q_\text{min}=0.5$ and a minimum population
of \SI{0.5}{\%} for contacts and \SI{0.1}{\%} for dihedrals, we obtain
12 metastable states for both contacts and
dihedral angles.\cite{note1} While overall the two dendrograms look
similar, we find that dihedrals appear to yield a more structured
native basin, while contacts seem to resolve the unfolded basin
better. In particular, contacts reveal that the unfolded basin splits
up in various well-characterized metastable states, demonstrating that
the unfolded basin exhibits nontrivial dynamical structure.
Overall, we wish to stress that the MPP dendrogram color-coded with
reaction coordinate Q shows clearly whether only structurally similar
states are merged, thus providing an insightful test of the quality of
the state partitioning. 
%\newpage
%
%%%%%%%%%%%%%%%%%%%%%%%%%%%%%%%%%%%%%%%%%%%%%%%%%%%%%%%%%%%%%%%%%%%%%%
%
\subsection{Structural characterization of states}

To obtain a useful state model, the conformational states
identified above should be structurally well-defined (to represent
distinct conformational ensembles) as well as long-lived or metastable
(to give a good MSM). As a structural characterization,
Fig.~\ref{fig:StateRep} shows the distribution of contact distances for
each state, as obtained for (a) contacts and (c) dihedral
angles. Similarly, Fig.~\SMclustering{b} shows the corresponding
dihedral angle distributions for the two cases.
Since contact distances exhibit a bimodal distribution (reflecting
formed and broken contacts) such that mean and variance do not well
describe the data, we use a box-plot representation with quartiles
$Q_i$ comprising the first $i \cdot \SI{25}{\%}$ of the data. The states are
ordered by decreasing fraction of native contacts, such that
state\,1 is the native state (all contact distances are shorter than
$d_{\rm c} =\SI{4.5}{\angstrom}$) and state\,12 is the completely
unfolded state with a broad distribution of large distances. The
contacts are ordered according to the seven main MoSAIC clusters
defined in Fig.~\ref{fig:mosaic}, which follow the protein backbone
from the N- to the C-terminus.

Interestingly, we find that the MoSAIC clusters provide a concise
characterization of the structure of the metastable states. The first
three states are structurally well-defined native-like states, which
combine \SI{62}{\%} and \SI{64}{\%} of the total population for contacts and
dihedrals, respectively. As discussed below, these states differ in
details of helix\,1. According to the MPP dendrogram
(Fig.~\ref{fig:dendrogram}), states 4 and 5 (for contacts) and states 4 to 6
(for dihedrals) also belong to the native energy basin. Compared to
state\,1, they are characterized by broken contacts on the C-terminal
side. The unfolded basin mainly consist of states 9 to 12, which show
different degrees of disorder. For contacts, hardly any contacts exist
in state\,12, while states\,9 and 10 at least exhibit formed contacts
in clusters 4 and 5, and (for state\,9) in cluster 6. For dihedrals,
this splitting of the unfolded states in different structures is less
obvious. Finally, there are several lowly populated ($\lesssim \SI{1}{\%}$)
intermediate states to be discussed below.

The state partitions obtained for contacts and dihedrals can be
directly compared in a Sankey plot (Fig.~\ref{fig:StateRep}b). We
find a simple correspondence between the main three native states, as
well as between the main two or three unfolded states. However, there
is no such clear relation for the lowly populated in-between states,
which are assigned differently for contacts and dihedrals.
To summarize, by focusing on the (clustered) features that describe
the considered process, the contact representation provides a concise
but sufficient structural characterization of the metastable states.

%
%%%%%%%%%%%%%%%%%%%%%%%%%%%%%%%%%%%%%%%%%%%%%%%%%%%%%%%%%%%%%%%%%%%%%%
%
\subsection{Essential coordinates of folding}

As a further state characterization, we now consider the essential
coordinates of the system, which are defined as the most important
features to discriminate the metastable states.\cite{Brandt18} To this
end, we adopt the decision tree-based program XGBoost,\cite{Chen16}
which employs a set of MD coordinates and a set of metastable states,
and trains a model to assign MD structures to the state they most
probably belong to. In a second step, this model is used to assess how
specific coordinates contribute to the identification of a metastable
state. That is, we define as accuracy the success rate of
assigning MD structures to the correct state, and monitor the
evolution of this score, when we iteratively remove features from the
training set. By discarding the least important coordinates first,
we readily filter out all nonessential coordinates (that do not
change the accuracy of the model when discarded) and thus obtain the
desired essential coordinates.

\begin{figure}[t!]
    \centering
    \includegraphics{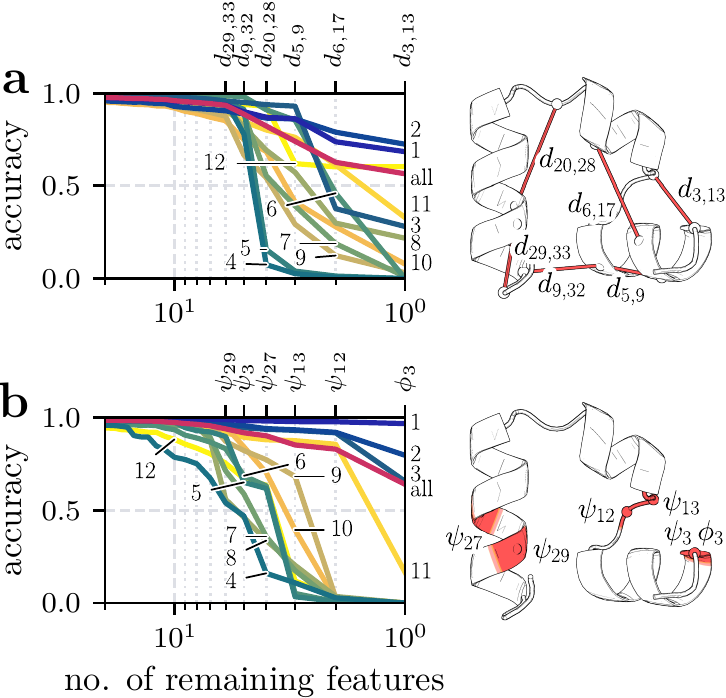}
    \caption{ Identification of essential coordinates by decision
      tree-based machine learning,\cite{Brandt18} obtained for (a)
      contacts and (b) dihedral angles. Employing
      XGBoost,\cite{Chen16} we used a learning rate $\eta=0.7$, a tree
      depth of 8, and 50 training rounds in every step. (Left)
      Accuracy loss of the assignment of all twelve metastable states
      (as well as their population-weighted mean labeled as ``all'')
      when we discard in every step the least important
      feature. (Right) Structural illustration of the most important
      essential coordinates.}
    \label{fig:xgbAnalysis}
\end{figure}

Displaying the accuracy of the assignment of all twelve metastable
states as a function of the remaining features included in the model,
Fig.~\ref{fig:xgbAnalysis} reveals that while only 6
contacts (out of 42) are necessary to discriminate all
metastable states with an accuracy of at least \SI{95}{\%},
nearly 20 dihedrals (out of 62) are required
for the same level of accuracy.
Somewhat surprisingly, the most important contacts
are between residue pairs $(3,13)$, $(6,17)$ and $(5,9)$ involving helix\,1,
because they define the three highly populated states of the native
basin. The main unfolded states are characterized by tertiary contact
$(20,28)$, which holds helix\,2 and helix\,3 together. The
lowly-populated intermediate states moreover require mainly contacts
$(6, 10)$ and $(9, 32)$ for their identification.
The situation is somewhat different for dihedral angles, where the
most important coordinate is $\phi_3$, which is sufficient to
discriminate the three main native states (see Fig.~\SMclustering{c}).
With the exception of the second most unfolded state 11 whose
definition mainly requires $\psi_{12}$, the discrimination of
intermediate and unfolded states requires multiple dihedral angles
located in all three helices.
Hence, we have learned that the substates of the native basin are best
described by dihedral angles; in fact a single angle, $\phi_3$,
suffices for HP35. On the other hand, the overall folding of the
protein and the substates of the unfolded basin are best described by
tertiary contacts connecting the helices.

%\newpage
%
%%%%%%%%%%%%%%%%%%%%%%%%%%%%%%%%%%%%%%%%%%%%%%%%%%%%%%%%%%%%%%%%%%%%%%
%%%%%%%%%%%%%%%%%%%%%%%%%%%%%%%%%%%%%%%%%%%%%%%%%%%%%%%%%%%%%%%%%%%%%%
%
\section{Construction of MSMs}

Employing the projection method of Hummer and Szabo,\cite{Hummer15} we
estimate the transition matrix of the metastable states for both
contacts and dihedrals. We discuss the dynamical properties and the
Markovianity of the resulting models, consider the resulting folding
times, and compare to previously published MSMs based on the same
trajectory.

\subsection{Implied timescales and Markovianity} \label{sec:DynPropStates}

\begin{figure*}%[ht!]
    \centering
    \includegraphics{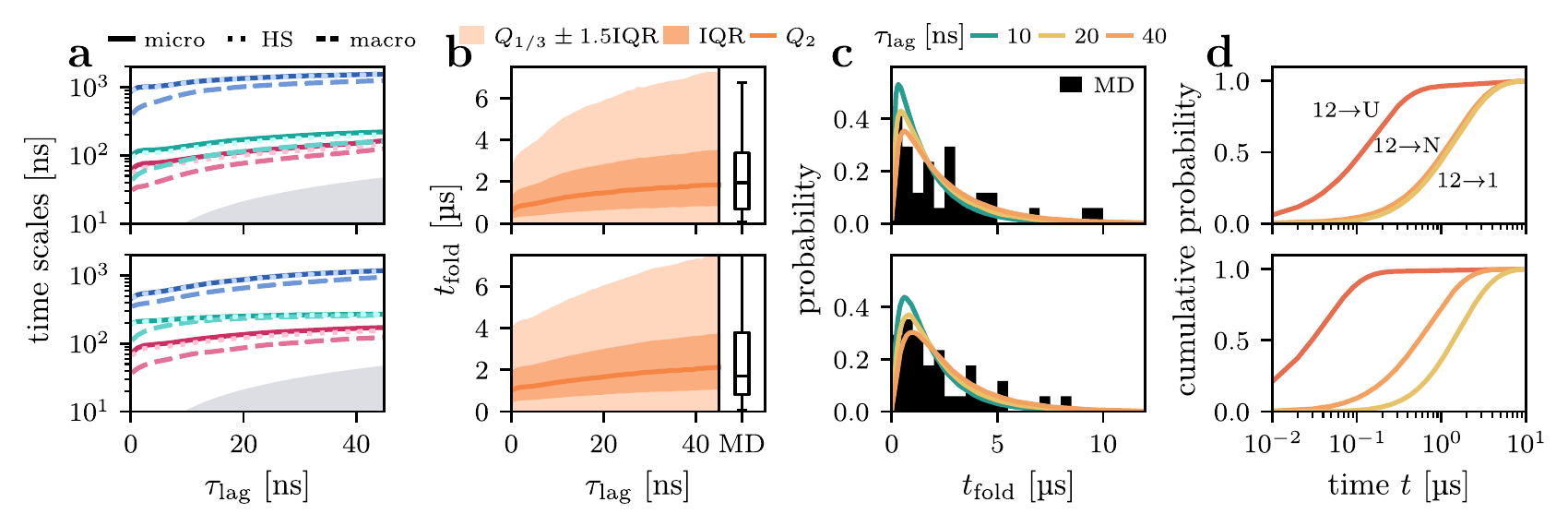}
    \caption{
      Timescales and folding dynamics
      exhibited by MSMs constructed from contacts (top) and dihedrals
      (bottom). (a) First three implied timescales $t_n$ shown as a
      function of the lag time $\tau_{\text{lag}}$, obtained for the
      microstates (full lines) and for the macrostates, either
      assuming local equilibration (dashed lines) or using the
      projection method of Hummer and Szabo\cite{Hummer15} (thin
      lines). Gray areas indicate that $t_n \le
      \tau_{\text{lag}}$. (b) Box-plot representation of the folding
      time distributions obtained from MSMs and MD data. (c) Histogram
      of the MD folding times $t_{\rm fold}$ from state\,12 to
      state\,1, compared to MSM folding time distributions for various
      lag times. (d) Time evolution of various cumulative population
      probabilities, obtained for $\tau_{\text{lag}}=\SI{10}{\nano\second}$.
      Starting in the completely unfolded state\,12, the
      probabilities $P_{12\rightarrow \text{U}}(t)$ reflects the
      hydrophobic collapse of the protein to the other states of the
      unfolded basin, $P_{12 \rightarrow \text{N}}(t)$ accounts for
      the time that the system remains in the unfolded basin, and
      $P_{12\rightarrow 1}(t)$ describes the overall folding process
      into the native state\,1.}
    \label{fig:timescales}
\end{figure*}

To construct a MSM from the metastable states obtained above, we
calculate the transition matrix $T_{ij}$, describing the probability
of a transition between states $i$ and $j$ during some chosen lag time
$\tau_{\text{lag}}$. By diagonalizing the transition matrix, we obtain
its eigenvalues $\lambda_n$ and the implied timescales
$t_{n} = - \tau_{\text{lag}}/\ln{\lambda_n}$.  For Markovian dynamics
these timescales should be constant [cf.\ Eq.~\eqref{eq:CKtest}];
since that is usually not the case for short lag times, constancy of
implied timescales can be used as a criterion to choose a suitable
$\tau_\text{lag}$.\cite{Prinz11}
Figure~\ref{fig:timescales}a shows the resulting implied timescales
$t_1$--$t_3$ reflecting the three slowest processes. We first consider
the timescales of the microstates, which are generally found to level
off for lag times $\tau_{\text{lag}} \gtrsim \SI{10}{\nano\second}$
for contacts and dihedrals. As anticipated above, this value of
$\tau_{\text{lag}}$ is used for MPP clustering to define macrostates.

Considering the macrostates, various methods exist to calculate the
corresponding transition matrix. Most straightforwardly, it can be
directly computed from the associated count matrix (as done for the
microstates), assuming that the macrostates are locally
equilibrated. However, this assumption rests on a timescale separation
between intrastate and interstate dynamics, which is often only
approximately true for macrostates constructed from MPP. As a remedy,
we may calculate the macrostate transition matrix by invoking an
approach to optimally project the microstate dynamics onto the
macrostate dynamics,\cite{Roeblitz13,Kells19,Cao20,Sharpe21} as
achieved by the method of Hummer and Szabo.\cite{Hummer15} Since the
direct calculation yields on average \SI{40}{\%} shorter timescales
while the latter virtually reproduces the timescales of the
microstates (Fig.~\ref{fig:timescales}a), we generally employ the
projection method in the following. Similarly, as found for the
microstates, the implied timescales of the macrostates level off for
lag times $\tau_{\text{lag}} \gtrsim \SI{10}{\nano\second}$. Hence, if
not noted otherwise, all further discussion will be based on this
value.  For further reference, Tab.~\SMTabTransMatrix{} lists the
interstate transition times and state lifetimes associated with the
resulting transition matrices for contacts and dihedrals.

The resulting macrostate timescales obtained for contacts and
dihedrals are compared in Fig.~\ref{fig:timescales}a. In the case of
contacts, the slowest timescale of about \SI{1.2}{\micro\second} is
clearly separated from the next ones ($\sim 0.10$ and
\SI{0.08}{\micro\second}). Inspecting the eigenvectors of the
transition matrix (Fig.~\SMCKtest{a}), we find that the slowest
process corresponds to the transition from unfolded states (12, 10 and
9) to folded states (1, 2 and 3), while the second slowest process
describes transitions from the completely unfolded state 12 to
partially unfolded states 6--10.
For dihedrals, the slowest timescale is significantly lower ($\sim
\SI{0.7}{\micro\second}$), followed by two clearly separated timescales
($\sim 0.2$ and \SI{0,1}{\micro\second}). The slowest process
corresponds again to the transition from unfolded (12, 11) to folded
states (1, 2 and 3), while the second eigenvector mostly accounts for
transitions from state 1 to state 2 and 3 in the native basin.
The significantly longer first implied timescale found for contacts
(\SI{1.2}{\micro\second}) compared to dihedrals
(\SI{0.7}{\micro\second}) is a consequence of the more apparent
timescale separation exhibited by the corresponding first PCs (Fig.~\SMPCA).

As a standard test of the quality of the resulting MSM, we next
check the validity of the Chapman-Kolmogorov equation\cite{Prinz11}
\begin{equation} \label{eq:CKtest}
    T(n\tau_{\text{lag}}) = T(\tau_{\text{lag}})^n
\end{equation}
with $n=1,2,3,\ldots$. Assuming that we start at time $t=0$ in a
specific macrostate, we can compare the MSM prediction of the decay of
this state to the corresponding results obtained from the MD data.
As shown in Fig.~\SMCKtest{b} for both contacts and dihedrals, we
obtain excellent agreement between MD and MSM results for most states
already for relatively short lag times
$\tau_{\text{lag}} \gtrsim \SI{10}{\nano\second}$.  For contacts,
exceptions include the lowly (\SI{0.9}{\%}) populated unfolded state 11,
and---to a much minor extent---state\,2 of the native basin, whose
evolution around \SI{100}{\nano\second} is only qualitatively reproduced. For
dihedrals, all states of the native basin pass the Chapman-Kolmogorov
test perfectly, while the main unfolded states 11 and 12 perform only
qualitatively.

%
%%%%%%%%%%%%%%%%%%%%%%%%%%%%%%%%%%%%%%%%%%%%%%%%%%%%%%%%%%%%%%%%%%%%%%
%
\subsection{Folding times}

To see how the implied timescales translate to measurable observables
of the folding process, we next consider the distribution of the
folding time $t_{\text{fold}}$, defined as the waiting time for the
transition of the completely unfolded state\,12 to the native
state\,1. To this end, we employed $10^9$ steps of a Markov chain
Monte Carlo propagation, where a trajectory is sampled from the given
transition matrix by drawing random numbers which determine the next
step. Shown in Fig.~\ref{fig:timescales}b as a function of the lag
time $\tau_{\text{lag}}$, the resulting folding time is found to
increase only little with $\tau_{\text{lag}}$. While the median of the
MSM prediction of $t_{\text{fold}}$ somewhat underestimates the MD
result, overall the MSM reproduces the rather broad MD folding-time
distributions for both contacts and dihedrals quite well.

Showing only about thirty folding events, the finite sampling of the
MD data also limits the prediction quality of the MSM. This
is demonstrated in Fig.~\ref{fig:timescales}c which compares the
histogram of folding times obtained from the MD data to
distributions obtained from MSMs using various lag times.
As the number of MD events is clearly too small to give a smooth
distribution, it is hard to assess the true deviation of the MSM from
(non-existent) statistically converged MD data.

%
%%%%%%%%%%%%%%%%%%%%%%%%%%%%%%%%%%%%%%%%%%%%%%%%%%%%%%%%%%%%%%%%%%%%%%
%
\subsection{Comparison to previous works}

In previous works, we have used the HP35 trajectory by Piana et
al.\cite{Piana12} to validate various methods of our MSM workflow,
including MPP, \cite{Jain14} robust density-based clustering,
\cite{Sittel16} dPCA+, \cite{Sittel17} and dynamic
coring. \cite{Nagel19} Moreover, the same MD data were employed by
Damjanovic et al.,\cite{Damjanovic21} who combined density-based
clustering with segment splitting, and Klem et al.\cite{Klem22} who
employed Gaussian mixture models for structural clustering. All these
works used backbone dihedral angles as features for the MSM, and can
therefore be directly compared to the dihedral-based MSM of this
work.

As an overview, Fig.~\SMPreviousModels{a} shows Sankey plots that compare the
metastable states obtained by the above-mentioned works to the states
of our present MSM. The 16-state MSM of Damjanovic et
al.\cite{Damjanovic21} yields a quite similar description of the
native basin by three main states, while the main unfolded states are
split up in several substates. Containing only four states, the rather
coarse-grained MSM of Klem et al.\cite{Klem22} cannot
resolve the main native states, but qualitatively reproduces the two
unfolded states of our model. We also considered the 12-state MSM
of Sittel et al.,\cite{Sittel17} which nicely resolved the free energy
landscape of the native basin and also yields several distinct
unfolded states. Keeping in mind that MSMs based on backbone dihedral angles
only allow for an approximate description of the unfolded basin, the
qualitative correspondence of the various state partitionings---in
particular for the native basin---appears satisfactory.
We also considered the first implied timescales and the folding time
of the various models (Fig.~\SMPreviousModels{b,c}), and found that
the various models underestimate the results of the new model on
average by a factor 2. Comparing the various MSM workflows, we
attribute this significant improvement of our model mainly to the
Hummer-Szabo projection (yielding on average \SI{40}{\%} longer
timescales, see Fig.~\ref{fig:timescales}a) and to the Gaussian
filtering, which was introduced in Sec.~\ref{sec:filter} to smooth
high-frequent fluctuations of the feature trajectory.

To discuss the effects of the Gaussian filtering in more detail, we
compare in Fig.~\SMfiltering{a} the results obtained from (i)
filtering (as discussed above), (ii) dynamic coring
\cite{Jain14,Nagel19} instead of filtering, and (iii) using no dynamic
correction at all. By requesting that the trajectory spends a minimum
time of \SI{3}{\nano\second} in the new state for the transition to be
counted, dynamic coring overall achieves somewhat shorter timescales
compared to filtering; e.g., we obtain $t_1=0.9$ and
$\SI{1.2}{\micro\second}$ in the case of contacts for
$\tau_\text{lag}=\SI{10}{\nano\second}$. Without any dynamic
correction, on the other hand, the results deteriorate considerably
(e.g., $t_1=0.7$ instead of $\SI{1.2}{\micro\second}$), particularly
for short lag times.

Apart from improving the implied timescales of the MSM, filtering can
be performed before clustering and therefore may avoid the
misclassification of points in the transition regions. To study this
aspect, Fig.~\SMfiltering{b} shows the contact representation of the
metastable states obtained from dynamic coring, which can be compared
to the results obtained for filtering (Fig.~\ref{fig:StateRep}).
Interestingly, we notice that the cored states are structurally less
clearly defined than the filtered states. For contacts, for example,
we find than the three main native states obtained from filtering are
merged into a single cored state, and that the cored states of the
unfolded basin are less distinct than their filtered counterparts. As
a further advantage of filtering, we mention that the reassignment of
trajectory frames carried out in dynamic coring makes it difficult to
subsequently apply the Hummer-Szabo projection, as this relies on a
clear correspondence of micro- and macrostates. Hence, with respect
to both structural characterization and slowest timescales, Gaussian
filtering represents a clear improvement over dynamic coring.

%\newpage
%
%%%%%%%%%%%%%%%%%%%%%%%%%%%%%%%%%%%%%%%%%%%%%%%%%%%%%%%%%%%%%%%%%%%%%%
%%%%%%%%%%%%%%%%%%%%%%%%%%%%%%%%%%%%%%%%%%%%%%%%%%%%%%%%%%%%%%%%%%%%%%
%
\section{Results on the folding of HP35}
\subsection{Ground truth observations}

As the above described procedure to construct an MSM comes with the
choice of a number of methods and associated metaparameters, it is
instructive to first discuss some results obtained directly from the
MD simulation. Shown in Fig.~\ref{fig:HP35}c, our first examples
include the RMSD of the MD trajectory from the crystal structure, the
percentage of native contacts $Q$, as well as the sum $\Psi$ over the
backbone dihedral angles $\psi$ of the three $\alpha$-helices.
Constructed directly from the MD data,\cite{Piana12} these
and related results may be considered as unbiased `ground
truth'.

We choose an upper (lower) threshold of \SI{2}{\angstrom}\
(\SI{6}{\angstrom}) for the RMSD of folded (unfolded) conformations
and employ Gaussian filtering ($\sigma =\SI{5}{ns}$), which yields 33
folding events and 32 unfolding events shown by the
\SI{300}{\micro\second} trajectory. The associated free energy profile
$\Delta G({\rm RMSD})$ consists of two states, showing a sharp minimum
for the native state and a shallow minimum reflecting the unfolded
state.
Interestingly, we find that the transition path time (i.e, the time
during a folding transition) of typically tens of nanoseconds is
significantly shorter than the folding time (i.e, the waiting time in
the unfolded state) of some microseconds. This indicates a concerted
or cooperative process, where all involved coordinates change at the
same time. Nonetheless, the wide barrier of $\sim 4 k_{\rm B}T$ height
as well as the shallow unfolded state seem to indicate the existence
of intermediate states.

Defining the percentage of formed native contacts as
$Q = \sum_i \xi_i(t)/N$ (with $\xi_i(t) =0,1$ if contact $i$ is broken
or formed, and $N$ being the total number of native contacts), the time
evolution of $1-Q$ is found to highly correlate with the RMSD. This
holds in particular for the unfolded parts of the trajectory, while
the folded parts match less. Consequently, the resulting free energy
profile $\Delta G(1-Q)$ reproduces well the unfolded part of
$\Delta G({\rm RMSD})$, while it shows a somewhat broader native
state.
The situation is the other way round for the time evolution of
$\Psi$ representing the normalized sum of helical dihedral angles
$\psi$, which accurately reproduces the RMSD in the native state,
but exhibits considerably more fluctuations in the unfolded
state. This is reflected in a sharp minimum of $\Delta G(\Psi)$ for
the native state and a largely unstructured unfolded region without a
minimum.

Apart from providing one-dimensional reaction coordinates, local
coordinates such as contact distances and dihedral angles allow for a
microscopic description of the folding process. As an example,
Fig.~\SMcoopTraj{a} shows the time evolution of the native contacts close
to a folding event. By partitioning the contacts in highly correlated
MoSAIC clusters (Fig.~\ref{fig:mosaic}), we find (as expected) that
the contacts in a cluster mostly move in a concerted manner. What is
more, we notice that at each folding event {\em all} still open
contacts are formed almost simultaneously, i.e., within the transition
path time of a few tens of nanoseconds. Hence, while the various
clusters in general evolve differently in time, a successful folding
transition involves the concerted forming of all contacts at the
same time.\cite{Chong21}
We note that this cooperative behavior
may explain the relatively long folding times ($\sim \SI{2}{\micro\second}$) in
spite of the modest energy barriers (a few $k_{\rm B}T$) shown by the
various one-dimensional energy landscapes in Fig.~\ref{fig:HP35}c.
Similar conclusions were also reached for the
open-close functional motion of T4 lysozyme \cite{Post22a} and the
allosteric transition in PDZ3 domain.\cite{Ali22}

While during a folding event all still open contacts form
cooperatively within a few tens of nanoseconds, it is nonetheless
interesting to study the order of the contact changes during that
short time, because this might indicate a potential causal relation
between the contact clusters. Figure~\ref{fig:mosaic}c shows that
typically clusters 4 and 5 (containing contacts connecting helices 2
and 3) form first, followed by cluster 6 (containing contacts at the
C-terminus). On the other hand, we find that usually either cluster 7
(that connects the two terminal ends) or clusters 1, 2 and 3
(containing contacts connecting helices 1 and 2) form lastly. That
is, we observe a preferred but not mandatory cluster formation order,
which indicates the existence of multiple folding pathways.

Cooperative behavior is also found for the $\psi$-dihedral
angles (Fig.~\SMcoopTraj{b}), which generally exhibit correlated
motion within a specific helix, and move all together during a folding
event. As indicated from the Leiden clustering of these angles
(Fig.~\SMdihedrals), however, the correlation of dihedral angles is overall
less distinctive than for contact distances.

Recalling that the metastable states of HP35 are well characterized in
terms of their contact clusters (Fig.~\ref{fig:StateRep}), the
question arises if we can turn the argument around and use the contact
clusters to construct conformational states. For example, we could
define a product state ($a,b,c,\ldots$) that indicates if the various
clusters $a$, $b$, $c$, etc. are formed or not. Alternatively, we could
use dihedral angles to define a product state ($\alpha_1,\, \alpha_2,\,\alpha_3$)
indicating which helices of HP35 are formed.
Similar approaches to employ sub-divisions of the protein structure
as features have been discussed previously.\cite{Elmer05II,Jain10}
Various attempts along these lines to build an MSM, however, showed
that representations built from contact clusters or helices are too
coarse grained to yield metastable conformational states that
accurately reproduce the dynamics of the system.

%\newpage
%
%%%%%%%%%%%%%%%%%%%%%%%%%%%%%%%%%%%%%%%%%%%%%%%%%%%%%%%%%%%%%%%%%%%%%%
%
\subsection{Kinetic network and folding pathways}

By constructing the above MSMs of HP35, we found 12 metastable states
with well-defined structures (Fig.~\ref{fig:StateRep}) and their
transition rates (Tab.~\SMTabTransMatrix). To connect these findings
with the underlying folding process, it is common practice to
illustrate the MSM via a network, where the node sizes correspond to the
population $\pi_i$ of the states and the edge weights $f_{ij}$ to the
transition probabilities $T_{ij}$. Since complex systems typically
exhibit too many edges to visualize both transition probabilities, it is
instructive to define a `kinetic distance' between each pair of
states.\cite{Noe16} Here we use the symmetric edge weight\cite{Nagel20}
\begin{equation} \label{eq:edgeweight}
f_{ij} = \pi_i T_{ij} = \pi_j T_{ji} = f_{ji},
\end{equation}
which exploits the detailed balance between the two states. To
optimally represent all resulting interstate distances in two
dimensions, we employ the force-directed algorithm
ForceAtlas2,\cite{ForceAtlas2} using a cut-off $f_\epsilon$ to
discard small fluxes (i.\,e., $f_{ij} \le f_\epsilon = \num{2e-5}$).
This leads to a kinetic network representation, in which high
transition rates correspond to closeness in the graph.

\begin{figure}[t!]
    \centering
    \includegraphics{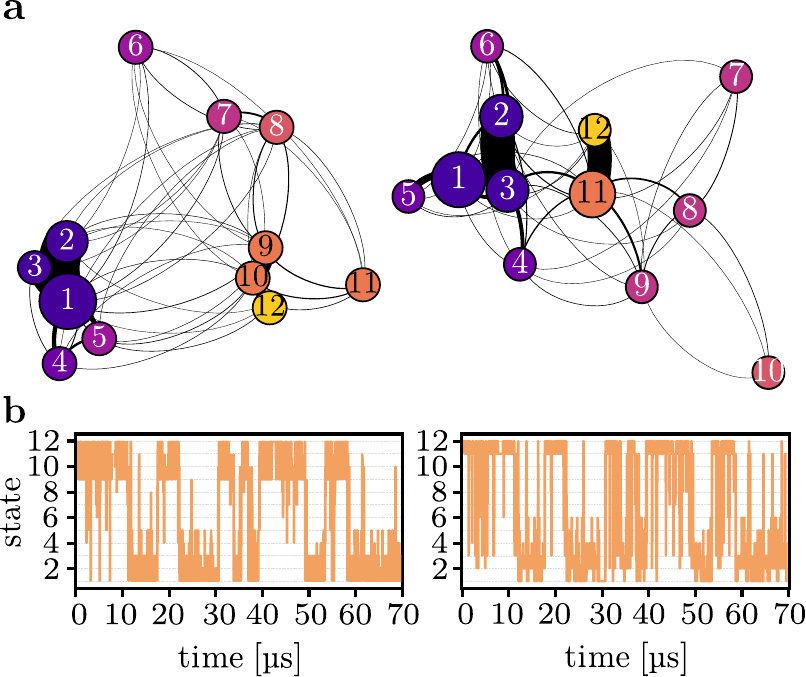}
    \caption{
      (a) Kinetic networks of the MSM and (b)
      exemplary state trajectories obtained
      for (left) contacts and (right) dihedrals. The node size
      indicates the population of the state and the color code
      reflects its fraction of native contacts $Q$ (cf.\
      Fig.~\ref{fig:dendrogram}).}
    \label{fig:networks}
\end{figure}

Figure~\ref{fig:networks}a shows the resulting kinetic networks
obtained for contacts and dihedrals. As anticipated from the MPP
dendrograms in Fig.~\ref{fig:dendrogram}, both networks show a native
basin comprising states 1 to 5 (1 to 6 for dihedrals), which is
clearly separated from the unfolded basin comprising states 9 to 12
(11 to 12 for dihedrals). Moreover, several lowly populated states
exist in the vicinity of the unfolded basin. To illustrate the
kinetics of the networks, it is instructive to inspect the
corresponding state trajectories obtained form the MD simulation
(Fig.~\ref{fig:networks}b). For contacts, we find fast
interconversion between states 1--5 of the native basin, as well as
between states 9--12 of the unfolded basin. The lowly populated
states 6--8 are either temporarily visited from the unfolded basin or
used as on-route intermediate states on the folding pathway. For
dihedrals, we also find fast interconversion between the states (1--6)
of the native basin and the states (11--12) of the unfolded
basin. The lowly populated states 7--10 are mostly approached from
the unfolded basin; only states 8 and 9 act in part as intermediate
states for folding.

As the MSM approximates well the timescales of the system
(Fig.~\ref{fig:timescales}a), we may illustrate various aspects of the
folding process by performing Markov chain Monte Carlo simulations of
the MSM. For example, by starting in the unfolded state\,12 and
calculating the time-dependent rise of the population of the native
state\,1, $P_{12\rightarrow 1}(t)$, we can infer from a single
exponential fit the overall folding time
$t_{\text{fold}}=\SI{1.7}{\micro\second}$ ($\SI{1.8}{\micro\second}$
for dihedrals), see Fig.~\ref{fig:timescales}d. (For clarity, we focus
on cumulative population probabilities, i.e., we disregard all
following reactions such as subsequent unfolding.) We may also study
the various subprocesses underlying folding. Starting in the
completely unfolded state\,12, the first step is the hydrophobic
collapse of the protein to the other states of the unfolded basin,
where (at least) contact clusters 4 and 5 (containing contacts
connecting helices 2 and 3) are formed. We note that this finding is
in line with our discussion of the formation of the contact clusters
in Fig.~\ref{fig:mosaic}c. Considering the time evolution of the
population of these states, $P_{12\rightarrow \text{U}}(t)$, we infer
the time $t_{12\rightarrow \text{U}}=\SI{160}{\nano\second}$
(\SI{43}{\nano\second} for dihedrals) for the hydrophobic
collapse. Next, we consider the time that the system remains in the
unfolded basin before leaving to the native basin. Deduced from the
rise of the sum of populations of all states of the native basin,
$P_{12 \rightarrow \text{N}}(t)$, we obtain
$t_{12 \rightarrow \text{N}}=\SI{1.6}{\micro\second}$
(\SI{0.9}{\micro\second} for dihedrals), indicating that the escape
from the unfolded basin clearly represents the slowest step of the
folding process.
As a final step, we consider the system's relaxation in the native
basin, i.e., the time $t_{\text{N}\rightarrow 1}$ it takes after
entering the native basin until state\,1 is reached. Interestingly, we
find that the native basin relaxation time obtained for contacts
($t_{\text{N}\rightarrow 1}=\SI{54}{\nano\second}$) is considerably
shorter than the result for dihedrals
($t_{\text{N}\rightarrow 1}=\SI{280}{\nano\second}$), which reflects
the higher metastability of the native states (in particular of
state\,2) found for dihedrals (Tab.~\SMTabTransMatrix).

Based on these general considerations, we now consider the main
folding pathways from the completely unfolded state\,12 to the native
state\,1, using MSMPathfinder\cite{Nagel20} for a systematic
construction of the path ensemble. Comparing the most frequented
folding paths as obtained from the MD trajectory and from the MSM,
Tab.~\SMTabPaths{} reveals good overall agreement between the results
of MD and MSM. For contacts, in most cases (26 out of 34 in MD) the
system first goes to state\,10. Moreover, we find six first
transitions to state\,9, as well as two direct transitions to state\,5
in the native basin. As discussed above, the system subsequently
spends most of the folding time $t_{\text{fold}}$ in the unfolded
basin, before it changes to the native basin via a concerted forming
of all still open contacts. That is, the broad distribution of folding
times shown in Figs.~\ref{fig:timescales}b,c originates mostly from
different escape times $t_{\text{U}\rightarrow \text{N}}$.
While we obtain similar results for dihedrals, we note that the
description of the unfolded basin is less well resolved. That is, from
completely unfolded state\,12 the first step goes with only a single
exception to state\,11, from where the native basin is approached.

Compared to the experimental result (\SI{0.73}{\micro\second} at
\SI{360}{\kelvin} for the considered Nle/Nle-mutant\cite{Kubelka06}),
the MD simulation of Piana et al.\cite{Piana12} overestimates the
overall folding time by a factor $2.5$. Moreover, various
experiments\cite{Brewer07, Kubelka06, Kubelka08} reported a fast
($\sim \SI{0.1}{\micro\second}$) transient, which was interpreted as
relaxation in the unfolded basin or in the folded basin. This compares
roughly to the associated timescales $t_{12\rightarrow \text{U}}$ and
$t_{\text{N}\rightarrow 1}$ discussed above.  Relating the factor 2.5
to a free energy difference $\Delta$ via $2.5=e^{\Delta/k_{\rm B}T}$,
we infer a quite low overall error
$\Delta \sim \SI{2.8}{\kilo\joule/\mole}$ of the employed force field,
which is in line with previous studies.\cite{Mittal10,Piana11}
Nonetheless, this accuracy of biomolecular force fields appears
surprisingly good, in particular when we consider energy barrier
heights associated with microsecond timescales.  This finding might be
a consequence of folding being eventually mediated by the forming of
interresidue contacts (which should be well described by common force
fields) and the relatively long folding times are related to the
finding that a successful folding transition involves the concerted
forming of all contacts that are still open.

%\newpage
%
%%%%%%%%%%%%%%%%%%%%%%%%%%%%%%%%%%%%%%%%%%%%%%%%%%%%%%%%%%%%%%%%%%%%%%
%%%%%%%%%%%%%%%%%%%%%%%%%%%%%%%%%%%%%%%%%%%%%%%%%%%%%%%%%%%%%%%%%%%%%%
%
\section{Discussion}
\subsection{Feature selection: Contacts vs.\ dihedrals}

Adopting the ultrafast folding of HP35 as well-established model
problem, we have studied the virtues and shortcomings of using
backbone dihedral angles or contact distances as features to construct
an MSM.
While dihedral angles are readily obtained from the MD trajectory,
they require an appropriate treatment of their periodicity
\cite{Altis07,Sittel17,Zoubouloglou22} and necessitate the exclusion
of uncorrelated dihedral motion. Following recent
work,\cite{Sittel17,Nagel20} here we used maximal-gap shifted
($\phi, \psi$) dihedral angles and excluded the terminal residues 1--2
and 34--35. With $\psi$-angles reflecting the
helicity of the protein and $\phi$-angles accounting for potential
left-to-right handed transitions (mainly in flexible loops), backbone
dihedral angles report directly on the local secondary structure. This
proves advantageous for the modeling of the conformational states in
the native basin of HP35, which can be mainly described by the single
angle $\phi_3$ (Fig.~\SMclustering{c}). However, dihedral angles
account only indirectly for the formation of tertiary structure during
folding, which hampers the modeling of the folding transition and the
conformational distribution in the unfolded basin.

Considering contacts, on the other hand, we have found that their
selection and appropriate calculation of the corresponding contact
distances requires some attention. By introducing a new definition of contact distances [Eq.~\eqref{eq:MinDist2}], we focused on the native contacts
of HP35, because they are expected to largely determine the folding
pathways.\cite{Sali94,Wolynes95,Best13}
Performing a correlation analysis \cite{Diez22} on the resulting 42
native contacts, we identified seven clusters (Fig.~\ref{fig:mosaic}),
whose contacts are highly correlated and change in
a concerted manner. These contact clusters were shown to directly
account for the cooperative folding of HP35 (Fig.~\SMcoopTraj{a}),
and also provide a concise characterization of the metastable states
of the system (Fig.~\ref{fig:StateRep}).
In particular, tertiary contacts
were shown to be key to the folding process, because they represent
the most direct descriptor of the origin of folding. As a consequence,
the first implied timescales of the MSMs obtained for contacts (e.g.,
$t_1 = \SI{1.2}{\micro\second}$ for $\tau_\text{lag} = \SI{10}{\nano\second}$) are
significantly slower than the corresponding results obtained for dihedrals
(\SI{0.7}{\micro\second}), indicating that contacts overall give a better
Markovian model than dihedrals.
Hence, reflecting different aspects of the structural dynamics,
contacts and dihedrals result in different collective variables and
metastable states, eventually leading to MSMs with different
timescales and pathways.

%\newpage
%
%%%%%%%%%%%%%%%%%%%%%%%%%%%%%%%%%%%%%%%%%%%%%%%%%%%%%%%%%%%%%%%%%%%%%%
%
\subsection{MSM workflow: What matters?}

While we have stressed the importance of the selection of features,
the discussion of the results obtained from the MSM inevitably rests
on the specific techniques employed for its construction, including
various methods of dynamic correction, dimensionality reduction, and
clustering. To briefly summarize what parts of the procedure made a
difference, we focus on the case of contacts in the following.

Although a number of more sophisticated methods is available,
\cite{Rohrdanz13, Wang20,Glielmo21} we suppose that standard linear
PCA is sufficient for the present case study (provided that
appropriate features are available). Including the first five
components, PCA was shown to explain the majority (\SI{80}{\%}) of the
correlation and to account well for the slowest timescales
(Fig.~\SMPCA). As a benefit, linear methods produce smoothly varying free
energy landscapes, which facilitate the subsequent clustering.
For the latter we adopted robust density-based clustering\cite{Sittel16} as an
accurate and efficient deterministic method to construct microstates. Presumably,
the popular $k$-means method should give clusterings of similar
overall quality, provided that the required metaparameters (such as
the number of states $k$) are optimized and that sufficiently many
iterations are run.\cite{Husic16}

Apart from these commonly used methods, we want to point out two
simple and powerful techniques, which are less widely known but
clearly improved the outcome of the MSM. First of all, we used Gaussian
low-pass filtering to smoothen high-frequent fluctuations of the feature
trajectory, which significantly reduced spurious transitions between
adjacent metastable states. While the idea is
similar to the definition of dynamic cores of the metastable
states\cite{Jain14,Nagel19} (where we request that the trajectory
spends a minimum time in the new state for the transition to be
counted), the filtering is performed before the clustering and
therefore helps to avoid the misclassification of points in the
transition regions.

Secondly, we want to highlight the explanatory power of the dendrograms
constructed from MPP clustering (Fig.~\ref{fig:dendrogram}).
Revealing how the metastable states emerge from
the microstates, the MPP dendrogram outlines the hierarchical
structure of the free energy landscape.\cite{Jain12} For example, we
learn that the unfolded part of the energy landscape of HP35 is
structured and reveals several metastable states with lifetimes of
about \SI{40}{\nano\second} (Tab.~\SMTabTransMatrix). Furthermore, we notice
that the dendrogram for dihedrals exhibits two rather metastable subsections of
the native basin, which explain the relatively long relaxation time
found for this basin (Fig.~\ref{fig:timescales}). By coloring the
microstates according to their mean number of native contacts $Q$, we
may assess if only structurally similar states are merged together,
thus illustrating the quality of the state partitioning. Last but not least,
the MPP dendrogram is closely related to the kinetic network
(Fig.~\ref{fig:networks}) that nicely illustrates the overall dynamics
exhibited by the MSM.

We finally wish to discuss the scalability of the presented MSM
workflow to larger systems, such as proteins with $N\!\approx\!10^3$
residues. This concerns particularly the first steps of the workflow,
that is, the definition of features and the dimensionality reduction,
while the computational effort for all subsequent steps essentially
remains the same (see the discussion at the beginning of
Sec.~\ref{sec:metastable_states}). Backbone dihedral angles naturally
scale linearly with $N$, and this holds approximately also for
interresidue contacts if local search algorithms are used (see the
discussion at the end of Sec.~\ref{sec:contacts}). As a consequence,
the definition of dihedrals requires about an hour and the definition
of contacts a few CPU days for a 1000 amino-acid protein (assuming
again $1.5 \times 10^6$ data points).
Using the resulting $\approx 10^3$ coordinates, the corresponding
correlation matrix is readily block-diagonalized via the Leiden
algorithm (Sec. 2.2), and the relevant main correlated clusters can be
diagonalized by a PCA, taking some CPU hours. This means that even for
large proteins the complete MSM workflow including all analyses takes
not more than a few CPU days on a standard desktop computer.

%
%%%%%%%%%%%%%%%%%%%%%%%%%%%%%%%%%%%%%%%%%%%%%%%%%%%%%%%%%%%%%%%%%%%%%%
%%%%%%%%%%%%%%%%%%%%%%%%%%%%%%%%%%%%%%%%%%%%%%%%%%%%%%%%%%%%%%%%%%%%%%
%
\section{Concluding remarks}

By designing an MSM, we aim to construct structurally well-defined
metastable states that provide a mechanistic understanding of the
considered biomolecular process. Adopting the folding of HP35 as a
textbook example, we have highlighted several important aspects for
the practical construction of an MSM.
\begin{itemize}
\item First, the overall quality of a dynamical model such as an MSM
  heavily relies on the selection of input coordinates or features,
  which faithfully account for the process under consideration.
  Inclusion of deceiving coordinates or the omission of important
  features will quite certainly corrupt subsequent analyses. We have
  employed a correlation analysis strategy to identify appropriate
  features,\cite{Diez22} and discussed in detail two main types, that
  is, interresidue contacts and backbone dihedral angles.
\item A successful state partitioning should provide conformational
  states that are clearly discriminated by some descriptors derived
  from the features, such as the mean and the variance of the contact
  cluster distributions (Fig.~\ref{fig:StateRep}). The
  quality of the state partitioning can be furthermore assessed by the
  MPP dendrogram (Fig.~\ref{fig:dendrogram}), which reveals potential
  conformational heterogeneity of the microstates being merged to a
  macrostate.
\item Dynamical corrections such as low-pass filtering of the feature
  trajectory and the optimal projection of microstate dynamics onto
macrostate dynamics\cite{Roeblitz13,Hummer15,Kells19,Cao20,Sharpe21}
may significantly improve the Markovianity of
  the MSM.
\end{itemize}

Proceeding this way, we have obtained MSMs describing the folding of
HP35, which correctly reproduce the slow timescales of the
process. This result is not achieved by design, but indicates a
consistent dynamical model. Alternatively, we may construct an MSM by
requesting long timescales from the outset.\cite{Nueske14,Wu20} While
this approach may simplify the construction of the MSM, it still needs
to be checked if the resulting metastable states are structurally well
characterized and do provide the desired mechanistic understanding of
the process.

\begin{acknowledgement}
  The authors thank Georg Diez, Matthias Post and Steffen Wolf for
  helpful comments and discussions, D.\ E.\ Shaw Research for
  sharing their trajectories of HP35, as well as Martin McCullagh and Yu-Shan
  Lin for sharing their MSMs of HP35. This work has been supported by
  the Deutsche Forschungsgemeinschaft (DFG) within the framework of
  the Research Unit FOR 5099 ''Reducing complexity of nonequilibrium''
  (project No.~431945604), the High Performance and Cloud Computing
  Group at the Zentrum f\"ur Datenverarbeitung of the University of
  T\"ubingen, the state of Baden-W\"urttemberg through bwHPC and the
  DFG through grant no INST 37/935-1 FUGG (RV bw16I016), and the Black
  Forest Grid Initiative.
%, and the Freiburg Institute for
%Advanced Studies (FRIAS) of the Albert-Ludwigs-University Freiburg.
\end{acknowledgement}

\subsection*{Data Availability Statement}
The simulation data and all intermediate results,
including our Python package msmhelper and detailed descriptions to
reproduce all steps of the analyses, can be downloaded from
https://github.com/moldyn/HP35. In particular, we provide trajectories
($\SI{300}{\micro\second}$ long, $1.5 \times 10^6$ data points) of (1)
the three discussed definitions of contact distances, (2) the
maximal-gap shifted dihedral angles, (3) the resulting principal
components, and (4) the resulting micro- and macrostates.

\begin{suppinfo}
Tables of the native contacts of HP35, as well as of
the transition matrices and folding pathways obtained for contacts and
dihedral angles. Figures of the cooperative transition of contacts and
dihedral angles, the correlation analysis for dihedral angels, details
on PCA and clustering, the Chapman-Kolmogorov tests, and comparisons
to other MSMs.
\end{suppinfo}

\bibliography{\dir/stock,\dir/md,new}

\providecommand{\latin}[1]{#1}
\makeatletter
\providecommand{\doi}
  {\begingroup\let\do\@makeother\dospecials
  \catcode`\{=1 \catcode`\}=2 \doi@aux}
\providecommand{\doi@aux}[1]{\endgroup\texttt{#1}}
\makeatother
\providecommand*\mcitethebibliography{\thebibliography}
\csname @ifundefined\endcsname{endmcitethebibliography}
  {\let\endmcitethebibliography\endthebibliography}{}
\begin{mcitethebibliography}{96}
\providecommand*\natexlab[1]{#1}
\providecommand*\mciteSetBstSublistMode[1]{}
\providecommand*\mciteSetBstMaxWidthForm[2]{}
\providecommand*\mciteBstWouldAddEndPuncttrue
  {\def\EndOfBibitem{\unskip.}}
\providecommand*\mciteBstWouldAddEndPunctfalse
  {\let\EndOfBibitem\relax}
\providecommand*\mciteSetBstMidEndSepPunct[3]{}
\providecommand*\mciteSetBstSublistLabelBeginEnd[3]{}
\providecommand*\EndOfBibitem{}
\mciteSetBstSublistMode{f}
\mciteSetBstMaxWidthForm{subitem}{(\alph{mcitesubitemcount})}
\mciteSetBstSublistLabelBeginEnd
  {\mcitemaxwidthsubitemform\space}
  {\relax}
  {\relax}

\bibitem[Berendsen(2007)]{Berendsen07}
Berendsen,~H. J.~C. \emph{Simulating the Physical World}; Cambridge University
  Press: Cambridge, 2007\relax
\mciteBstWouldAddEndPuncttrue
\mciteSetBstMidEndSepPunct{\mcitedefaultmidpunct}
{\mcitedefaultendpunct}{\mcitedefaultseppunct}\relax
\EndOfBibitem
\bibitem[Lange and {Grubm\"uller}(2006)Lange, and {Grubm\"uller}]{Lange06b}
Lange,~O.~F.; {Grubm\"uller},~H. Collective {Langevin} dynamics of
  conformational motions in proteins. \emph{J. Chem. Phys.} \textbf{2006},
  \emph{124}, 214903\relax
\mciteBstWouldAddEndPuncttrue
\mciteSetBstMidEndSepPunct{\mcitedefaultmidpunct}
{\mcitedefaultendpunct}{\mcitedefaultseppunct}\relax
\EndOfBibitem
\bibitem[Hegger and Stock(2009)Hegger, and Stock]{Hegger09}
Hegger,~R.; Stock,~G. Multidimensional {Langevin} modeling of biomolecular
  dynamics. \emph{J. Chem. Phys.} \textbf{2009}, \emph{130}, 034106\relax
\mciteBstWouldAddEndPuncttrue
\mciteSetBstMidEndSepPunct{\mcitedefaultmidpunct}
{\mcitedefaultendpunct}{\mcitedefaultseppunct}\relax
\EndOfBibitem
\bibitem[Ayaz \latin{et~al.}(2021)Ayaz, Tepper, Br\"unig, Kappler, Daldrop, and
  Netz]{Ayaz21}
Ayaz,~C.; Tepper,~L.; Br\"unig,~F.~N.; Kappler,~J.; Daldrop,~J.~O.; Netz,~R.~R.
  {Non-Markovian} modeling of protein folding. \emph{Proc. Natl. Acad. Sci.
  USA} \textbf{2021}, \emph{118}, e2023856118\relax
\mciteBstWouldAddEndPuncttrue
\mciteSetBstMidEndSepPunct{\mcitedefaultmidpunct}
{\mcitedefaultendpunct}{\mcitedefaultseppunct}\relax
\EndOfBibitem
\bibitem[Buchete and Hummer(2008)Buchete, and Hummer]{Buchete08}
Buchete,~N.-V.; Hummer,~G. Coarse master equations for peptide folding
  dynamics. \emph{J. Phys. Chem. B} \textbf{2008}, \emph{112}, 6057--6069\relax
\mciteBstWouldAddEndPuncttrue
\mciteSetBstMidEndSepPunct{\mcitedefaultmidpunct}
{\mcitedefaultendpunct}{\mcitedefaultseppunct}\relax
\EndOfBibitem
\bibitem[Bowman \latin{et~al.}(2009)Bowman, Beauchamp, Boxer, and
  Pande]{Bowman09}
Bowman,~G.~R.; Beauchamp,~K.~A.; Boxer,~G.; Pande,~V.~S. Progress and
  challenges in the automated construction of {Markov} state models for full
  protein systems. \emph{J. Chem. Phys.} \textbf{2009}, \emph{131},
  124101\relax
\mciteBstWouldAddEndPuncttrue
\mciteSetBstMidEndSepPunct{\mcitedefaultmidpunct}
{\mcitedefaultendpunct}{\mcitedefaultseppunct}\relax
\EndOfBibitem
\bibitem[Prinz \latin{et~al.}(2011)Prinz, Wu, Sarich, Keller, Senne, Held,
  Chodera, {Sch\"utte}, and No{\'e}]{Prinz11}
Prinz,~J.-H.; Wu,~H.; Sarich,~M.; Keller,~B.; Senne,~M.; Held,~M.;
  Chodera,~J.~D.; {Sch\"utte},~C.; No{\'e},~F. Markov models of molecular
  kinetics: generation and validation. \emph{J. Chem. Phys.} \textbf{2011},
  \emph{134}, 174105\relax
\mciteBstWouldAddEndPuncttrue
\mciteSetBstMidEndSepPunct{\mcitedefaultmidpunct}
{\mcitedefaultendpunct}{\mcitedefaultseppunct}\relax
\EndOfBibitem
\bibitem[Bowman \latin{et~al.}(2013)Bowman, Pande, and No{\'e}]{Bowman13a}
Bowman,~G.~R.; Pande,~V.~S.; No{\'e},~F. \emph{An Introduction to Markov State
  Models}; Springer: Heidelberg, 2013\relax
\mciteBstWouldAddEndPuncttrue
\mciteSetBstMidEndSepPunct{\mcitedefaultmidpunct}
{\mcitedefaultendpunct}{\mcitedefaultseppunct}\relax
\EndOfBibitem
\bibitem[Wang \latin{et~al.}(2018)Wang, Cao, Zhu, and Huang]{Wang17a}
Wang,~W.; Cao,~S.; Zhu,~L.; Huang,~X. Constructing {Markov} State Models to
  elucidate the functional conformational changes of complex biomolecules.
  \emph{WIREs Comp. Mol. Sci.} \textbf{2018}, \emph{8}, e1343\relax
\mciteBstWouldAddEndPuncttrue
\mciteSetBstMidEndSepPunct{\mcitedefaultmidpunct}
{\mcitedefaultendpunct}{\mcitedefaultseppunct}\relax
\EndOfBibitem
\bibitem[Scherer \latin{et~al.}(2015)Scherer, Trendelkamp-Schroer, Paul,
  Perez-Hernandez, Hoffmann, Plattner, Wehmeyer, Prinz, and No{\'e}]{Scherer15}
Scherer,~M.~K.; Trendelkamp-Schroer,~B.; Paul,~F.; Perez-Hernandez,~G.;
  Hoffmann,~M.; Plattner,~N.; Wehmeyer,~C.; Prinz,~J.-H.; No{\'e},~F. {PyEMMA}
  2: A Software Package for Estimation, Validation, and Analysis of {Markov}
  Models. \emph{J. Chem. Theory Comput.} \textbf{2015}, \emph{11}, 5525\relax
\mciteBstWouldAddEndPuncttrue
\mciteSetBstMidEndSepPunct{\mcitedefaultmidpunct}
{\mcitedefaultendpunct}{\mcitedefaultseppunct}\relax
\EndOfBibitem
\bibitem[Beauchamp \latin{et~al.}(2011)Beauchamp, Bowman, Lane, Maibaum, Haque,
  and Pande]{MSMBuilder}
Beauchamp,~K.~A.; Bowman,~G.~R.; Lane,~T.~J.; Maibaum,~L.; Haque,~I.~S.;
  Pande,~V.~S. {MSMBuilder2}: Modeling Conformational Dynamics on the
  Picosecond to Millisecond Scale. \emph{J. Chem. Theory Comput.}
  \textbf{2011}, \emph{7}, 3412--3419\relax
\mciteBstWouldAddEndPuncttrue
\mciteSetBstMidEndSepPunct{\mcitedefaultmidpunct}
{\mcitedefaultendpunct}{\mcitedefaultseppunct}\relax
\EndOfBibitem
\bibitem[N\"uske \latin{et~al.}(2014)N\"uske, Keller, Perez-Hernández, Mey,
  and No{\'e}]{Nueske14}
N\"uske,~F.; Keller,~B.~G.; Perez-Hernández,~G.; Mey,~A. S. J.~S.; No{\'e},~F.
  Variational Approach to Molecular Kinetics. \emph{J. Chem. Theory Comput.}
  \textbf{2014}, \emph{10}, 1739--1752\relax
\mciteBstWouldAddEndPuncttrue
\mciteSetBstMidEndSepPunct{\mcitedefaultmidpunct}
{\mcitedefaultendpunct}{\mcitedefaultseppunct}\relax
\EndOfBibitem
\bibitem[Wu and No{\'e}(2020)Wu, and No{\'e}]{Wu20}
Wu,~H.; No{\'e},~F. Learning {Markov} Processes from Time Series Data. \emph{J.
  Nonlinear. Sci.} \textbf{2020}, \emph{30}, 23–66\relax
\mciteBstWouldAddEndPuncttrue
\mciteSetBstMidEndSepPunct{\mcitedefaultmidpunct}
{\mcitedefaultendpunct}{\mcitedefaultseppunct}\relax
\EndOfBibitem
\bibitem[Sittel and Stock(2018)Sittel, and Stock]{Sittel18}
Sittel,~F.; Stock,~G. Perspective: Identification of Collective Coordinates and
  Metastable States of Protein Dynamics. \emph{J. Chem. Phys.} \textbf{2018},
  \emph{149}, 150901\relax
\mciteBstWouldAddEndPuncttrue
\mciteSetBstMidEndSepPunct{\mcitedefaultmidpunct}
{\mcitedefaultendpunct}{\mcitedefaultseppunct}\relax
\EndOfBibitem
\bibitem[Scherer \latin{et~al.}(2019)Scherer, Husic, Hoffmann, Paul, Wu, and
  {No\'e}]{Scherer19}
Scherer,~M.~K.; Husic,~B.~E.; Hoffmann,~M.; Paul,~F.; Wu,~H.; {No\'e},~F.
  Variational selection of features for molecular kinetics. \emph{J. Chem.
  Phys.} \textbf{2019}, \emph{150}, 194108\relax
\mciteBstWouldAddEndPuncttrue
\mciteSetBstMidEndSepPunct{\mcitedefaultmidpunct}
{\mcitedefaultendpunct}{\mcitedefaultseppunct}\relax
\EndOfBibitem
\bibitem[Ravindra \latin{et~al.}(2020)Ravindra, Smith, and Tiwary]{Ravindra20}
Ravindra,~P.; Smith,~Z.; Tiwary,~P. Automatic mutual information noise omission
  {(AMINO):} generating order parameters for molecular systems. \emph{Mol.
  Syst. Des. Eng.} \textbf{2020}, \emph{5}, 339--348\relax
\mciteBstWouldAddEndPuncttrue
\mciteSetBstMidEndSepPunct{\mcitedefaultmidpunct}
{\mcitedefaultendpunct}{\mcitedefaultseppunct}\relax
\EndOfBibitem
\bibitem[Konovalov \latin{et~al.}(2021)Konovalov, Unarta, Cao, Goonetilleke,
  and Huang]{Konovalov21}
Konovalov,~K.~A.; Unarta,~I.~C.; Cao,~S.; Goonetilleke,~E.~C.; Huang,~X.
  {Markov} State Models to Study the Functional Dynamics of Proteins in the
  Wake of Machine Learning. \emph{JACS Au} \textbf{2021}, \emph{1},
  1330--1341\relax
\mciteBstWouldAddEndPuncttrue
\mciteSetBstMidEndSepPunct{\mcitedefaultmidpunct}
{\mcitedefaultendpunct}{\mcitedefaultseppunct}\relax
\EndOfBibitem
\bibitem[Husic \latin{et~al.}(2016)Husic, McGibbon, Sultan, and Pande]{Husic16}
Husic,~B.~E.; McGibbon,~R.~T.; Sultan,~M.~M.; Pande,~V.~S. Optimized parameter
  selection reveals trends in {Markov} state models for protein folding.
  \emph{J. Chem. Phys.} \textbf{2016}, \emph{145}, 194103\relax
\mciteBstWouldAddEndPuncttrue
\mciteSetBstMidEndSepPunct{\mcitedefaultmidpunct}
{\mcitedefaultendpunct}{\mcitedefaultseppunct}\relax
\EndOfBibitem
\bibitem[not()]{note2}
{As the commonly employed rotational fit of the Cartesian trajectory to a
  reference structure cannot entirely remove the overall rotation of a flexible
  system, the residual overall rotation may completely destroy the outcome of a
  subsequent dimensionality reduction.\cite{Sittel14} Interestingly, this
  break-down may be not obvious, when we select input coordinates by maximizing
  the slowest timescales\cite{Scherer19}}\relax
\mciteBstWouldAddEndPuncttrue
\mciteSetBstMidEndSepPunct{\mcitedefaultmidpunct}
{\mcitedefaultendpunct}{\mcitedefaultseppunct}\relax
\EndOfBibitem
\bibitem[Mu \latin{et~al.}(2005)Mu, Nguyen, and Stock]{Mu05}
Mu,~Y.; Nguyen,~P.~H.; Stock,~G. Energy Landscape of a Small Peptide Revealed
  by Dihedral Angle Principal Component Analysis. \emph{Proteins}
  \textbf{2005}, \emph{58}, 45 -- 52\relax
\mciteBstWouldAddEndPuncttrue
\mciteSetBstMidEndSepPunct{\mcitedefaultmidpunct}
{\mcitedefaultendpunct}{\mcitedefaultseppunct}\relax
\EndOfBibitem
\bibitem[Sittel \latin{et~al.}(2014)Sittel, Jain, and Stock]{Sittel14}
Sittel,~F.; Jain,~A.; Stock,~G. Principal component analysis of molecular
  dynamics: On the use of {Cartesian} vs. internal coordinates. \emph{J. Chem.
  Phys.} \textbf{2014}, \emph{141}, 014111\relax
\mciteBstWouldAddEndPuncttrue
\mciteSetBstMidEndSepPunct{\mcitedefaultmidpunct}
{\mcitedefaultendpunct}{\mcitedefaultseppunct}\relax
\EndOfBibitem
\bibitem[Altis \latin{et~al.}(2008)Altis, Otten, Nguyen, Hegger, and
  Stock]{Altis08}
Altis,~A.; Otten,~M.; Nguyen,~P.~H.; Hegger,~R.; Stock,~G. Construction of the
  free energy landscape of biomolecules via dihedral angle principal component
  analysis. \emph{J. Chem. Phys.} \textbf{2008}, \emph{128}, 245102\relax
\mciteBstWouldAddEndPuncttrue
\mciteSetBstMidEndSepPunct{\mcitedefaultmidpunct}
{\mcitedefaultendpunct}{\mcitedefaultseppunct}\relax
\EndOfBibitem
\bibitem[Maisuradze \latin{et~al.}(2009)Maisuradze, Liwo, and
  Scheraga]{Maisuradze09a}
Maisuradze,~G.~G.; Liwo,~A.; Scheraga,~H.~A. Principal Component Analysis for
  Protein Folding Dynamics. \emph{J. Mol. Biol.} \textbf{2009}, \emph{385}, 312
  -- 329\relax
\mciteBstWouldAddEndPuncttrue
\mciteSetBstMidEndSepPunct{\mcitedefaultmidpunct}
{\mcitedefaultendpunct}{\mcitedefaultseppunct}\relax
\EndOfBibitem
\bibitem[Riccardi \latin{et~al.}(2009)Riccardi, Nguyen, and Stock]{Riccardi09}
Riccardi,~L.; Nguyen,~P.~H.; Stock,~G. Free energy landscape of an {RNA}
  hairpin constructed via dihedral angle principal component analysis. \emph{J.
  Phys. Chem. B} \textbf{2009}, \emph{113}, 16660 -- 16668\relax
\mciteBstWouldAddEndPuncttrue
\mciteSetBstMidEndSepPunct{\mcitedefaultmidpunct}
{\mcitedefaultendpunct}{\mcitedefaultseppunct}\relax
\EndOfBibitem
\bibitem[Fenwick \latin{et~al.}(2014)Fenwick, Orellana, Esteban-Martín,
  Orozco, and Salvatella]{Fenwick14}
Fenwick,~R.~B.; Orellana,~L.; Esteban-Martín,~S.; Orozco,~M.; Salvatella,~X.
  Correlated motions are a fundamental property of $\beta$-sheets. \emph{Nat.
  Commun.} \textbf{2014}, \emph{5}, 4070\relax
\mciteBstWouldAddEndPuncttrue
\mciteSetBstMidEndSepPunct{\mcitedefaultmidpunct}
{\mcitedefaultendpunct}{\mcitedefaultseppunct}\relax
\EndOfBibitem
\bibitem[{L\"atzer} \latin{et~al.}(2008){L\"atzer}, Shen, and
  Wolynes]{Laetzer08}
{L\"atzer},~J.; Shen,~T.; Wolynes,~P.~G. Conformational Switching upon
  Phosphorylation: A Predictive Framework Based on Energy Landscape Principles.
  \emph{Biochem.} \textbf{2008}, \emph{47}, 2110--2122\relax
\mciteBstWouldAddEndPuncttrue
\mciteSetBstMidEndSepPunct{\mcitedefaultmidpunct}
{\mcitedefaultendpunct}{\mcitedefaultseppunct}\relax
\EndOfBibitem
\bibitem[Hori \latin{et~al.}(2009)Hori, Chikenji, Berry, and Takada]{Hori09}
Hori,~N.; Chikenji,~G.; Berry,~R.~S.; Takada,~S. Folding energy landscape and
  network dynamics of small globular proteins. \emph{Proc. Natl. Acad. Sci.
  USA} \textbf{2009}, \emph{106}, 73--78\relax
\mciteBstWouldAddEndPuncttrue
\mciteSetBstMidEndSepPunct{\mcitedefaultmidpunct}
{\mcitedefaultendpunct}{\mcitedefaultseppunct}\relax
\EndOfBibitem
\bibitem[Kalgin \latin{et~al.}(2013)Kalgin, Caflisch, Chekmarev, and
  Karplus]{Kalgin13}
Kalgin,~I.~V.; Caflisch,~A.; Chekmarev,~S.~F.; Karplus,~M. New Insights into
  the Folding of a beta-Sheet Miniprotein in a Reduced Space of Collective
  Hydrogen Bond Variables: Application to a Hydrodynamic Analysis of the
  Folding Flow. \emph{J. Phys. Chem. B} \textbf{2013}, \emph{117},
  6092--6105\relax
\mciteBstWouldAddEndPuncttrue
\mciteSetBstMidEndSepPunct{\mcitedefaultmidpunct}
{\mcitedefaultendpunct}{\mcitedefaultseppunct}\relax
\EndOfBibitem
\bibitem[Swope \latin{et~al.}(2004)Swope, Pitera, Suits, Pitman, Eleftheriou,
  Fitch, Germain, Rayshubski, Ward, Zhestkov, and Zhou]{Swope04II}
Swope,~W.~C.; Pitera,~J.~W.; Suits,~F.; Pitman,~M.; Eleftheriou,~M.;
  Fitch,~B.~G.; Germain,~R.~S.; Rayshubski,~A.; Ward,~T. J.~C.; Zhestkov,~Y.
  \latin{et~al.}  Describing Protein Folding Kinetics by Molecular Dynamics
  Simulations. 2. Example Applications to Alanine Dipeptide and a
  $\beta$-Hairpin Peptide. \emph{J. Phys. Chem. B} \textbf{2004}, \emph{108},
  6582--6594\relax
\mciteBstWouldAddEndPuncttrue
\mciteSetBstMidEndSepPunct{\mcitedefaultmidpunct}
{\mcitedefaultendpunct}{\mcitedefaultseppunct}\relax
\EndOfBibitem
\bibitem[Best and Hummer(2005)Best, and Hummer]{Best05}
Best,~R.~B.; Hummer,~G. Reaction coordinates and rates from transition paths.
  \emph{Proc. Natl. Acad. Sci. USA} \textbf{2005}, \emph{102}, 6732--6737\relax
\mciteBstWouldAddEndPuncttrue
\mciteSetBstMidEndSepPunct{\mcitedefaultmidpunct}
{\mcitedefaultendpunct}{\mcitedefaultseppunct}\relax
\EndOfBibitem
\bibitem[Ernst \latin{et~al.}(2015)Ernst, Sittel, and Stock]{Ernst15}
Ernst,~M.; Sittel,~F.; Stock,~G. Contact- and distance-based principal
  component analysis of protein dynamics. \emph{J. Chem. Phys.} \textbf{2015},
  \emph{143}, 244114\relax
\mciteBstWouldAddEndPuncttrue
\mciteSetBstMidEndSepPunct{\mcitedefaultmidpunct}
{\mcitedefaultendpunct}{\mcitedefaultseppunct}\relax
\EndOfBibitem
\bibitem[Ernst \latin{et~al.}(2017)Ernst, Wolf, and Stock]{Ernst17}
Ernst,~M.; Wolf,~S.; Stock,~G. Identification and validation of reaction
  coordinates describing protein functional motion: Hierarchical dynamics of
  {T4 Lysozyme}. \emph{J. Chem. Theory Comput.} \textbf{2017}, \emph{13}, 5076
  -- 5088\relax
\mciteBstWouldAddEndPuncttrue
\mciteSetBstMidEndSepPunct{\mcitedefaultmidpunct}
{\mcitedefaultendpunct}{\mcitedefaultseppunct}\relax
\EndOfBibitem
\bibitem[Oide and Sugita(2022)Oide, and Sugita]{Oide22}
Oide,~M.; Sugita,~Y. Protein folding intermediates on the dimensionality
  reduced landscape with {UMAP} and native contact likelihood. \emph{J. Chem.
  Phys.} \textbf{2022}, \emph{157}, 075101\relax
\mciteBstWouldAddEndPuncttrue
\mciteSetBstMidEndSepPunct{\mcitedefaultmidpunct}
{\mcitedefaultendpunct}{\mcitedefaultseppunct}\relax
\EndOfBibitem
\bibitem[Amadei \latin{et~al.}(1993)Amadei, Linssen, and Berendsen]{Amadei93}
Amadei,~A.; Linssen,~A. B.~M.; Berendsen,~H. J.~C. Essential dynamics of
  proteins. \emph{Proteins} \textbf{1993}, \emph{17}, 412--425\relax
\mciteBstWouldAddEndPuncttrue
\mciteSetBstMidEndSepPunct{\mcitedefaultmidpunct}
{\mcitedefaultendpunct}{\mcitedefaultseppunct}\relax
\EndOfBibitem
\bibitem[Perez-Hernandez \latin{et~al.}(2013)Perez-Hernandez, Paul, Giorgino,
  De~Fabritiis, and No{\'e}]{Perez-Hernandez13}
Perez-Hernandez,~G.; Paul,~F.; Giorgino,~T.; De~Fabritiis,~G.; No{\'e},~F.
  Identification of slow molecular order parameters for {Markov} model
  construction. \emph{J. Chem. Phys.} \textbf{2013}, \emph{139}, 015102\relax
\mciteBstWouldAddEndPuncttrue
\mciteSetBstMidEndSepPunct{\mcitedefaultmidpunct}
{\mcitedefaultendpunct}{\mcitedefaultseppunct}\relax
\EndOfBibitem
\bibitem[Diez \latin{et~al.}(2022)Diez, Nagel, and Stock]{Diez22}
Diez,~G.; Nagel,~D.; Stock,~G. Correlation-based feature selection to identify
  functional dynamics in proteins. \emph{J. Chem. Theory Comput.}
  \textbf{2022}, \emph{18}, 5079 – 5088\relax
\mciteBstWouldAddEndPuncttrue
\mciteSetBstMidEndSepPunct{\mcitedefaultmidpunct}
{\mcitedefaultendpunct}{\mcitedefaultseppunct}\relax
\EndOfBibitem
\bibitem[Chiu \latin{et~al.}(2005)Chiu, Kubelka, Herbst-Irmer, Eaton,
  Hofrichter, and Davies]{Chiu05}
Chiu,~T.~K.; Kubelka,~J.; Herbst-Irmer,~R.; Eaton,~W.~A.; Hofrichter,~J.;
  Davies,~D.~R. High-resolution x-ray crystal structures of the villin
  headpiece subdomain, an ultrafast folding protein. \emph{Proc. Natl. Acad.
  Sci. USA} \textbf{2005}, \emph{102}, 7517--7522\relax
\mciteBstWouldAddEndPuncttrue
\mciteSetBstMidEndSepPunct{\mcitedefaultmidpunct}
{\mcitedefaultendpunct}{\mcitedefaultseppunct}\relax
\EndOfBibitem
\bibitem[Brewer \latin{et~al.}(2007)Brewer, Song, Raleigh, and Dyer]{Brewer07}
Brewer,~S.; Song,~B.; Raleigh,~D.; Dyer,~R. Residue Specific Resolution of
  Protein Folding Dynamics Using Isotope-Edited Infrared Temperature Jump
  Spectroscopy. \emph{Biochem.} \textbf{2007}, \emph{46}, 3279--3285\relax
\mciteBstWouldAddEndPuncttrue
\mciteSetBstMidEndSepPunct{\mcitedefaultmidpunct}
{\mcitedefaultendpunct}{\mcitedefaultseppunct}\relax
\EndOfBibitem
\bibitem[Kubelka \latin{et~al.}(2006)Kubelka, Chiu, Davies, Eaton, and
  Hofrichter]{Kubelka06}
Kubelka,~J.; Chiu,~T.~K.; Davies,~D.~R.; Eaton,~W.~A.; Hofrichter,~J.
  {Sub-microsecond protein folding}. \emph{J. Mol. Biol.} \textbf{2006},
  \emph{359}, 546--553\relax
\mciteBstWouldAddEndPuncttrue
\mciteSetBstMidEndSepPunct{\mcitedefaultmidpunct}
{\mcitedefaultendpunct}{\mcitedefaultseppunct}\relax
\EndOfBibitem
\bibitem[Kubelka \latin{et~al.}(2008)Kubelka, Henry, Cellmer, Hofrichter, and
  Eaton]{Kubelka08}
Kubelka,~J.; Henry,~E.~R.; Cellmer,~T.; Hofrichter,~J.; Eaton,~W.~A. {Chemical,
  physical, and theoretical kinetics of an ultrafast folding protein}.
  \emph{Proc. Natl. Acad. Sci. USA} \textbf{2008}, \emph{105},
  18655--18662\relax
\mciteBstWouldAddEndPuncttrue
\mciteSetBstMidEndSepPunct{\mcitedefaultmidpunct}
{\mcitedefaultendpunct}{\mcitedefaultseppunct}\relax
\EndOfBibitem
\bibitem[Reiner \latin{et~al.}(2010)Reiner, Henklein, and Kiefhaber]{Reiner10}
Reiner,~A.; Henklein,~P.; Kiefhaber,~T. An unlocking/relocking barrier in
  conformational fluctuations of villin headpiece subdomain. \emph{Proc. Natl.
  Acad. Sci. USA} \textbf{2010}, \emph{107}, 4955 -- 4960\relax
\mciteBstWouldAddEndPuncttrue
\mciteSetBstMidEndSepPunct{\mcitedefaultmidpunct}
{\mcitedefaultendpunct}{\mcitedefaultseppunct}\relax
\EndOfBibitem
\bibitem[Duan and Kollman(1998)Duan, and Kollman]{Duan98}
Duan,~Y.; Kollman,~P.~A. Pathways to a protein folding intermediate observed in
  a 1-microsecond simulation in aqueous solution. \emph{Science} \textbf{1998},
  \emph{282}, 740--744\relax
\mciteBstWouldAddEndPuncttrue
\mciteSetBstMidEndSepPunct{\mcitedefaultmidpunct}
{\mcitedefaultendpunct}{\mcitedefaultseppunct}\relax
\EndOfBibitem
\bibitem[Snow \latin{et~al.}(2002)Snow, Nguyen, Pande, and Gruebele]{Snow02}
Snow,~C.~D.; Nguyen,~H.; Pande,~V.~S.; Gruebele,~M. Absolute comparison of
  simulated and experimenta protein folding dynamics. \emph{Nature (London)}
  \textbf{2002}, \emph{420}, 102\relax
\mciteBstWouldAddEndPuncttrue
\mciteSetBstMidEndSepPunct{\mcitedefaultmidpunct}
{\mcitedefaultendpunct}{\mcitedefaultseppunct}\relax
\EndOfBibitem
\bibitem[Ensign \latin{et~al.}(2007)Ensign, Kasson, and Pande]{Ensign07}
Ensign,~D.~L.; Kasson,~P.~M.; Pande,~V.~S. Heterogeneity even at the speed
  limit of folding: large-scale molecular dynamics study of a fast-folding
  variant of the villin headpiece. \emph{J. Mol. Biol.} \textbf{2007},
  \emph{374}, 806--816\relax
\mciteBstWouldAddEndPuncttrue
\mciteSetBstMidEndSepPunct{\mcitedefaultmidpunct}
{\mcitedefaultendpunct}{\mcitedefaultseppunct}\relax
\EndOfBibitem
\bibitem[Rajan \latin{et~al.}(2010)Rajan, Freddolino, and Schulten]{Rajan10}
Rajan,~A.; Freddolino,~P.~L.; Schulten,~K. Going beyond clustering in MD
  trajectory analysis: an application to villin headpiece folding. \emph{PLoS
  One} \textbf{2010}, \emph{5}, e9890\relax
\mciteBstWouldAddEndPuncttrue
\mciteSetBstMidEndSepPunct{\mcitedefaultmidpunct}
{\mcitedefaultendpunct}{\mcitedefaultseppunct}\relax
\EndOfBibitem
\bibitem[Beauchamp \latin{et~al.}(2011)Beauchamp, Ensign, Das, and
  Pande]{Beauchamp11a}
Beauchamp,~K.~A.; Ensign,~D.~L.; Das,~R.; Pande,~V.~S. Quantitative comparison
  of villin headpiece subdomain simulations and triplet–triplet energy
  transfer experiments. \emph{Proc. Natl. Acad. Sci. USA} \textbf{2011},
  \emph{108}, 12734 -- 12739\relax
\mciteBstWouldAddEndPuncttrue
\mciteSetBstMidEndSepPunct{\mcitedefaultmidpunct}
{\mcitedefaultendpunct}{\mcitedefaultseppunct}\relax
\EndOfBibitem
\bibitem[Piana \latin{et~al.}(2012)Piana, Lindorff-Larsen, and Shaw]{Piana12}
Piana,~S.; Lindorff-Larsen,~K.; Shaw,~D.~E. {Protein folding kinetics and
  thermodynamics from atomistic simulation}. \emph{Proc. Natl. Acad. Sci. USA}
  \textbf{2012}, \emph{109}, 17845--17850\relax
\mciteBstWouldAddEndPuncttrue
\mciteSetBstMidEndSepPunct{\mcitedefaultmidpunct}
{\mcitedefaultendpunct}{\mcitedefaultseppunct}\relax
\EndOfBibitem
\bibitem[Jain and Stock(2014)Jain, and Stock]{Jain14}
Jain,~A.; Stock,~G. Hierarchical folding free energy landscape of {HP35}
  revealed by most probable path clustering. \emph{J. Phys. Chem. B}
  \textbf{2014}, \emph{118}, 7750 -- 7760\relax
\mciteBstWouldAddEndPuncttrue
\mciteSetBstMidEndSepPunct{\mcitedefaultmidpunct}
{\mcitedefaultendpunct}{\mcitedefaultseppunct}\relax
\EndOfBibitem
\bibitem[Nagel \latin{et~al.}(2019)Nagel, Weber, Lickert, and Stock]{Nagel19}
Nagel,~D.; Weber,~A.; Lickert,~B.; Stock,~G. Dynamical coring of {Markov} state
  models. \emph{J. Chem. Phys.} \textbf{2019}, \emph{150}, 094111\relax
\mciteBstWouldAddEndPuncttrue
\mciteSetBstMidEndSepPunct{\mcitedefaultmidpunct}
{\mcitedefaultendpunct}{\mcitedefaultseppunct}\relax
\EndOfBibitem
\bibitem[Nagel \latin{et~al.}(2020)Nagel, Weber, and Stock]{Nagel20}
Nagel,~D.; Weber,~A.; Stock,~G. {MSMPathfinder}: Identification of pathways in
  {Markov} state models. \emph{J. Chem. Theory Comput.} \textbf{2020},
  \emph{16}, 7874 -- 7882\relax
\mciteBstWouldAddEndPuncttrue
\mciteSetBstMidEndSepPunct{\mcitedefaultmidpunct}
{\mcitedefaultendpunct}{\mcitedefaultseppunct}\relax
\EndOfBibitem
\bibitem[Sormani \latin{et~al.}(2020)Sormani, Rodriguez, and Laio]{Sormani20}
Sormani,~G.; Rodriguez,~A.; Laio,~A. Explicit Characterization of the
  Free-Energy Landscape of a Protein in the Space of All Its {C$_\alpha$}
  Carbons. \emph{J. Chem. Theory Comput.} \textbf{2020}, \emph{16},
  80--87\relax
\mciteBstWouldAddEndPuncttrue
\mciteSetBstMidEndSepPunct{\mcitedefaultmidpunct}
{\mcitedefaultendpunct}{\mcitedefaultseppunct}\relax
\EndOfBibitem
\bibitem[Damjanovic \latin{et~al.}(2021)Damjanovic, Murphy, and
  Lin]{Damjanovic21}
Damjanovic,~J.; Murphy,~J.~M.; Lin,~Y.-S. CATBOSS: Cluster Analysis of
  Trajectories Based on Segment Splitting. \emph{J. Chem. Inf. Model.}
  \textbf{2021}, \emph{61}, 5066--5081\relax
\mciteBstWouldAddEndPuncttrue
\mciteSetBstMidEndSepPunct{\mcitedefaultmidpunct}
{\mcitedefaultendpunct}{\mcitedefaultseppunct}\relax
\EndOfBibitem
\bibitem[Klem \latin{et~al.}(2022)Klem, Hocky, and McCullagh]{Klem22}
Klem,~H.; Hocky,~G.~M.; McCullagh,~M. Size-and-Shape Space Gaussian Mixture
  Models for Structural Clustering of Molecular Dynamics Trajectories. \emph{J.
  Chem. Theory Comput.} \textbf{2022}, \emph{18}, 3218--3230\relax
\mciteBstWouldAddEndPuncttrue
\mciteSetBstMidEndSepPunct{\mcitedefaultmidpunct}
{\mcitedefaultendpunct}{\mcitedefaultseppunct}\relax
\EndOfBibitem
\bibitem[Chong and Ham(2021)Chong, and Ham]{Chong21}
Chong,~S.-H.; Ham,~S. Time-dependent communication between multiple amino acids
  during protein folding. \emph{Chem. Sci.} \textbf{2021}, \emph{12},
  5944--5951\relax
\mciteBstWouldAddEndPuncttrue
\mciteSetBstMidEndSepPunct{\mcitedefaultmidpunct}
{\mcitedefaultendpunct}{\mcitedefaultseppunct}\relax
\EndOfBibitem
\bibitem[Hornak \latin{et~al.}(2006)Hornak, Abel, Okur, Strockbine, Roitberg,
  and Simmerling]{Hornak06}
Hornak,~V.; Abel,~R.; Okur,~A.; Strockbine,~B.; Roitberg,~A.; Simmerling,~C.
  Comparison of multiple {Amber} force fields and development of improved
  protein backbone parameters. \emph{Proteins} \textbf{2006}, \emph{65},
  712--725\relax
\mciteBstWouldAddEndPuncttrue
\mciteSetBstMidEndSepPunct{\mcitedefaultmidpunct}
{\mcitedefaultendpunct}{\mcitedefaultseppunct}\relax
\EndOfBibitem
\bibitem[Best and Hummer(2009)Best, and Hummer]{Best09}
Best,~R.~B.; Hummer,~G. Optimized Molecular Dynamics Force Fields Applied to
  the Helix-Coil Transition of Polypeptides. \emph{J. Phys. Chem. B}
  \textbf{2009}, \emph{113}, 9004--9015\relax
\mciteBstWouldAddEndPuncttrue
\mciteSetBstMidEndSepPunct{\mcitedefaultmidpunct}
{\mcitedefaultendpunct}{\mcitedefaultseppunct}\relax
\EndOfBibitem
\bibitem[Lindorff-Larsen \latin{et~al.}(2010)Lindorff-Larsen, Piana, Palmo,
  Maragakis, Klepeis, Dror, and Shaw]{Lindorff-Larsen10}
Lindorff-Larsen,~K.; Piana,~S.; Palmo,~K.; Maragakis,~P.; Klepeis,~J.~L.;
  Dror,~R.~O.; Shaw,~D.~E. Improved side-chain torsion potentials for the
  {Amber} ff99SB protein force field. \emph{Proteins} \textbf{2010}, \emph{78},
  1950 -- 1958\relax
\mciteBstWouldAddEndPuncttrue
\mciteSetBstMidEndSepPunct{\mcitedefaultmidpunct}
{\mcitedefaultendpunct}{\mcitedefaultseppunct}\relax
\EndOfBibitem
\bibitem[Jorgensen \latin{et~al.}(1983)Jorgensen, Chandrasekhar, Madura, Impey,
  and Klein]{Jorgensen83}
Jorgensen,~W.~L.; Chandrasekhar,~J.; Madura,~J.~D.; Impey,~R.~W.; Klein,~M.~L.
  Comparison of simple potential functions for simulating liquid water.
  \emph{J. Chem. Phys.} \textbf{1983}, \emph{79}, 926\relax
\mciteBstWouldAddEndPuncttrue
\mciteSetBstMidEndSepPunct{\mcitedefaultmidpunct}
{\mcitedefaultendpunct}{\mcitedefaultseppunct}\relax
\EndOfBibitem
\bibitem[not()]{note4}
{As shown in Tab. S1, we get 42 contacts from Eq. (2) and 50 from Eq.
  (1).}\relax
\mciteBstWouldAddEndPunctfalse
\mciteSetBstMidEndSepPunct{\mcitedefaultmidpunct}
{}{\mcitedefaultseppunct}\relax
\EndOfBibitem
\bibitem[Heringa and Argos(1991)Heringa, and Argos]{Heringa91}
Heringa,~J.; Argos,~P. Side-chain clusters in protein structures and their role
  in protein folding. \emph{J. Mol. Biol.} \textbf{1991}, \emph{220}, 151 --
  171\relax
\mciteBstWouldAddEndPuncttrue
\mciteSetBstMidEndSepPunct{\mcitedefaultmidpunct}
{\mcitedefaultendpunct}{\mcitedefaultseppunct}\relax
\EndOfBibitem
\bibitem[Yao \latin{et~al.}(2019)Yao, Momin, and Hamelberg]{Yao19}
Yao,~X.-Q.; Momin,~M.; Hamelberg,~D. Establishing a Framework of Using
  Residue–Residue Interactions in Protein Difference Network Analysis.
  \emph{J. Chem. Inf. Model.} \textbf{2019}, \emph{59}, 3222--3228\relax
\mciteBstWouldAddEndPuncttrue
\mciteSetBstMidEndSepPunct{\mcitedefaultmidpunct}
{\mcitedefaultendpunct}{\mcitedefaultseppunct}\relax
\EndOfBibitem
\bibitem[Sali \latin{et~al.}(1994)Sali, Shakhnovich, and Karplus]{Sali94}
Sali,~A.; Shakhnovich,~E.; Karplus,~M. How does a protein fold? \emph{Nature
  (London)} \textbf{1994}, \emph{369}, 248 -- 251\relax
\mciteBstWouldAddEndPuncttrue
\mciteSetBstMidEndSepPunct{\mcitedefaultmidpunct}
{\mcitedefaultendpunct}{\mcitedefaultseppunct}\relax
\EndOfBibitem
\bibitem[Wolynes \latin{et~al.}(1995)Wolynes, Onuchic, and
  Thirumalai]{Wolynes95}
Wolynes,~P.~G.; Onuchic,~J.~N.; Thirumalai,~D. Navigating the Folding Routes.
  \emph{Science} \textbf{1995}, \emph{267}, 1619--1620\relax
\mciteBstWouldAddEndPuncttrue
\mciteSetBstMidEndSepPunct{\mcitedefaultmidpunct}
{\mcitedefaultendpunct}{\mcitedefaultseppunct}\relax
\EndOfBibitem
\bibitem[Best \latin{et~al.}(2013)Best, Hummer, and Eaton]{Best13}
Best,~R.~B.; Hummer,~G.; Eaton,~W.~A. Native contacts determine protein folding
  mechanisms in atomistic simulations. \emph{Proc. Natl. Acad. Sci. USA}
  \textbf{2013}, \emph{110}, 17874--17879\relax
\mciteBstWouldAddEndPuncttrue
\mciteSetBstMidEndSepPunct{\mcitedefaultmidpunct}
{\mcitedefaultendpunct}{\mcitedefaultseppunct}\relax
\EndOfBibitem
\bibitem[Best and Hummer(2010)Best, and Hummer]{Best10}
Best,~R.~B.; Hummer,~G. Coordinate-dependent diffusion in protein folding.
  \emph{Proc. Natl. Acad. Sci. USA} \textbf{2010}, \emph{107}, 1088 --
  1093\relax
\mciteBstWouldAddEndPuncttrue
\mciteSetBstMidEndSepPunct{\mcitedefaultmidpunct}
{\mcitedefaultendpunct}{\mcitedefaultseppunct}\relax
\EndOfBibitem
\bibitem[Lange and {Grubm\"uller}(2006)Lange, and {Grubm\"uller}]{Lange06}
Lange,~O.~F.; {Grubm\"uller},~H. Generalized Correlation for Biomolecular
  Dynamics. \emph{Proteins} \textbf{2006}, \emph{62}, 1053--1061\relax
\mciteBstWouldAddEndPuncttrue
\mciteSetBstMidEndSepPunct{\mcitedefaultmidpunct}
{\mcitedefaultendpunct}{\mcitedefaultseppunct}\relax
\EndOfBibitem
\bibitem[Traag \latin{et~al.}(2019)Traag, Waltman, and {van Eck}]{Traag19}
Traag,~V.; Waltman,~L.; {van Eck},~N. From {Louvain} to {Leiden}: guaranteeing
  well-connected communities. \emph{Sci. Rep.} \textbf{2019}, \emph{9},
  5233\relax
\mciteBstWouldAddEndPuncttrue
\mciteSetBstMidEndSepPunct{\mcitedefaultmidpunct}
{\mcitedefaultendpunct}{\mcitedefaultseppunct}\relax
\EndOfBibitem
\bibitem[Buchenberg \latin{et~al.}(2015)Buchenberg, Schaudinnus, and
  Stock]{Buchenberg15}
Buchenberg,~S.; Schaudinnus,~N.; Stock,~G. Hierarchical biomolecular dynamics:
  Picosecond hydrogen bonding regulates microsecond conformational transitions.
  \emph{J. Chem. Theory Comput.} \textbf{2015}, \emph{11}, 1330--1336\relax
\mciteBstWouldAddEndPuncttrue
\mciteSetBstMidEndSepPunct{\mcitedefaultmidpunct}
{\mcitedefaultendpunct}{\mcitedefaultseppunct}\relax
\EndOfBibitem
\bibitem[Post \latin{et~al.}(2022)Post, Lickert, Diez, Wolf, and
  Stock]{Post22a}
Post,~M.; Lickert,~B.; Diez,~G.; Wolf,~S.; Stock,~G. Cooperative protein
  allosteric transition mediated by a fluctuating transmission network.
  \emph{J. Mol. Bio.} \textbf{2022}, \emph{434}, 167679\relax
\mciteBstWouldAddEndPuncttrue
\mciteSetBstMidEndSepPunct{\mcitedefaultmidpunct}
{\mcitedefaultendpunct}{\mcitedefaultseppunct}\relax
\EndOfBibitem
\bibitem[Ali \latin{et~al.}(2022)Ali, Gulzar, Wolf, and Stock]{Ali22}
Ali,~A. A. A.~I.; Gulzar,~A.; Wolf,~S.; Stock,~G. Nonequilibrium modeling of
  the elementary step in {PDZ3} allosteric communication. \emph{J. Phys. Chem.
  Lett.} \textbf{2022}, \emph{13}, 9862 – 9868\relax
\mciteBstWouldAddEndPuncttrue
\mciteSetBstMidEndSepPunct{\mcitedefaultmidpunct}
{\mcitedefaultendpunct}{\mcitedefaultseppunct}\relax
\EndOfBibitem
\bibitem[Altis \latin{et~al.}(2007)Altis, Nguyen, Hegger, and Stock]{Altis07}
Altis,~A.; Nguyen,~P.~H.; Hegger,~R.; Stock,~G. Dihedral angle principal
  component analysis of molecular dynamics simulations. \emph{J. Chem. Phys.}
  \textbf{2007}, \emph{126}, 244111\relax
\mciteBstWouldAddEndPuncttrue
\mciteSetBstMidEndSepPunct{\mcitedefaultmidpunct}
{\mcitedefaultendpunct}{\mcitedefaultseppunct}\relax
\EndOfBibitem
\bibitem[Sittel \latin{et~al.}(2017)Sittel, Filk, and Stock]{Sittel17}
Sittel,~F.; Filk,~T.; Stock,~G. Principal component analysis on a torus: Theory
  and application to protein dynamics. \emph{J. Chem. Phys.} \textbf{2017},
  \emph{147}, 244101\relax
\mciteBstWouldAddEndPuncttrue
\mciteSetBstMidEndSepPunct{\mcitedefaultmidpunct}
{\mcitedefaultendpunct}{\mcitedefaultseppunct}\relax
\EndOfBibitem
\bibitem[Zoubouloglou \latin{et~al.}(2022)Zoubouloglou, García-Portugués, and
  Marron]{Zoubouloglou22}
Zoubouloglou,~P.; García-Portugués,~E.; Marron,~J.~S. Scaled Torus Principal
  Component Analysis. \emph{J. Comput. Graph. Stat.} \textbf{2022}, \emph{0},
  1--12\relax
\mciteBstWouldAddEndPuncttrue
\mciteSetBstMidEndSepPunct{\mcitedefaultmidpunct}
{\mcitedefaultendpunct}{\mcitedefaultseppunct}\relax
\EndOfBibitem
\bibitem[Sittel and Stock(2016)Sittel, and Stock]{Sittel16}
Sittel,~F.; Stock,~G. Robust Density-Based Clustering to Identify Metastable
  Conformational States of Proteins. \emph{J. Chem. Theory Comput.}
  \textbf{2016}, \emph{12}, 2426--2435\relax
\mciteBstWouldAddEndPuncttrue
\mciteSetBstMidEndSepPunct{\mcitedefaultmidpunct}
{\mcitedefaultendpunct}{\mcitedefaultseppunct}\relax
\EndOfBibitem
\bibitem[Rohrdanz \latin{et~al.}(2013)Rohrdanz, Zheng, and
  Clementi]{Rohrdanz13}
Rohrdanz,~M.~A.; Zheng,~W.; Clementi,~C. Discovering Mountain Passes via
  Torchlight: Methods for the Definition of Reaction Coordinates and Pathways
  in Complex Macromolecular Reactions. \emph{Annu. Rev. Phys. Chem.}
  \textbf{2013}, \emph{64}, 295--316\relax
\mciteBstWouldAddEndPuncttrue
\mciteSetBstMidEndSepPunct{\mcitedefaultmidpunct}
{\mcitedefaultendpunct}{\mcitedefaultseppunct}\relax
\EndOfBibitem
\bibitem[Wang \latin{et~al.}(2020)Wang, {Lamim Ribeiro}, and Tiwary]{Wang20}
Wang,~Y.; {Lamim Ribeiro},~J.~M.; Tiwary,~P. Machine learning approaches for
  analyzing and enhancing molecular dynamics simulations. \emph{Curr. Opin.
  Struct. Biol.} \textbf{2020}, \emph{61}, 139--145\relax
\mciteBstWouldAddEndPuncttrue
\mciteSetBstMidEndSepPunct{\mcitedefaultmidpunct}
{\mcitedefaultendpunct}{\mcitedefaultseppunct}\relax
\EndOfBibitem
\bibitem[Glielmo \latin{et~al.}(2021)Glielmo, Husic, Rodriguez, Clementi,
  No{\'e}, and Laio]{Glielmo21}
Glielmo,~A.; Husic,~B.~E.; Rodriguez,~A.; Clementi,~C.; No{\'e},~F.; Laio,~A.
  Unsupervised Learning Methods for Molecular Simulation Data. \emph{Chem.
  Rev.} \textbf{2021}, \emph{121}, 9722--9758\relax
\mciteBstWouldAddEndPuncttrue
\mciteSetBstMidEndSepPunct{\mcitedefaultmidpunct}
{\mcitedefaultendpunct}{\mcitedefaultseppunct}\relax
\EndOfBibitem
\bibitem[Jain and Stock(2012)Jain, and Stock]{Jain12}
Jain,~A.; Stock,~G. Identifying metastable states of folding proteins. \emph{J.
  Chem. Theory Comput.} \textbf{2012}, \emph{8}, 3810 -- 3819\relax
\mciteBstWouldAddEndPuncttrue
\mciteSetBstMidEndSepPunct{\mcitedefaultmidpunct}
{\mcitedefaultendpunct}{\mcitedefaultseppunct}\relax
\EndOfBibitem
\bibitem[{Sch\"utte} \latin{et~al.}(2011){Sch\"utte}, No{\'e}, Lu, Sarich, and
  Vanden-Eijnden]{Schuette11}
{Sch\"utte},~C.; No{\'e},~F.; Lu,~J.; Sarich,~M.; Vanden-Eijnden,~E. Markov
  state models based on milestoning. \emph{J. Chem. Phys.} \textbf{2011},
  \emph{134}, 204105\relax
\mciteBstWouldAddEndPuncttrue
\mciteSetBstMidEndSepPunct{\mcitedefaultmidpunct}
{\mcitedefaultendpunct}{\mcitedefaultseppunct}\relax
\EndOfBibitem
\bibitem[Lemke and Keller(2016)Lemke, and Keller]{Lemke16}
Lemke,~O.; Keller,~B.~G. Density-based cluster algorithms for the
  identification of core sets. \emph{J. Chem. Phys.} \textbf{2016}, \emph{145},
  164104\relax
\mciteBstWouldAddEndPuncttrue
\mciteSetBstMidEndSepPunct{\mcitedefaultmidpunct}
{\mcitedefaultendpunct}{\mcitedefaultseppunct}\relax
\EndOfBibitem
\bibitem[not()]{note3}
{In general, we find that different energy basins have different
  metastabilities, making it inappropriate to split the dendrogram at a
  constant value of $Q_{\text{min}}$. As a remedy, the MPP algorithm was
  extended to automatically detect branches, by considering both a minimum
  population $P_{\text{min}}$ and a minimum metastability $Q_{\text{min}}$.
  Branches are evaluated in a top-down fashion, with sub-branches being
  considered in case they meet the minimum criteria. Once all branches have
  been processed, any remaining microstates are assigned to their dynamically
  nearest macrostate}\relax
\mciteBstWouldAddEndPuncttrue
\mciteSetBstMidEndSepPunct{\mcitedefaultmidpunct}
{\mcitedefaultendpunct}{\mcitedefaultseppunct}\relax
\EndOfBibitem
\bibitem[not()]{note1}
{In fact, these conditions produced 13 (instead of 12) contacts-based states.
  Since the additional state 13 (on the very left side of the dendrogram) is
  completely unfolded just like state 12, however, we simply may merge them
  into a single state when we discuss folding. This facilitates the comparison
  with the 12 dihedral-based states }\relax
\mciteBstWouldAddEndPuncttrue
\mciteSetBstMidEndSepPunct{\mcitedefaultmidpunct}
{\mcitedefaultendpunct}{\mcitedefaultseppunct}\relax
\EndOfBibitem
\bibitem[Brandt \latin{et~al.}(2018)Brandt, Sittel, Ernst, and Stock]{Brandt18}
Brandt,~S.; Sittel,~F.; Ernst,~M.; Stock,~G. Machine Learning of Biomolecular
  Reaction Coordinates. \emph{J. Phys. Chem. Lett.} \textbf{2018}, \emph{9},
  2144 -- 2150\relax
\mciteBstWouldAddEndPuncttrue
\mciteSetBstMidEndSepPunct{\mcitedefaultmidpunct}
{\mcitedefaultendpunct}{\mcitedefaultseppunct}\relax
\EndOfBibitem
\bibitem[Chen and Guestrin(2016)Chen, and Guestrin]{Chen16}
Chen,~T.; Guestrin,~C. XGBoost: {A} Scalable Tree Boosting System. \emph{CoRR}
  \textbf{2016}, abs/1603.02754\relax
\mciteBstWouldAddEndPuncttrue
\mciteSetBstMidEndSepPunct{\mcitedefaultmidpunct}
{\mcitedefaultendpunct}{\mcitedefaultseppunct}\relax
\EndOfBibitem
\bibitem[Hummer and Szabo(2015)Hummer, and Szabo]{Hummer15}
Hummer,~G.; Szabo,~A. Optimal Dimensionality Reduction of Multistate Kinetic
  and {Markov}-State Models. \emph{J. Phys. Chem. B} \textbf{2015}, \emph{119},
  9029--9037\relax
\mciteBstWouldAddEndPuncttrue
\mciteSetBstMidEndSepPunct{\mcitedefaultmidpunct}
{\mcitedefaultendpunct}{\mcitedefaultseppunct}\relax
\EndOfBibitem
\bibitem[R{\"o}blitz and Weber(2013)R{\"o}blitz, and Weber]{Roeblitz13}
R{\"o}blitz,~S.; Weber,~M. {Fuzzy spectral clustering by PCCA+: application to
  Markov state models and data classification}. \emph{Adv. Data Anal. Classi.}
  \textbf{2013}, \emph{7}, 147--179\relax
\mciteBstWouldAddEndPuncttrue
\mciteSetBstMidEndSepPunct{\mcitedefaultmidpunct}
{\mcitedefaultendpunct}{\mcitedefaultseppunct}\relax
\EndOfBibitem
\bibitem[Kells \latin{et~al.}(2019)Kells, Mihalka, Annibale, and
  Rosta]{Kells19}
Kells,~A.; Mihalka,~Z.~E.; Annibale,~A.; Rosta,~E. Mean first passage times in
  variational coarse graining using {Markov} state models. \emph{J. Chem.
  Phys.} \textbf{2019}, \emph{150}, 134107\relax
\mciteBstWouldAddEndPuncttrue
\mciteSetBstMidEndSepPunct{\mcitedefaultmidpunct}
{\mcitedefaultendpunct}{\mcitedefaultseppunct}\relax
\EndOfBibitem
\bibitem[Cao \latin{et~al.}(2020)Cao, Montoya-Castillo, Wang, Markland, and
  Huang]{Cao20}
Cao,~S.; Montoya-Castillo,~A.; Wang,~W.; Markland,~T.~E.; Huang,~X. On the
  advantages of exploiting memory in {Markov} state models for biomolecular
  dynamics. \emph{J. Chem. Phys.} \textbf{2020}, \emph{153}, 014105\relax
\mciteBstWouldAddEndPuncttrue
\mciteSetBstMidEndSepPunct{\mcitedefaultmidpunct}
{\mcitedefaultendpunct}{\mcitedefaultseppunct}\relax
\EndOfBibitem
\bibitem[Sharpe and Wales(2021)Sharpe, and Wales]{Sharpe21}
Sharpe,~D.~J.; Wales,~D.~J. Nearly reducible finite {Markov} chains: Theory and
  algorithms. \emph{J. Chem. Phys.} \textbf{2021}, \emph{155}, 140901\relax
\mciteBstWouldAddEndPuncttrue
\mciteSetBstMidEndSepPunct{\mcitedefaultmidpunct}
{\mcitedefaultendpunct}{\mcitedefaultseppunct}\relax
\EndOfBibitem
\bibitem[Elmer \latin{et~al.}(2005)Elmer, Park, and Pande]{Elmer05II}
Elmer,~S.~P.; Park,~S.; Pande,~V.~S. Foldamer dynamics expressed via Markov
  state models. II. State space decomposition. \emph{J. Chem. Phys.}
  \textbf{2005}, \emph{123}, 114903\relax
\mciteBstWouldAddEndPuncttrue
\mciteSetBstMidEndSepPunct{\mcitedefaultmidpunct}
{\mcitedefaultendpunct}{\mcitedefaultseppunct}\relax
\EndOfBibitem
\bibitem[Jain \latin{et~al.}(2010)Jain, Hegger, and Stock]{Jain10}
Jain,~A.; Hegger,~R.; Stock,~G. Hidden complexity of protein energy landscape
  revealed by principal component analysis by parts. \emph{J. Phys. Chem.
  Lett.} \textbf{2010}, \emph{1}, 2769--2773\relax
\mciteBstWouldAddEndPuncttrue
\mciteSetBstMidEndSepPunct{\mcitedefaultmidpunct}
{\mcitedefaultendpunct}{\mcitedefaultseppunct}\relax
\EndOfBibitem
\bibitem[No{\'e} \latin{et~al.}(2016)No{\'e}, Banisch, and Clementi]{Noe16}
No{\'e},~F.; Banisch,~R.; Clementi,~C. Commute Maps: Separating Slowly Mixing
  Molecular Configurations for Kinetic Modeling. \emph{J. Chem. Theory Comput.}
  \textbf{2016}, \emph{12}, 5620--5630\relax
\mciteBstWouldAddEndPuncttrue
\mciteSetBstMidEndSepPunct{\mcitedefaultmidpunct}
{\mcitedefaultendpunct}{\mcitedefaultseppunct}\relax
\EndOfBibitem
\bibitem[Jacomy \latin{et~al.}(2014)Jacomy, Venturini, Heymann, and
  Bastian]{ForceAtlas2}
Jacomy,~M.; Venturini,~T.; Heymann,~S.; Bastian,~M. ForceAtlas2, a Continuous
  Graph Layout Algorithm for Handy Network Visualization Designed for the
  {Gephi} Software. \emph{PLOS ONE} \textbf{2014}, \emph{9}, 1--12\relax
\mciteBstWouldAddEndPuncttrue
\mciteSetBstMidEndSepPunct{\mcitedefaultmidpunct}
{\mcitedefaultendpunct}{\mcitedefaultseppunct}\relax
\EndOfBibitem
\bibitem[Mittal and Best(2010)Mittal, and Best]{Mittal10}
Mittal,~J.; Best,~R.~B. Tackling Force-Field Bias in Protein Folding
  Simulations: Folding of Villin {HP35 and Pin WW Domains} in Explicit Water.
  \emph{Biophys. J.} \textbf{2010}, \emph{99}, L26--L28\relax
\mciteBstWouldAddEndPuncttrue
\mciteSetBstMidEndSepPunct{\mcitedefaultmidpunct}
{\mcitedefaultendpunct}{\mcitedefaultseppunct}\relax
\EndOfBibitem
\bibitem[Piana \latin{et~al.}(2011)Piana, Lindorff-Larsen, and Shaw]{Piana11}
Piana,~S.; Lindorff-Larsen,~K.; Shaw,~D.~E. How Robust Are Protein Folding
  Simulations with Respect to Force Field Parameterization? \emph{Biophys. J.}
  \textbf{2011}, \emph{100}, L47 -- L49\relax
\mciteBstWouldAddEndPuncttrue
\mciteSetBstMidEndSepPunct{\mcitedefaultmidpunct}
{\mcitedefaultendpunct}{\mcitedefaultseppunct}\relax
\EndOfBibitem
\end{mcitethebibliography}


\providecommand{\latin}[1]{#1}
\makeatletter
\providecommand{\doi}
  {\begingroup\let\do\@makeother\dospecials
  \catcode`\{=1 \catcode`\}=2 \doi@aux}
\providecommand{\doi@aux}[1]{\endgroup\texttt{#1}}
\makeatother
\providecommand*\mcitethebibliography{\thebibliography}
\csname @ifundefined\endcsname{endmcitethebibliography}
  {\let\endmcitethebibliography\endthebibliography}{}
\begin{mcitethebibliography}{10}
\providecommand*\natexlab[1]{#1}
\providecommand*\mciteSetBstSublistMode[1]{}
\providecommand*\mciteSetBstMaxWidthForm[2]{}
\providecommand*\mciteBstWouldAddEndPuncttrue
  {\def\EndOfBibitem{\unskip.}}
\providecommand*\mciteBstWouldAddEndPunctfalse
  {\let\EndOfBibitem\relax}
\providecommand*\mciteSetBstMidEndSepPunct[3]{}
\providecommand*\mciteSetBstSublistLabelBeginEnd[3]{}
\providecommand*\EndOfBibitem{}
\mciteSetBstSublistMode{f}
\mciteSetBstMaxWidthForm{subitem}{(\alph{mcitesubitemcount})}
\mciteSetBstSublistLabelBeginEnd
  {\mcitemaxwidthsubitemform\space}
  {\relax}
  {\relax}

\bibitem[Kubelka \latin{et~al.}(2006)Kubelka, Chiu, Davies, Eaton, and
  Hofrichter]{Kubelka06}
Kubelka,~J.; Chiu,~T.~K.; Davies,~D.~R. \latin{et~al.}  {Sub-microsecond
  protein folding}. \emph{J. Mol. Biol.} \textbf{2006}, \emph{359},
  546--553\relax
\mciteBstWouldAddEndPuncttrue
\mciteSetBstMidEndSepPunct{\mcitedefaultmidpunct}
{\mcitedefaultendpunct}{\mcitedefaultseppunct}\relax
\EndOfBibitem
\bibitem[Piana \latin{et~al.}(2012)Piana, Lindorff-Larsen, and Shaw]{Piana12}
Piana,~S.; Lindorff-Larsen,~K.; Shaw,~D.~E. {Protein folding kinetics and
  thermodynamics from atomistic simulation}. \emph{Proc. Natl. Acad. Sci. USA}
  \textbf{2012}, \emph{109}, 17845--17850\relax
\mciteBstWouldAddEndPuncttrue
\mciteSetBstMidEndSepPunct{\mcitedefaultmidpunct}
{\mcitedefaultendpunct}{\mcitedefaultseppunct}\relax
\EndOfBibitem
\bibitem[Sittel and Stock(2016)Sittel, and Stock]{Sittel16}
Sittel,~F.; Stock,~G. Robust Density-Based Clustering to Identify Metastable
  Conformational States of Proteins. \emph{J. Chem. Theory Comput.}
  \textbf{2016}, \emph{12}, 2426--2435\relax
\mciteBstWouldAddEndPuncttrue
\mciteSetBstMidEndSepPunct{\mcitedefaultmidpunct}
{\mcitedefaultendpunct}{\mcitedefaultseppunct}\relax
\EndOfBibitem
\bibitem[Hummer and Szabo(2015)Hummer, and Szabo]{Hummer15}
Hummer,~G.; Szabo,~A. Optimal Dimensionality Reduction of Multistate Kinetic
  and {Markov}-State Models. \emph{J. Phys. Chem. B} \textbf{2015}, \emph{119},
  9029--9037\relax
\mciteBstWouldAddEndPuncttrue
\mciteSetBstMidEndSepPunct{\mcitedefaultmidpunct}
{\mcitedefaultendpunct}{\mcitedefaultseppunct}\relax
\EndOfBibitem
\bibitem[Nagel \latin{et~al.}(2019)Nagel, Weber, Lickert, and Stock]{Nagel19}
Nagel,~D.; Weber,~A.; Lickert,~B. \latin{et~al.}  Dynamical coring of {Markov}
  state models. \emph{J. Chem. Phys.} \textbf{2019}, \emph{150}, 094111\relax
\mciteBstWouldAddEndPuncttrue
\mciteSetBstMidEndSepPunct{\mcitedefaultmidpunct}
{\mcitedefaultendpunct}{\mcitedefaultseppunct}\relax
\EndOfBibitem
\bibitem[Sittel \latin{et~al.}(2017)Sittel, Filk, and Stock]{Sittel17}
Sittel,~F.; Filk,~T.; Stock,~G. Principal component analysis on a torus: Theory
  and application to protein dynamics. \emph{J. Chem. Phys.} \textbf{2017},
  \emph{147}, 244101\relax
\mciteBstWouldAddEndPuncttrue
\mciteSetBstMidEndSepPunct{\mcitedefaultmidpunct}
{\mcitedefaultendpunct}{\mcitedefaultseppunct}\relax
\EndOfBibitem
\bibitem[Klem \latin{et~al.}(2022)Klem, Hocky, and McCullagh]{Klem22}
Klem,~H.; Hocky,~G.~M.; McCullagh,~M. Size-and-Shape Space Gaussian Mixture
  Models for Structural Clustering of Molecular Dynamics Trajectories. \emph{J.
  Chem. Theory Comput.} \textbf{2022}, \emph{18}, 3218--3230\relax
\mciteBstWouldAddEndPuncttrue
\mciteSetBstMidEndSepPunct{\mcitedefaultmidpunct}
{\mcitedefaultendpunct}{\mcitedefaultseppunct}\relax
\EndOfBibitem
\bibitem[Damjanovic \latin{et~al.}(2021)Damjanovic, Murphy, and
  Lin]{Damjanovic21}
Damjanovic,~J.; Murphy,~J.~M.; Lin,~Y.-S. CATBOSS: Cluster Analysis of
  Trajectories Based on Segment Splitting. \emph{J. Chem. Inf. Model.}
  \textbf{2021}, \emph{61}, 5066--5081\relax
\mciteBstWouldAddEndPuncttrue
\mciteSetBstMidEndSepPunct{\mcitedefaultmidpunct}
{\mcitedefaultendpunct}{\mcitedefaultseppunct}\relax
\EndOfBibitem
\bibitem[Nagel \latin{et~al.}(2020)Nagel, Weber, and Stock]{Nagel20}
Nagel,~D.; Weber,~A.; Stock,~G. {MSMPathfinder}: Identification of pathways in
  {Markov} state models. \emph{J. Chem. Theory Comput.} \textbf{2020},
  \emph{16}, 7874 -- 7882\relax
\mciteBstWouldAddEndPuncttrue
\mciteSetBstMidEndSepPunct{\mcitedefaultmidpunct}
{\mcitedefaultendpunct}{\mcitedefaultseppunct}\relax
\EndOfBibitem
\end{mcitethebibliography}

\end{document}

\begin{table}[hb!]
  \caption{Native contacts of HP35, as obtained from the
  crystal structure\cite{Kubelka06} (left) and from the MD simulation by Piana
  et al.,\cite{Piana12} where the minimum contact distance is defined via
  Eq.~\MainEqDistance{} of the main text (middle, MD contacts I), and via
  Eq.~\MainEqAtomDistance{} but omitting atom pairs not satisfying the population
  cutoff of 30\,\% (right, MD contacts II).  For each contact definition and
  each contact, we list the percentage of the simulation time when the contact
  is formed, as well as the decay time $\tau$ of the contact autocorrelation
  function $C_{ij}(t) = \langle \delta d_{ij}(t) \delta d_{ij}(0)
  \rangle/\langle \delta d_{ij}^2\rangle$, i.e.
    $C_{ij}(\tau) = 1/e$.  The contact type is assigned as helical if
    it is a helix-stabilizing ($n, n+4$) contact, as hydrogen bond
    if both residues have oppositely charged side chains, and as
    hydrophobic else.}
    \label{SI:tab:contacts}
    {\tiny
    	\hspace{-0.5cm}
    	\begin{tabular}{llcrrrrrr}
    		\toprule
	        \multirow{2}*{residues} &
    		\multirow{2}*{contact type} &
    		MoSAIC &
    		\multicolumn{2}{c}{crystal contacts}&
    		\multicolumn{2}{c}{MD contacts I}&
    		\multicolumn{2}{c}{MD contacts II}\\
    		&&
    		cluster&
    		$\text{formed}\;[\%]$ & $\tau\;[\si{\micro\second}] $&
    		$\text{formed}\;[\%]$ & $\tau\;[\si{\micro\second}] $&
    		$\text{formed}\;[\%]$ & $\tau\;[\si{\micro\second}] $\\
    		\midrule
    		LEU1--ASP5 & helical  & -- &11.5 & 0.019 & 39.8 & 0.009& -- & --\\
    		LEU1--PHE6 & hydrophobic & -- & 2.1 & 0.027& -- & --& -- & --\\
    		LEU1--VAL9 & hydrophobic & -- & 3.2 & 0.034& -- & --& -- & --\\
    		LEU1--PHE10 & hydrophobic & -- & 1.9 & 0.030& -- & --& -- & --\\
    		LEU1--ARG14 & hydrophobic & -- & 7.3 & 0.527& -- & --& -- & --\\
    		LEU1--LEU34 & hydrophobic & -- & 3.7 & 0.030& -- & --& -- & --\\
    		SER2--PHE6 & hydrogen bond & -- & 29.7 & 0.039 & 35.7 & 0.048& -- & --\\
    		ASP3--LYS7 & helical & 8 & 71.0 & 0.092 & 75.8 & 0.050 & 74.7 & 0.072\\
    		ASP3--THR13 & hydrophobic & 1 & -- & -- & 56.9 & 1.380 & 46.5 & 1.574\\
    		ASP3--ARG14 & hydrogen bond & 1 & 29.2 & 1.104 & 69.6 & 1.210 & 61.6 & 1.466\\
    		\midrule
    		GLU4--ALA8 & helical & 8 & 63.0 & 0.055 & 65.6 & 0.048 & 65.0 & 0.055\\
    		ASP5--VAL9 & helical & 8 & 54.7 & 0.080 & 69.6 & 0.108 & 67.1 & 0.150\\
    		ASP5--ARG14 & hydrogen bond & 1 & -- & -- & 49.9 & 0.241 & 47.2 & 0.471\\
    		PHE6--PHE10 & helical & 8 & 74.8 & 0.679 & 80.4 & 0.413 & 75.7 & 0.589\\
    		PHE6--GLY11 & hydrophobic & 2 & 56.9 & 0.639 & 60.4 & 0.348 & 56.9 & 0.639\\
    		PHE6--MET12 & hydrophobic & 2 & 25.9 & 1.481 & 74.1 & 0.998 & 63.6 & 1.558\\
    		PHE6--THR13 & hydrophobic & -- & 16.4 & 1.366 & 61.5 & 0.806& -- & --\\
    		PHE6--ARG14 & hydrophobic & 1 & 24.2 & 1.481 & 73.7 & 0.948 & 69.4 & 1.339\\
    		PHE6--PHE17 & hydrophobic & 2 & 42.9 & 1.408 & 70.0 & 1.071 & 57.5 & 1.352\\
    		LYS7--GLY11 & hydrogen bond & 2 & 71.6 & 0.929 & 74.1 & 0.798 & 72.7 & 0.857\\
    		\midrule
    		LYS7--MET12 & hydrophobic  & 2 & 66.0 & 1.725 & 73.3 & 0.958 & 70.1 & 1.222\\
    		LYS7--THR13 & hydrophobic  & 2 & 16.4 & 1.481 & 55.0 & 0.672 & 33.5 & 1.481\\
    		VAL9--LYS32 & hydrophobic  & 7 & 50.7 & 1.380 & 60.0 & 1.222 & 54.7 & 1.380\\
    		VAL9--LEU34 & hydrophobic  & -- & 22.1 & 0.840 & 36.2 & 0.452& -- & --\\
    		PHE10--PHE17 & hydrophobic & -- & 20.6 & 1.352 & 62.0 & 0.874& -- & --\\
    		PHE10--LEU28 & hydrophobic & -- & 27.6 & 1.380 & 67.5 & 1.138& -- & --\\
    		PHE10--NLE29 & hydrophobic & 7 & 16.4 & 1.352 & 59.6 & 0.892 & 36.0 & 1.380\\
    		PHE10--LYS32 & hydrophobic & -- & 20.2 & 1.352 & 55.2 & 1.150& -- & --\\
    		PHE10--LEU34 & hydrophobic & 7 & 33.8 & 1.060 & 47.3 & 0.554 & 33.8 & 1.060\\
    		MET12--ALA16 & hydrophobic & 3 & 49.4 & 1.298 & 72.8 & 0.141 & 61.9 & 1.071\\
    		\midrule
    		MET12--PHE17 & hydrophobic & 3 & 52.9 & 1.394 & 74.3 & 0.560 & 61.5 & 1.234\\
    		MET12--LEU20 & hydrophobic & 3 & 42.3 & 1.272 & 56.4 & 0.766 & 52.4 & 1.272\\
    		MET12--LEU28 & hydrophobic & -- & 18.8 & 1.173 & 32.5 & 0.743& -- & --\\
    		THR13--PHE17 & hydrogen bond & 3 & 78.0 & 1.312 & 80.9 & 1.060 & 78.4 & 1.312\\
    		ARG14--ALA18 & helical & 8 & 77.5 & 1.298 & 79.4 & 0.939 & 78.1 & 1.222\\
    		SER15--ASN19 & helical & 8 & 69.1 & 0.613 & 78.0 & 0.256 & 73.4 & 0.589\\
    		ALA16--LEU20 & hydrogen bond & 8 & 72.3 & 0.120 & 80.4 & 0.066 & 79.5 & 0.092\\
    		PHE17--GLN25 & hydrophobic & 4 & 83.2 & 0.665 & 87.7 & 0.537 & 86.7 & 0.626\\
    		PHE17--LEU28 & hydrophobic & -- & 24.9 & 0.645 & 71.0 & 0.665& -- & --\\
    		PHE17--NLE29 & hydrophobic & 8 & 32.3 & 0.883 & 58.3 & 0.543 & 48.2 & 0.848\\
    		\midrule
    		ALA18--GLN25 & hydrophobic & 4 & 62.6 & 0.506 & 80.3 & 0.501 & 79.1 & 0.537\\
    		LEU20--NLE24 & hydrogen bond & 8 & 27.5 & 0.036 & 81.3 & 0.012 & 56.5 & 0.053\\
    		LEU20--GLN25 & hydrophobic & 4 & 86.8 & 0.249 & 93.2 & 0.105 & 89.9 & 0.244\\
    		LEU20--LEU28 & hydrophobic & 5 & 54.2 & 0.532 & 70.8 & 0.405 & 54.2 & 0.532\\
    		PRO21--GLN25 & hydrogen bond & 8 & 97.2 & 0.107 & 98.0 & 0.081 & 97.8 & 0.088\\
    		LEU22--GLN26 & helical & 8 & 95.4 & 0.099 & 96.7 & 0.074 & 96.5 & 0.081\\
    		TRP23--HIS27 & helical & 8 & 90.8 & 0.121 & 93.1 & 0.088 & 91.5 & 0.117\\
    		NLE24--LEU28 & helical & 5 & 89.3 & 0.287 & 90.9 & 0.225 & 90.2 & 0.262\\
    		GLN25--NLE29 & helical & 5 & 87.5 & 0.645 & 88.8 & 0.589 & 88.1 & 0.639\\
    		GLN26--LYS30 & helical & 8 & 81.6 & 0.751 & 84.1 & 0.658 & 83.0 & 0.729\\
    		\midrule
    		GLN26--PHE35 & hydrophobic & -- & -- & -- & 56.3 & 0.086& -- & --\\
    		HIS27--GLU31 & helical & 8 & 77.9 & 0.413 & 83.3 & 0.134 & 81.1 & 0.222\\
    		LEU28--LYS32 & helical & 8 & 77.0 & 0.147 & 80.4 & 0.108 & 78.7 & 0.142\\
    		NLE29--GLY33 & hydrogen bond & 6 & 74.2 & 0.112 & 77.2 & 0.082 & 74.6 & 0.108\\
    		NLE29--LEU34 & hydrogen bond & 6 & 66.5 & 0.161 & 75.1 & 0.088 & 67.9 & 0.130\\
    		NLE29--PHE35 & hydrophobic & 6 & 4.9 & 0.081 & 67.2 & 0.073 & 60.3 & 0.119\\
    		LYS30--PHE35 & hydrophobic & 6 & -- & -- & 66.3 & 0.054 & 52.9 & 0.113\\
    		\bottomrule
    	\end{tabular}
    }
\end{table}
\clearpage

%%%%%%%%%%%%%%%%%%%%%%%%%
%%%% FIG S1 %%%%%%%%%%%%%
%%%%%%%%%%%%%%%%%%%%%%%%%
\begin{figure}[ht!]
    \centering
    \includegraphics[width=\textwidth]{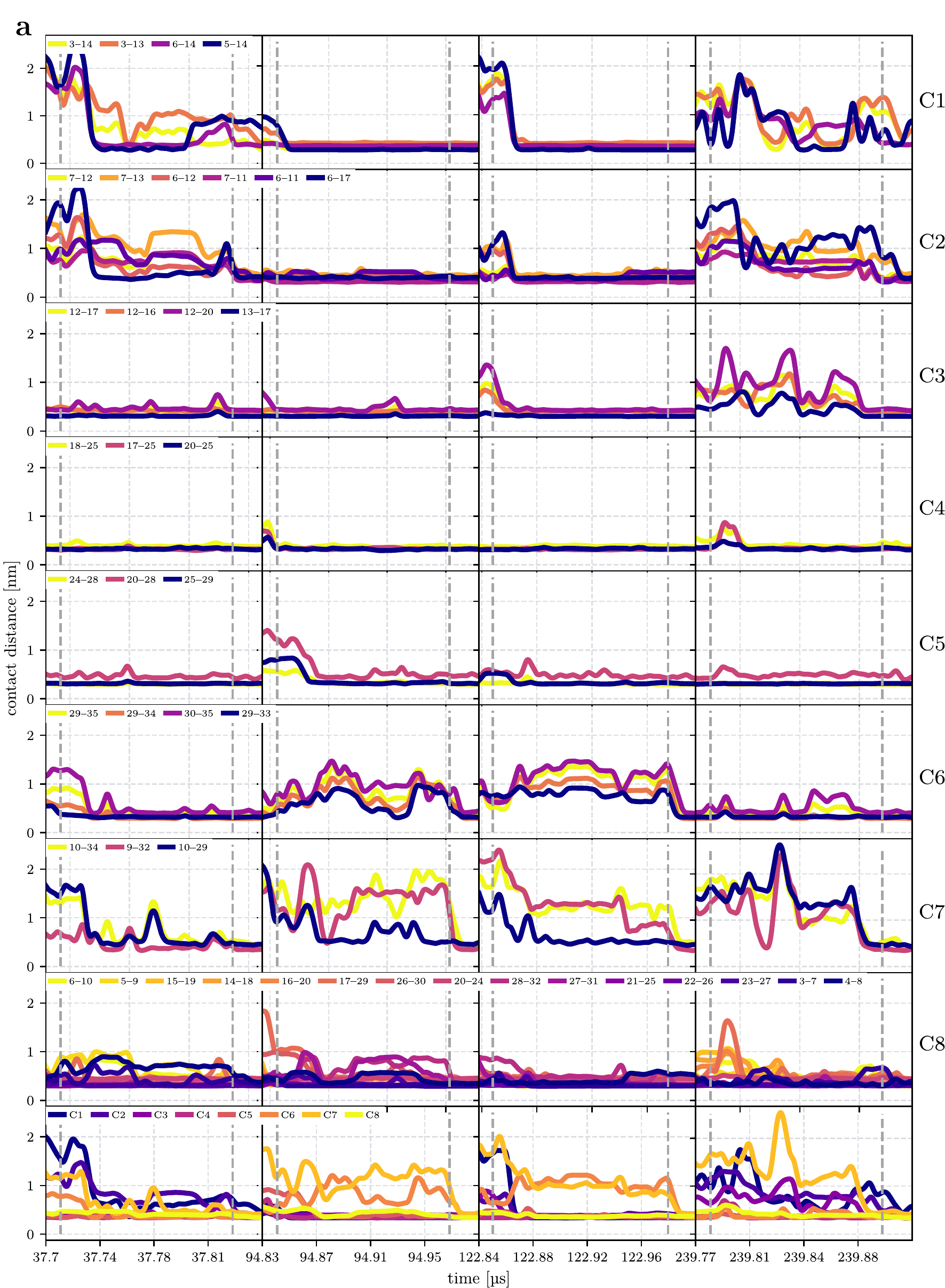}
\end{figure}

\begin{figure}[ht!]
    \centering
    \includegraphics[width=\textwidth]{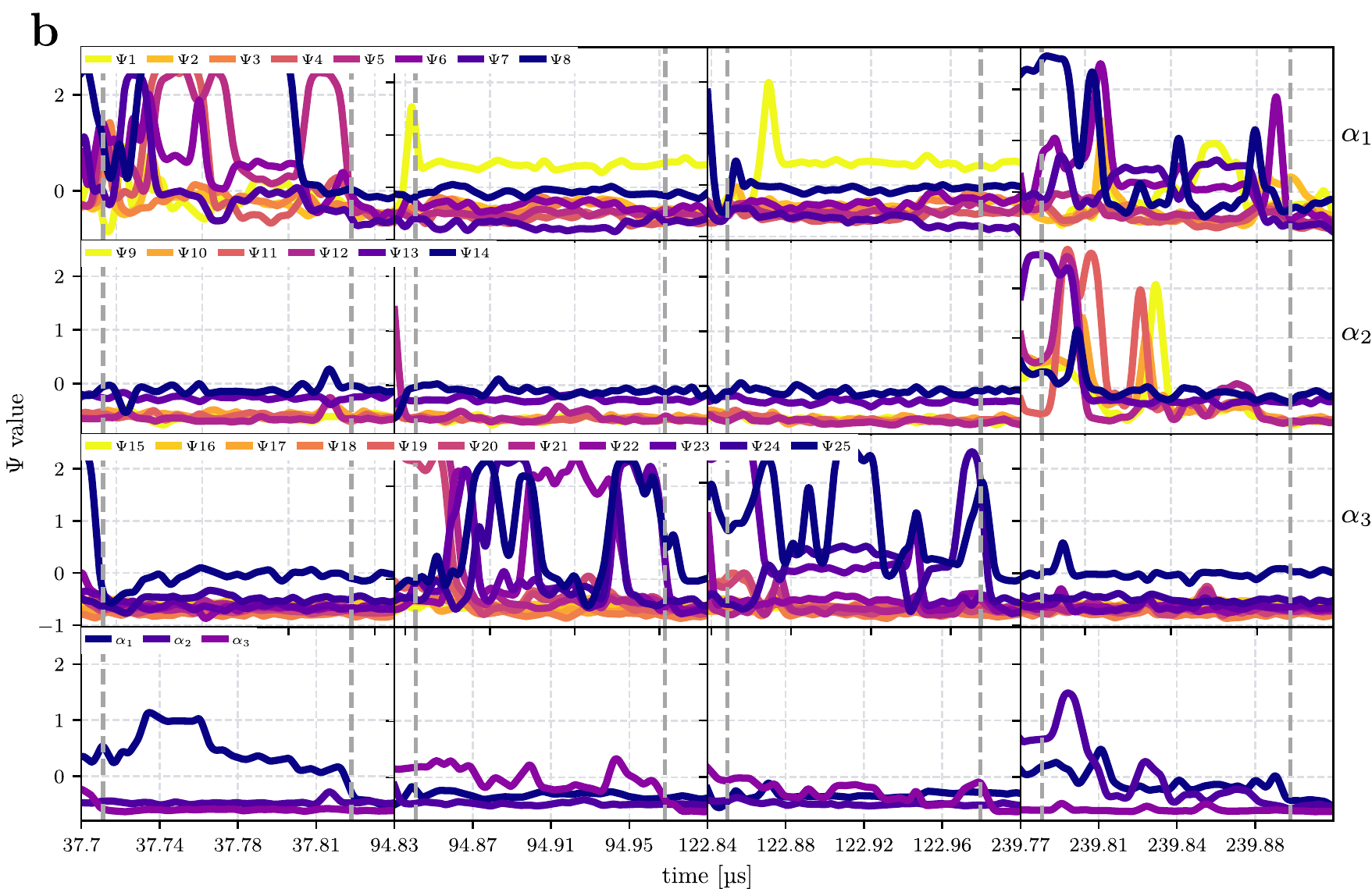}
    \caption{Cooperative behavior of input features during 4 randomly selected folding events, identified with RMSD thresholds, for (a) contacts and (b) dihedral angles. The contacts are grouped by MoSAIC clusters and the figure shows that distances belonging to the same cluster move together and 'jump' at the same time below the contact threshold of \SI{4.5}{\angstrom}. Moreover, one can see that some clusters (cluster 4 and cluster 5, featuring contacts connecting helix 2 and helix 3 and contacts belonging to the third helix) are most times already formed very early on. This is in good agreement with Fig. 2c showing the order of clusters formation. Dihedrals are grouped by helices and they also show cooperativity within the secondary structure they belong to, although they are not as correlated as in the contacts case.}
    \label{SI:fig:coop_clusters}
\end{figure}

%%%%%%%%%%%%%%%%%%%%%%%%%
%%%% FIG S2 %%%%%%%%%%%%%
%%%%%%%%%%%%%%%%%%%%%%%%%
\begin{figure}[ht!]
    \centering
    \includegraphics{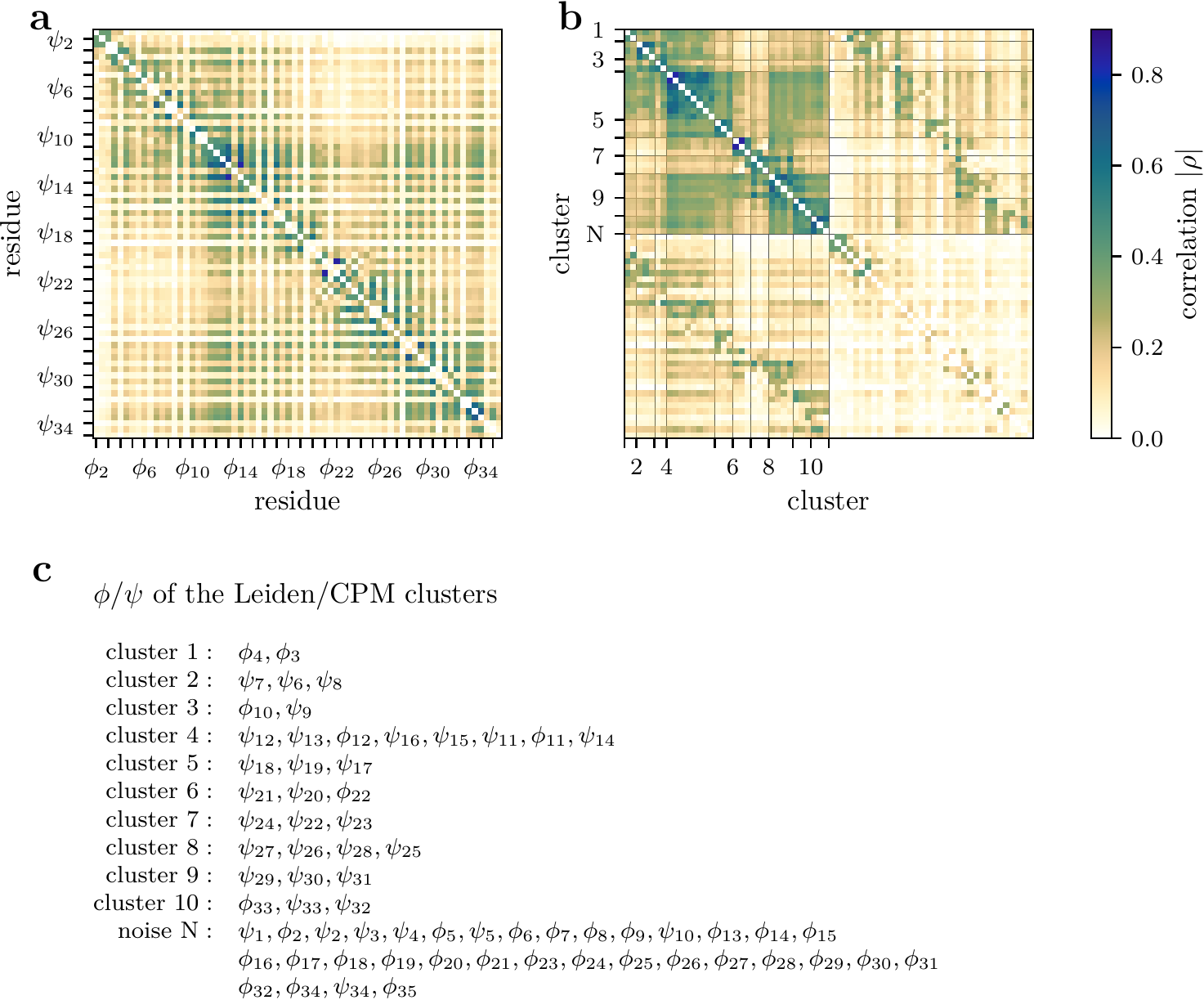}
    \caption{Preselection of backbone dihedral angles. For all dihedral angles
    the (a) pairwise correlations with all other dihedral angles, and the (b)
    corresponding Leiden/CPM clusters including (c) the tabular representation.
    The resolution parameter $\gamma=0.45$ was optimized by silhouette score
    using cross validation, and all clusters containing only a single
    coordinate were sorted into the cluster \emph{N}. Gaussian filtered
    angles with $\sigma=\SI{2}{\nano\second}$ were used.}
    \label{SI:fig:dihedral_preselection}
\end{figure}

%%%%%%%%%%%%%%%%%%%%%%%%%
%%%% FIG S3 %%%%%%%%%%%%%
%%%%%%%%%%%%%%%%%%%%%%%%%
\begin{figure}[ht!]
    \centering
    \includegraphics{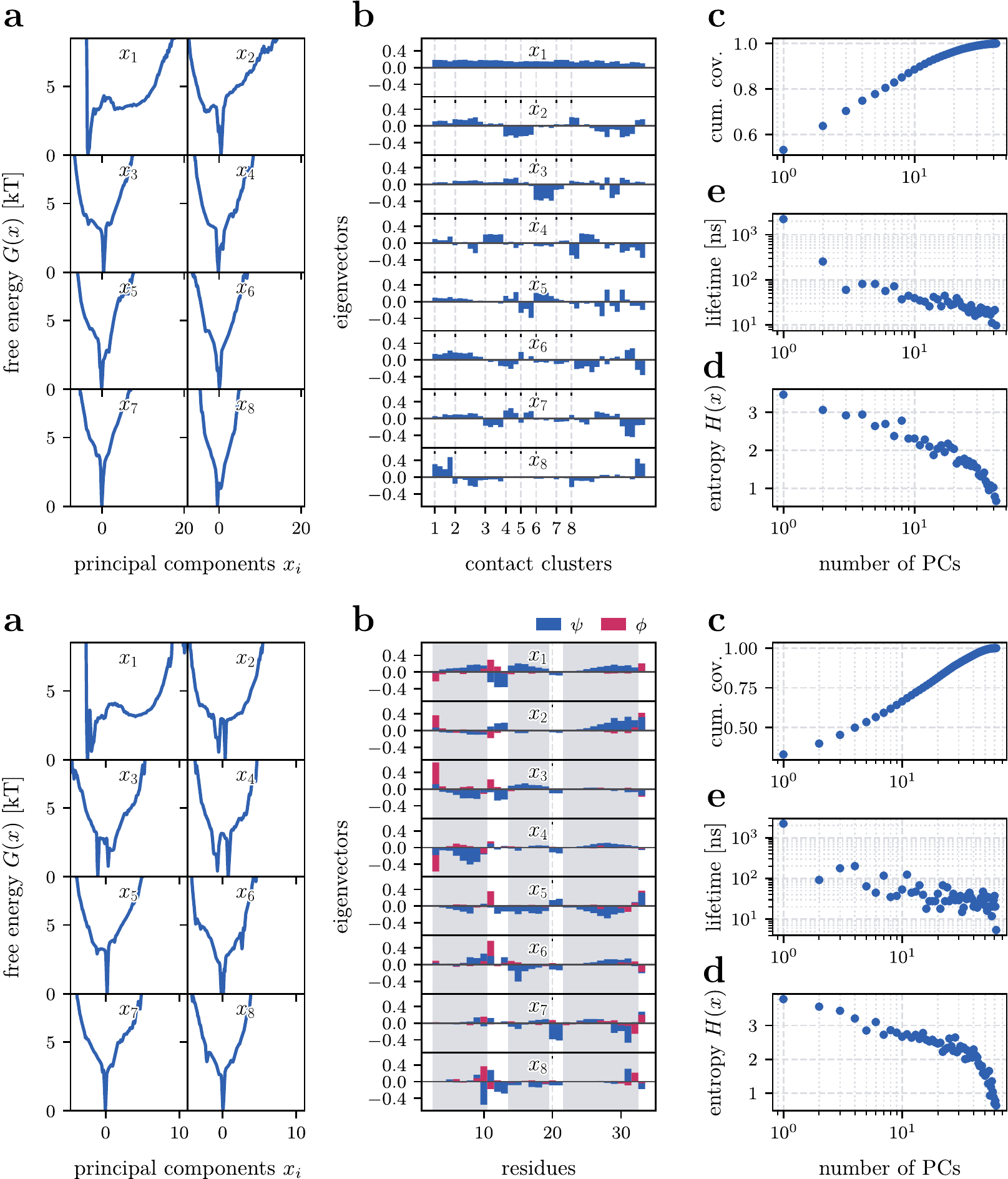}
    \caption{Selection of principal components $x_i$ for (top) contact
    distances, and (bottom) backbone dihedral angles. (a) Free energy
    curves, (b) corresponding eigenvectors, (c) cumulative covariance
    of the first $n$ principal components, and (d) lifetime $\tau:
    \text{ACF}(\tau) = e^{-1}$ of the principal components. (d) Shannon entropy
    $H(X) = - \sum_{x\in X} p(x) \ln p(x)$, which provides an information
    theoretic approach to rank the principal components by their importance.
    This approach is based on the fact that entropy is a measure to measure the
    amount of information contained in a component.}
    \label{SI:fig:pca}
\end{figure}

%%%%%%%%%%%%%%%%%%%%%%%%%
%%%% FIG S4 %%%%%%%%%%%%%
%%%%%%%%%%%%%%%%%%%%%%%%%
\begin{figure}[ht!]
    \centering
    \includegraphics{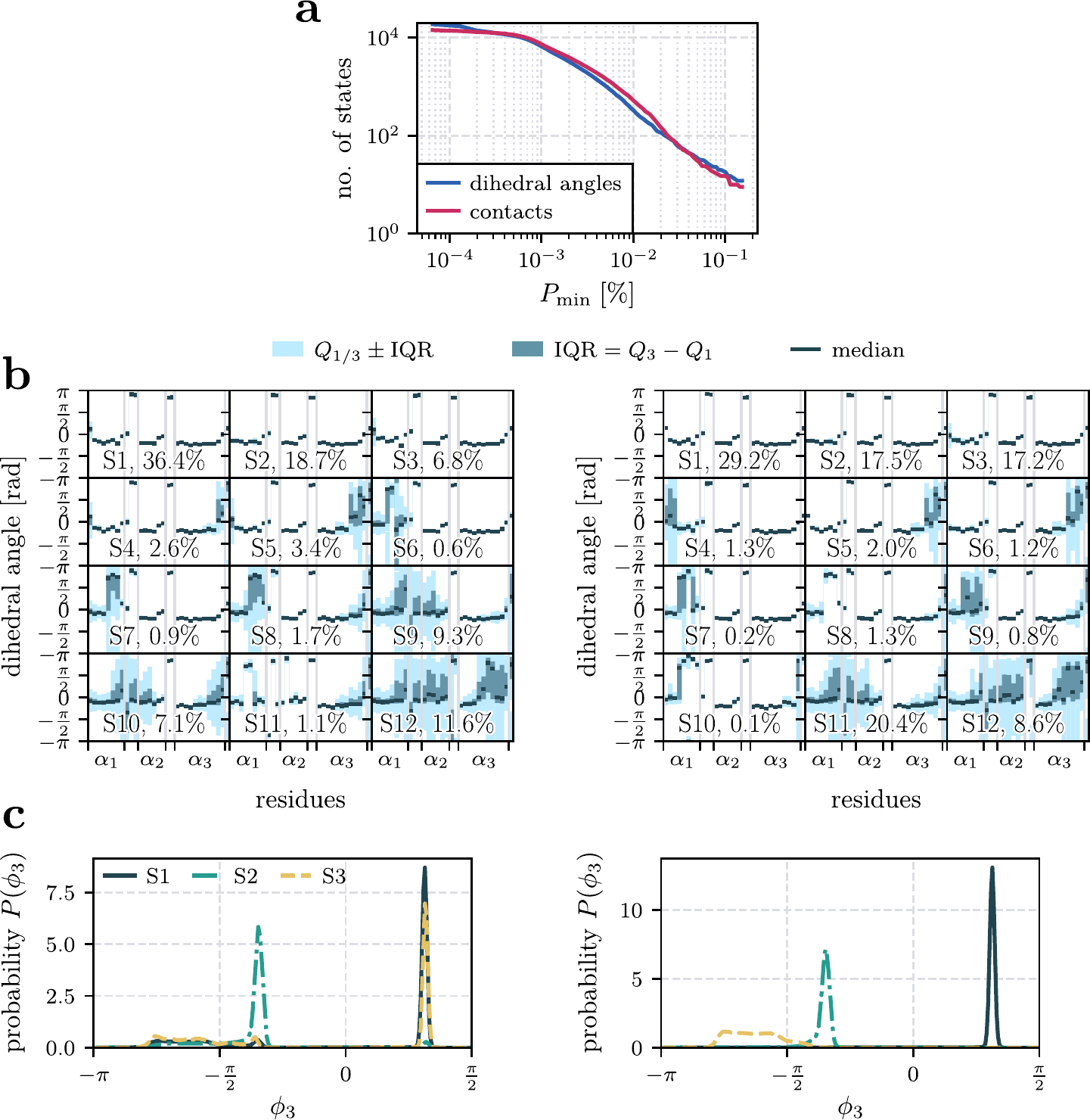}
    \caption{Further characterization of metastable states.
    (a) Resulting number of states for varying minimal population
    $P_\text{min}$, used for the density-based clustering algorithm.\cite{Sittel16}
    (b) Structural characterization of the twelve metastable states of HP35,
    obtained for (left) contacts and (right) dihedral angles. The states are
    ordered by decreasing fraction of native contacts $Q$.
    As for most residues the $\psi$ angles are more important than the
    corresponding $\phi$ (see Fig.~\ref{SI:fig:dihedral_preselection}), we
    restrict the representation to $\psi$ angles only for the sake of clarity.
    Since the $\psi$ angles exhibit a bimodal distribution (reflecting
    $\alpha\leftrightarrow\beta$) such that mean and variance do not well
    describe the data, we use a box-plot representation with quartiles $Q_i$
    (comprising the first $i \cdot 25\,\%$ of the data) that define the median
    $Q_2$, the interquartile range $\text{IQR}=Q_3- Q_1$ and the lower (upper)
    bound as the smallest (largest) data point in $Q_{1/3} \pm \text{IQR}$.
    (c) Comparing the $\phi_3$ distribution of the 3 major natives states based
    on (left) contacts and (right) dihedral angles.
    }
    \label{SI:fig:state_chracterization}
\end{figure}

%%%%%%%%%%%%%%%%%%%%%%%%%
%%%% FIG S5 %%%%%%%%%%%%%
%%%%%%%%%%%%%%%%%%%%%%%%%
\begin{figure}[ht!]
    \centering
    \includegraphics{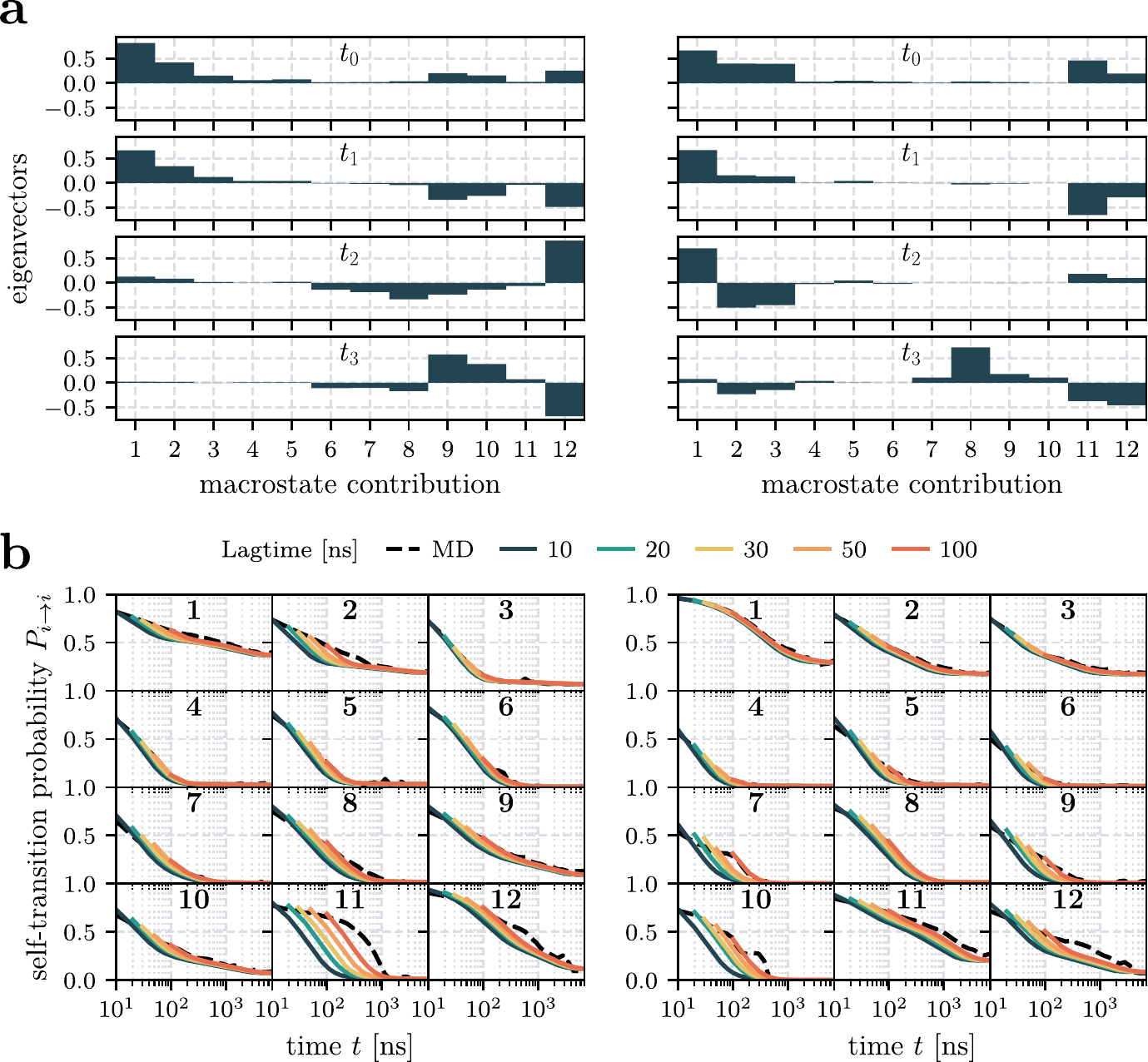}
    \caption{Further analysis and validation of the Markov state models,
    obtained by using the Hummer-Szabo projection.\cite{Hummer15}
    (a) Left eigenvector corresponding to the four slowest processes $t_i$ of
    the twelve-state Markov model, obtained for (left) contacts and (right)
    dihedral angles, where $t_0$ corresponds to the stationary process and
    $t_i$ with $i\ge1$ to the $i$-th implied timescale.
    (b) Chapman-Kolmogorov tests of the twelve states for varying lag times,
    obtained for (left) contacts and (right) dihedral angles.
    }
    \label{SI:fig:model_validation}
\end{figure}

%%%%%%%%%%%%%%%%%%%%%%%%%
%%%% FIG S6 %%%%%%%%%%%%%
%%%%%%%%%%%%%%%%%%%%%%%%%
\begin{figure}[ht!]
	\centering
	\includegraphics{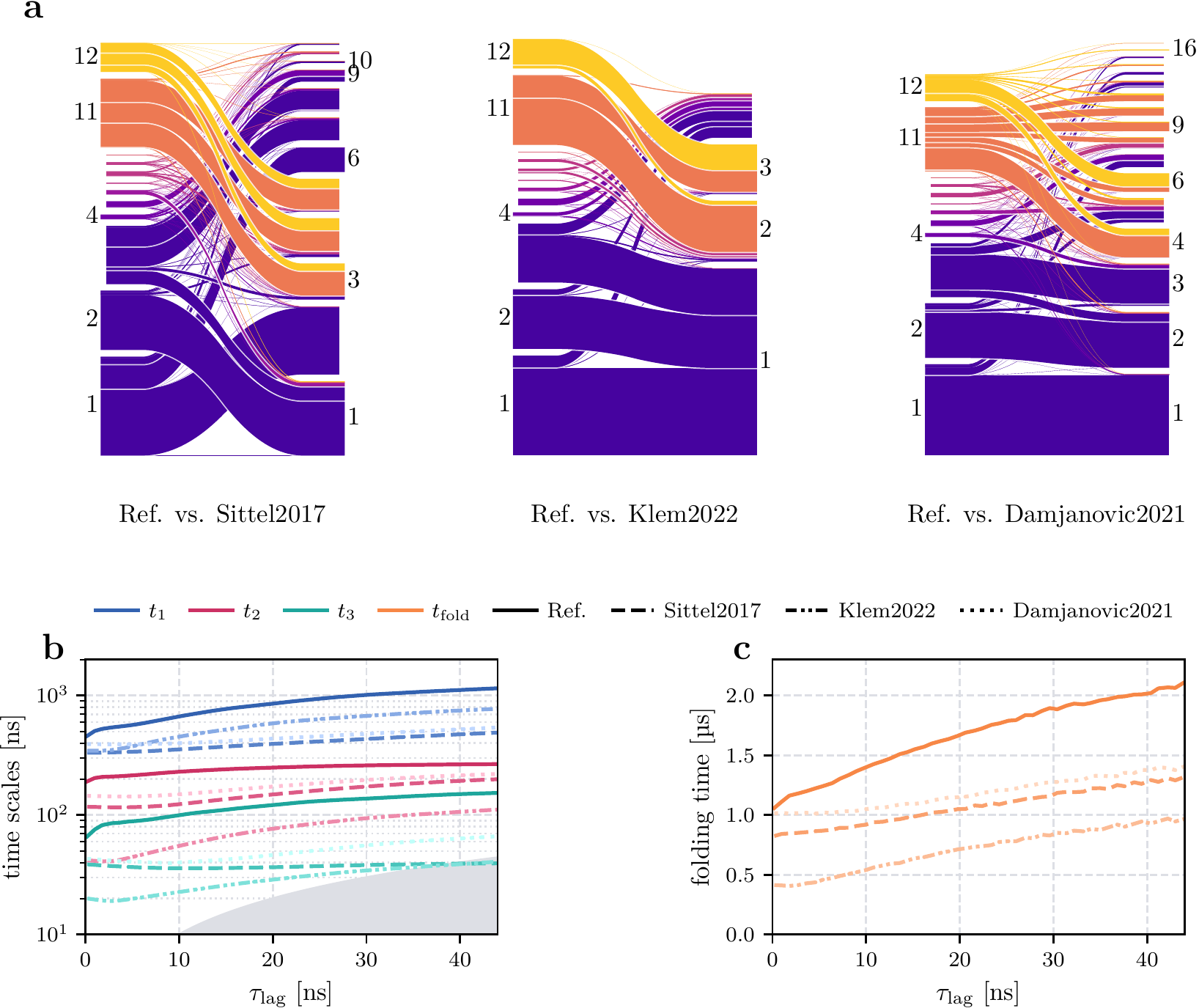}
	\caption{Comparison of our dihedral-based MSM (Ref) to previously
      existing models, including the work of
      Sittel2017,\cite{Sittel17} Klem2022,\cite{Klem22} and
      Damjanovic2021.\cite{Damjanovic21} (a) Sankey diagrams
      illustrating the relationship between the state partitioning of
      our MSM (left) and the state definitions of the other models
      (right). (b) First three implied timescales $t_n$ shown as a
      function of the lag time $\tau_\text{lag}$ for all models.  (c)
      Median folding times $t_\text{fold}$, that is the waiting time
      from the main unfolded state to the main native state, i.e.,
      transitions ($12\to1$) for our model, ($5\to2$) for Sittel2017,
      ($6\to1$) for Damjanovic2021, and ($3\to1$) for Klem2022.}
	\label{SI:fig:comparison}
      \end{figure}

%%%%%%%%%%%%%%%%%%%%%%%%%
%%%% FIG S7 %%%%%%%%%%%%%
%%%%%%%%%%%%%%%%%%%%%%%%%
\begin{figure}[ht!]
	\centering
	\includegraphics{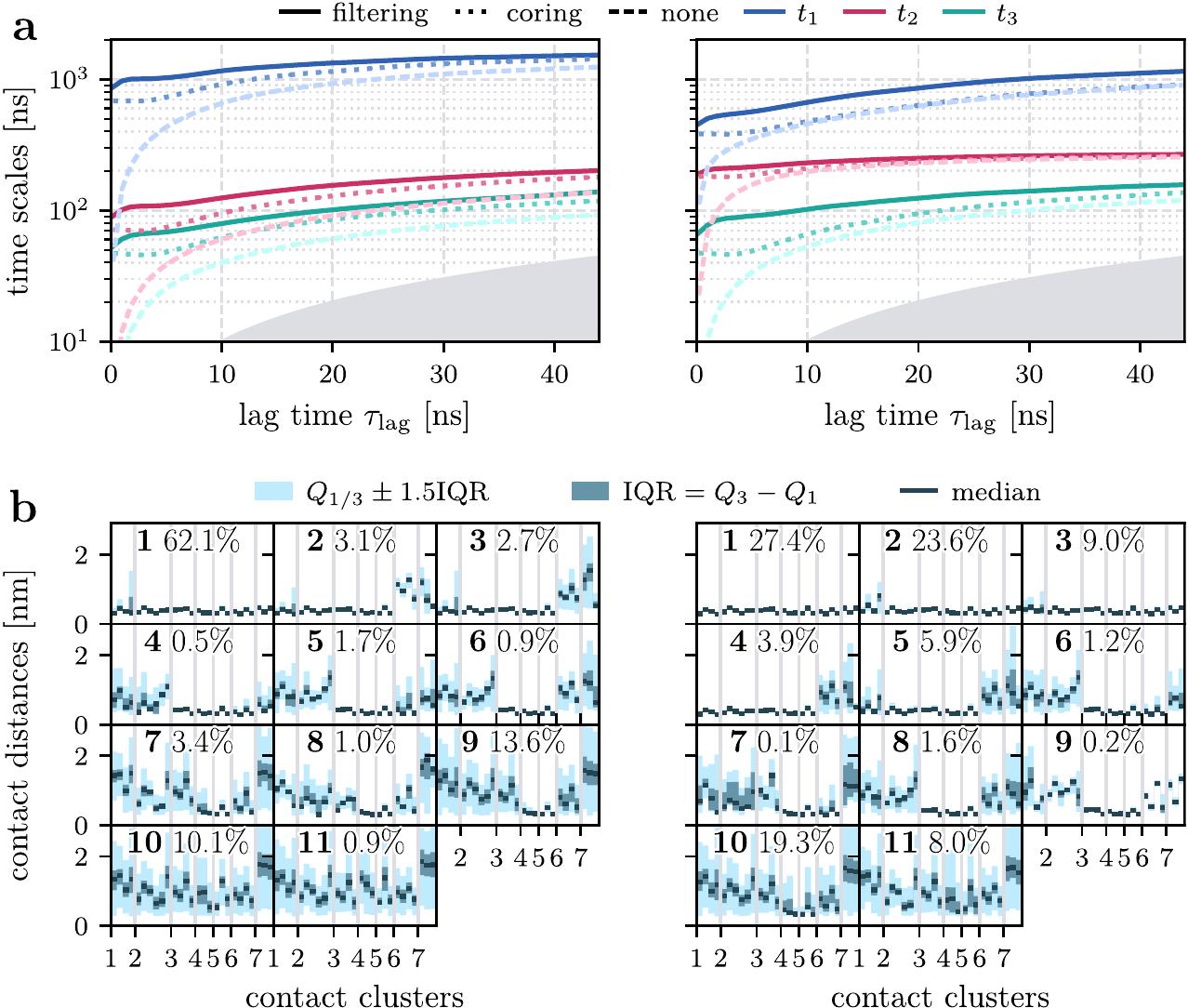}
	\caption{Effects of Gaussian filtering of the feature
          trajectory, compared to dynamic coring \cite{Nagel19} and
          using no dynamical correction. To this end, we repeated the
          MSM workflow described in the main paper (PCA, density-based
          clustering, and MPP), using the same parameters except for a
          lower minimum population of $P_\text{min}=15$, which ensures
          a comparable number of microstates when no dynamical
          correction is employed. In the case of dynamic coring, we
          used the previously determined \cite{Nagel19, Nagel20}
          coring time $\tau_\text{cor}=\SI{3}{\nano\second}$, which is
          expected to result in a similar effect as the Gaussian
          filtering with $\sigma = 2\,$ns. (a) Three slowest implied
          timescales $t_i$ of the various MSMs, obtained via the
          Hummer-Szabo projection\cite{Hummer15} for (left) contacts
          and (right) dihedral angles. (b) Structural characterization
          of the associated eleven metastable states
          obtained for dynamical coring. As in the main paper, the
          states are ordered by decreasing fraction of native contacts
          $Q$, and the contacts are ordered according to the seven
          main MoSAIC clusters (Fig.~\MainFigContactRep{}). For each
          state, the distribution of contact distances are represented
          by the median $Q_2$, the interquartile range
          $\text{IQR}=Q_3- Q_1$ and the lower (upper) bound as the
          smallest (largest) data point in
          $Q_{1/3} \pm 1.5 \,\mathrm{IQR}$.  }
	\label{SI:fig:gaussian_filtering}
\end{figure}

%%%%%%%%%%%%%%%%%%%%%%%%%
%%%% TAB S2 %%%%%%%%%%%%%
%%%%%%%%%%%%%%%%%%%%%%%%%
\clearpage
\begin{table}[hb!]
  \centering
  \caption{Timescales $t_{ij} = R_{ij}^{-1}$ (in units of ns) obtained from the
  transition rate matrix $R \equiv I - T$, based on the transition matrix $T$
  obtained via the Hummer-Szabo projection\cite{Hummer15} and a lag time of
  $\tau_\text{lag}=\SI{10}{\nano\second}$.
  }
    \label{SI:tab:rates}
    {\scriptsize
    \begin{tabular}{lrrrrrrrrrrrr}
      &&&&&&&&&&&&\\
      \multicolumn{13}{c}{{\small (a) contact-based states}}\\
      &&&&&&&&&&&&\\
      \toprule
      \multirow{2}*{state from} & \multicolumn{12}{c}{state to}\\
      & 1 & 2 & 3 & 4 & 5 & 6 & 7 & 8 & 9 & 10 & 11 & 12\\
      \midrule
1 & \textbf{56} & 85 & 282 & 874 & 902 & 22076 & 18179 & 128850 & 11973 & 43900 & 953654 & --\\
2 & 43 & \textbf{38} & 389 & 2730 & 4463 & 344049 & 178748 & -- & 12458 & 53762 & -- & 287691\\
3 & 53 & 140 & \textbf{37} & 2170 & 2543 & 7930 & 13857 & 13400 & 22701 & 39305 & 186616 & --\\
4 & 62 & 486 & 638 & \textbf{35} & 126 & -- & -- & 3421 & 8336 & 1363 & -- & --\\
5 & 85 & 620 & 1445 & 173 & \textbf{45} & 52520 & -- & 2718 & 8710 & 880 & 29755 & 1080\\
6 & 296 & -- & 720 & 19055 & -- & \textbf{58} & 153 & 210 & 1059 & -- & 1631 & 141069\\
7 & 476 & 1484 & 1478 & -- & -- & 238 & \textbf{34} & 63 & 171 & 2486 & 2376 & --\\
8 & 1964 & 23078 & 8785 & 2887 & 3004 & 638 & 115 & \textbf{50} & 379 & 231 & 959 & 3610\\
9 & 3005 & 4927 & 28548 & 21389 & -- & 8548 & 1838 & 2325 & \textbf{51} & 72 & 1048 & 320\\
10 & 11715 & 99758 & 43659 & 4587 & 1777 & 15544 & -- & 811 & 55 & \textbf{37} & 920 & 190\\
11 & -- & -- & 9886 & 57008 & 14118 & 4888 & 8349 & 647 & 136 & 116 & \textbf{51} & 569\\
12 & -- & 29706 & -- & -- & 3334 & -- & -- & -- & 394 & 309 & 6398 & \textbf{167}\\
    \bottomrule
      &&&&&&&&&&&&\\
      &&&&&&&&&&&&\\
      \multicolumn{13}{c}{{\small (b) dihedral-based states}}\\
      &&&&&&&&&&&&\\
    \toprule
      \multirow{2}*{state from} & \multicolumn{12}{c}{state to}\\
      & 1 & 2 & 3 & 4 & 5 & 6 & 7 & 8 & 9 & 10 & 11 & 12\\
      \midrule
1 & \textbf{275} & 1327 & 889 & 19718 & 590 & 5414249 & 200861 & 229587 & 305607 & -- & 54517 & --\\
2 & 810 & \textbf{49} & 55 & 3212 & 45726 & 1340 & 1878297 & -- & -- & -- & 12595 & --\\
3 & 517 & 54 & \textbf{40} & 495 & 7959 & 709 & 208208 & 29965 & 4314 & 669580 & 1339 & 15658\\
4 & 937 & 178 & 42 & \textbf{25} & 65713 & 1011 & 1661 & 1628 & 361 & -- & 177 & --\\
5 & 41 & -- & 710 & -- & \textbf{35} & 739 & -- & 32480 & 14780 & 665793 & 1597 & 3009\\
6 & 2381 & 96 & 50 & 5432 & 532 & \textbf{24} & -- & 4734 & 532 & 62074 & 231 & 469\\
7 & 29399 & 2182 & 621 & 290 & 17636 & -- & \textbf{26} & 40 & 219 & -- & 203 & --\\
8 & -- & 15254 & 2069 & 1416 & 158232 & 7699 & 282 & \textbf{55} & 195 & 420 & 131 & --\\
9 & 6535 & -- & 255 & 230 & 27738 & 399 & 416 & 110 & \textbf{29} & 2307 & 70 & --\\
10 & -- & -- & 2351 & -- & 48749 & 4588 & -- & 49 & 367 & \textbf{38} & 247 & --\\
11 & 32138 & 13107 & 1506 & 2657 & 12679 & 4060 & 16373 & 2234 & 1645 & 85612 & \textbf{87} & 112\\
12 & 417111 & -- & 9576 & -- & 44254 & 3063 & -- & -- & -- & -- & 47 & \textbf{48}\\
    \bottomrule
    \end{tabular}
    }

\end{table}

%%%%%%%%%%%%%%%%%%%%%%%%%
%%%% TAB S3 %%%%%%%%%%%%%
%%%%%%%%%%%%%%%%%%%%%%%%%
\begin{table}[!ht]
  \centering
  \caption{List of the main folding pathways assuming that we start in state~12 and end in state~1. Compared are (left) the paths found in the MD trajectory and (right) the most probable folding pathways identified by MSMPathfinder\cite{Nagel20} based on the transition matrix $T$ obtained via the Hummer-Szabo projection\cite{Hummer15} and a lag time of $\tau_\text{lag}=\SI{10}{\nano\second}$.}
{\scriptsize
	\centering
	{\small (a) contacts-based states}
	\begin{minipage}{.5\linewidth}
		\centering
	\begin{tabular}{lrr}

      	&&\\
	\toprule
		 MD Pathways & Freq & Length [\si{\micro\second}] \\
		 \midrule
		 (12, 10, 9, 1) & 6 & 3.25  \\
		 (12, 10, 5, 1) & 4 & 2.29 \\
		 (12, 9, 10, 4, 1) & 2 & 1.74  \\
		 (12, 5, 1) & 2 & 2.78  \\
		 (12, 10, 8, 1) & 2 & 0.75  \\
		 (12, 10, 1) & 2 & 1.97  \\
		 (12, 10, 9, 8, 1) & 2 & 2.91  \\
		 (12, 10, 8, 6, 1) & 2 & 4.86  \\
		 (12, 10, 11, 8, 1) & 1 & 0.66  \\
		 (12, 9, 10, 1) & 1 & 1.92 \\
		 (12, 10, 5, 4, 3, 1) & 1 & 1.95  \\
		 (12, 5, 2, 1) & 1 & 0.05 \\
		 (12, 10, 9, 8, 7, 1) & 1 & 0.78  \\
		 (12, 10, 7, 1) & 1 & 9.85  \\
		 (12, 9, 7, 1)& 1 & 3.18  \\
		 (12, 10, 9, 7, 1) & 1 & 0.14 \\
		 (12, 10, 8, 3, 1) & 1 & 2.83 \\
		 (12, 9, 8, 7, 1) & 1 & 3.94  \\
		 (12, 9, 10, 7, 8, 1) & 1 & 1.85  \\
		 (12, 9, 10, 5, 1) & 1 & 0.61  \\
		 \midrule
		Total events & 34 &  \\

	\bottomrule
	\end{tabular}
	\end{minipage}%
	\begin{minipage}{.5\linewidth}
		\centering
	\begin{tabular}{lrr}

      	&&\\
	\toprule
		  MSM Pathways & Prob [\%] & Length [\si{\micro\second}]\\
		 \midrule
		(12, 5, 1) & 9.13 & 1.58 \\
		(12, 9, 1) & 8.06 & 1.74 \\
		(12, 10, 9, 1) & 7.25 & 1.78 \\
		(12, 10, 5, 1) & 7.24 & 1.78 \\
		(12, 9, 2, 1) & 4.41 & 1.78 \\
		(12, 9, 10, 5, 1) & 4.18 & 1.83 \\
		(12, 10, 9, 2, 1) & 3.40 & 1.83 \\
		(12, 10, 4, 1) & 3.04 & 1.76 \\
		(12, 5, 4, 1) & 2.54 & 1.63 \\
		(12, 10, 5, 4, 1) & 2.01 & 1.81 \\
		(12, 10, 1) & 1.93 & 1.74 \\
		(12, 9, 10, 4, 1) & 1.75 & 1.83 \\
		(12, 9, 7, 1) & 1.45 & 1.84 \\
		(12, 10, 8, 7, 1) & 1.39 & 1.89 \\
		(12, 10, 9, 7, 1) & 1.30 & 1.88 \\
		(12, 9, 10, 5, 4, 1) & 1.16 & 1.87 \\
		(12, 5, 2, 1) & 1.14 & 1.63 \\
		(12, 9, 10, 1) & 1.12 & 1.78 \\
		(12, 10, 8, 1) & 1.06 & 1.84 \\
		(12, 10, 5, 2, 1) & 0.90 & 1.80 \\

		 \midrule
		Total probability & 64.46 & \\
	\bottomrule
	\end{tabular}
\end{minipage}}

\end{table}

\begin{table}[!ht]
	{\scriptsize
		\centering
		{\small (b) dihedrals-based states}
		\begin{minipage}{.5\linewidth}
			\centering
	\begin{tabular}{lrr}

		&&\\
		\toprule
		MD Pathways & Freq & Length [\si{\micro\second}] \\
		\midrule

		(12, 11, 3, 2, 1) & 7 & 4.19 \\
		(12, 11, 4, 3, 2, 1) & 6 & 2.78 \\
		(12, 11, 2, 1) & 4 & 2.80 \\
		(12, 11, 1) & 4 & 2.30 \\
		(12, 6, 2, 1) & 2 & 0.76 \\
		(12, 11, 6, 2, 1) & 2 & 0.46 \\
		(12, 11, 5, 1) & 2 & 7.12 \\
		(12, 3, 2, 4, 1) & 1 & 0.75 \\
		(12, 9, 3, 2, 1) & 1 & 1.21 \\
		(12, 11, 9, 3, 6, 2, 5, 1) & 1 & 4.33 \\
		(12, 3, 6, 5, 1) & 1 & 2.11 \\
		(12, 11, 8, 7, 9, 4, 3, 2, 1) & 1 & 0.76 \\
		(12, 11, 2, 5, 1) & 1 & 3.73 \\
		(12, 11, 8, 9, 3, 2, 1) & 1 & 5.17 \\
		(12, 11, 6, 3, 4, 2, 1) & 1 & 0.08 \\
		(12, 11, 6, 5, 1) & 1 & 4.70 \\

		\midrule
		Total events & 36 &\\
		\bottomrule
\end{tabular}
	\end{minipage}%
\begin{minipage}{.5\linewidth}
	\centering
\begin{tabular}{lrr}
	&&\\
	\toprule
	MSM Pathways & Prob [\%] & Length [\si{\micro\second}]\\
	\midrule

	(12, 11, 3, 1) & 22.40 & 2.02 \\
	(12, 11, 3, 2, 1) & 12.96 & 2.07 \\
	(12, 11, 4, 3, 1) & 7.61 & 2.06 \\
	(12, 11, 5, 1) & 6.35 & 1.69 \\
	(12, 11, 4, 3, 2, 1) & 4.40 & 2.11 \\
	(12, 11, 6, 3, 1) & 4.09 & 2.04 \\
	(12, 11, 1) & 2.87 & 1.65 \\
	(12, 11, 9, 3, 1) & 2.67 & 2.07 \\
	(12, 11, 6, 3, 2, 1) & 2.35 & 2.10 \\
	(12, 11, 2, 3, 1) & 2.31 & 2.08 \\
	(12, 11, 6, 2, 3, 1) & 1.89 & 2.09 \\
	(12, 11, 2, 1) & 1.87 & 2.05 \\
	(12, 11, 9, 4, 3, 1) & 1.77 & 2.09 \\
	(12, 11, 4, 2, 3, 1) & 1.61 & 2.11 \\
	(12, 11, 6, 2, 1) & 1.56 & 2.05 \\

	\midrule
	Total probability & 76.71& \\
	\bottomrule
\end{tabular}	\end{minipage}
}
\end{table}

\clearpage
\bibliography{\dir/stock, \dir/md, ../new}